\documentclass[11pt]{article}
\usepackage{jheppub}

\usepackage{epsfig}
\usepackage{amssymb}
\usepackage{amsmath}
\usepackage{amsfonts}

\newcommand{\be}{\begin{equation}}
\newcommand{\ee}{\end{equation}}
\newcommand{\bea}{\begin{eqnarray}}
\newcommand{\eea}{\end{eqnarray}}
\newcommand{\bi}{\begin{itemize}}
\newcommand{\ei}{\end{itemize}}
\newcommand{\ben}{\begin{enumerate}}
\newcommand{\een}{\end{enumerate}}

\newcommand{\ep}{\epsilon}
\newcommand{\eps}{\epsilon}

\newcommand{\nn}{\nonumber}

\def\gsim{\mathrel{\rlap{\lower4pt\hbox{\hskip1pt$\sim$}}
    \raise1pt\hbox{$>$}}}         %greater than or approx. symbol
\def\lsim{\mathrel{\rlap{\lower4pt\hbox{\hskip1pt$\sim$}}
    \raise1pt\hbox{$<$}}}         %less than or approx. symbol

\def \ep {\epsilon}

\begin{document}

\hfill{\today}

\newcommand{\picturepage}[2]{
\begin{minipage}{1cm}
      \includegraphics[width=20mm, height=20mm]{#1}
    \end{minipage} \quad &  
        \begin{minipage}[c]{15cm}    
            #2 
        \end{minipage}
   }

\title{Two-loop planar master integrals for the production of off-shell 
vector bosons  in hadron collisions
}

\author[1]{Johannes M. Henn,}
\author[2]{Kirill Melnikov}
\author[3]{and Vladimir A. Smirnov}

\affiliation[1]{Institute for Advanced Study, Princeton, NJ 08540, USA, USA}
\affiliation[2]{Department of Physics and Astronomy, Johns Hopkins University, Baltimore, USA}
\affiliation[3]{
Skobeltsyn Institute of Nuclear Physics of Moscow State University, 
119991 Moscow, Russia}

\emailAdd{jmhenn@ias.edu}
\emailAdd{melnikov@pha.jhu.edu}
\emailAdd{smirnov@theory.sinp.msu.ru}

\abstract{We describe the calculation  of all planar master integrals that are needed  for the computation of NNLO QCD 
corrections to the production of two off-shell vector bosons in hadron collisions.  The most complicated representatives 
of integrals in this class  are the two-loop four-point functions  where two 
external lines are on the light-cone and 
two other external lines have different invariant masses.  We compute these and other 
relevant integrals analytically using differential equations 
in external kinematic variables and express our results in terms of Goncharov polylogarithms.  
The case of two equal off-shellnesses,  recently considered in Ref.~\cite{tancredi}, appears 
as a particular case of our general solution.
}

\maketitle

\section{Introduction} 

Production of pairs of vector bosons in hadron collisions is an important process that is used 
by ATLAS and CMS collaborations 
to study QCD dynamics,  understand fine details of electroweak interactions  and 
validate Monte Carlo event generators 
that are employed for estimating backgrounds in searches for physics beyond 
the Standard Model \cite{atlas,cms}. 
For this reason, high-quality  theoretical predictions for  these processes 
are warranted.  Currently, the theoretical description 
of $pp \to V_1 V_2 $ processes  includes next-to-leading order (NLO) QCD corrections
\cite{dixon1,dixon2}, electroweak corrections \cite{bier}, threshold resummation \cite{thr} 
and consistent  matching of these processes to parton showers 
\cite{shower}.
Upgrading theoretical predictions for vector boson pair production to next-to-next-to-leading order  (NNLO) in 
perturbative QCD, represents  a natural step towards an even better 
understanding of these processes.    To show how such  an improved understanding 
may be helpful, we describe three concrete examples 
where further advances  in  theory predictions for vector boson production  
are  extremely valuable.  

The first one  is related to persistent and significant discrepancies between 
theoretical predictions and measured cross-sections and kinematic distributions for 
$pp \to W^+W^-$ production, observed 
both at $7$ and at $8~{\rm TeV}$ LHC by ATLAS and CMS collaborations \cite{atlas,cms}. 
It is important to compute NNLO QCD corrections 
to this process in order to exclude them once and for all as a potential reason for 
that  discrepancy.  It is also important to explore other vector boson production processes, 
such as $ZZ$ and $ZW$. In case of the latter one, calculation of  NNLO QCD 
virtual corrections requires dealing with the situation where two vector bosons have close, but 
different, masses. 

The second example is related to precise measurements of the Higgs coupling to electroweak bosons at the LHC. 
Such measurements,
important for understanding the mechanism of electroweak symmetry breaking, 
 require good control of  
 backgrounds from continuous vector boson production, a particularly 
pressing issue in case of $pp \to W^+W^- \to l_1^+l_2^- \nu_1 \bar \nu_2$ since in this case  
$W$-bosons  can not be fully reconstructed. 
NNLO QCD predictions for $q \bar q \to VV^*$, where one vector boson is on the mass-shell and 
the other one is off the mass-shell,   will be extremely helpful for this purpose. 

To explain the  third example, we remind the reader about the 
 recent suggestion to measure the Higgs boson width at the LHC,  by counting the number of 
$ZZ$ events above the $2 m_Z$ threshold \cite{Caola:2013yja} (see also \cite{Campbell:2013una}). 
It is estimated \cite{Caola:2013yja,Campbell:2013una}
that the Higgs bosons width as small as ten to twenty times its Standard Model value 
can be probed.  However, since 
this is a counting experiment,  an accurate prediction for all processes that produce pairs 
of $Z$-bosons at high invariant mass is crucial. 
The challenge therefore is to compute  $q \bar q \to ZZ$, $gg \to ZZ$ as well as the 
interference of $gg \to H^* \to ZZ$ and $ gg \to ZZ$ amplitudes to the highest precision possible, 
to facilitate the model-independent measurement of the Higgs boson width at the LHC. 

Having argued that extending theoretical description of  vector boson pair production to
NNLO QCD is important, we note that 
computing NNLO QCD corrections to hadron collider processes in general is difficult for several reasons. 
A practical framework for such computations did not exist until very recently, but it appears that, after almost ten years of 
research, we   finally have it.   Indeed, as  recent  NNLO QCD results for 
$pp \to 2j$ \cite{Ridder:2013mf,Currie:2013dwa}, 
$pp \to t \bar t$ \cite{Czakon:2013goa,Baernreuther:2012ws} 
and 
$pp \to H+j$ \cite{Boughezal:2013uia}
show, we now understand quite  well how to combine  infra-red divergent virtual and real emission corrections 
to arrive  at physical  results.  
The main bottleneck in extending available  NNLO QCD predictions to  other, more complex, 
processes is the lack of known two-loop {\it virtual}  amplitudes.  Indeed, absence of 
two-loop scattering amplitudes  for $q\bar q \to VV$ and $gg \to VV$ 
is the {\it only} reason why no 
NNLO QCD predictions  are  available for $pp \to ZZ$ and  $pp \to W^+W^-$, both on- and  off- the mass-shell. 

The standard modern technology for multi-loop computations consists of   
three primary steps: re-writing scattering amplitudes 
through a minimal set of tensor integrals, 
reduction of this set to a few master integrals using integration-by-parts 
identities \cite{ibp} and, finally, computation of the master integrals.  For a long time, 
the computation of the master integrals could have been  considered to be  
the least-understood part of this process since very often it  is performed on a case-by-case basis.   A relatively 
systematic way to study master integrals is provided by the differential equations in external kinematic variables 
that can easily be derived \cite{kotikov,remiddi}  using integration-by-parts identities.
However,  while the differential equation method  was  applied 
to a large number  of various master integrals (see, e.g., \cite{caffo,gr}), 
its  systematic applicability 
for  finding master integrals that depend on a {\it large } number of 
kinematic variables was not always clear. 

 Recently, it was suggested \cite{jhenn}  that, for a generic 
multi-loop problem, a choice of master integrals can be made that 
transforms  differential equations in such a way that their   iterative solution  in dimensional 
regularization parameter $\epsilon = (4-D)/2$  becomes straightforward.  While 
this conjecture was never proven in full generality, the technique of Ref.~\cite{jhenn}  
was successfully applied to compute highly non-trivial Feynman  integrals~\cite{Henn:2013tua,Henn:2013woa,Henn:2013nsa}, 
suggesting  its tremendous utility for practical computations. 
In this paper we will use this technique to compute all {\it planar} master 
integrals for $ q \bar q \to V_1 V_2 $ and $gg \to V_1 V_2$ processes, where $V_{1,2}$ stands for  
vector bosons with 
different invariant masses.  We will show that all integrals that belong to this class 
can be computed in a streamlined manner using the technique  of Ref.~\cite{jhenn}. 

Before proceeding to the main body of the paper, we will comment on related  results 
for two-loop four-point integrals with all internal particles massless, that are 
available in the literature.  The two-loop four-point functions 
with all, or all but one, external particles on the light cone are known since long ago
\cite{Smirnov:1999gc,Smirnov:1999wz,Tausk:1999vh,Anastasiou:2000mf,gr1,gr2}.
Recently, these results were extended to the case where 
two external particles have equal invariant masses \cite{tancredi}.   The calculation 
reported in Ref.~\cite{tancredi} is the limiting case of the general results that we 
report here and we use it extensively to cross-check our calculation.  Finally, very recently 
some master integrals that belong to the same class that we consider in this paper were computed in 
Ref.~\cite{papa} using a variant of the differential equation method.  We did not compare our results 
with that reference since results presented in Ref.~\cite{papa} are for  unphysical Euclidean kinematics 
while we compute those integrals directly in the physical region. 

The remainder of the paper is organized as follows. In the next Section,
 we introduce  our notation and explain the basic strategy.  In Section \ref{sec:diffeqs}
we discuss the differential equations and point out their general properties 
that are used later.  In Section \ref{sec:solution} we explain how we constructed 
the analytic solutions  of these differential equations  in terms of multiple polylogarithms 
in the physical region. 
In Section \ref{sec:boundary} we explain how boundary conditions in the physical region 
were computed. 
In Section \ref{sec:analytic} we point out a simple way to perform the analytic continuation 
for a certain class of integrals relevant for our analysis. 
In Section \ref{sec:masters}, we  list all the master integrals and give their  boundary asymptotic behaviour
in the physical region. 
In Section \ref{sec:checks} we describe checks of our results. 
We conclude in Section \ref{sec:concl}.
Finally, in attached files, we give  matrices that are needed to construct the differential equations 
for our basis of master integrals and the analytic results  for all the planar two-loop four-point 
integrals in terms of Goncharov polylogarithms. 

\section{Notation} 
\label{sec:notation}

We consider two-loop QCD corrections to  the
process $q(q_1) \bar q(q_2) \to V^*(q_3) V^*(q_4)$.  The four-momenta 
of external particles satisfy $q_1^2 =0, q_2^2 = 0$ and $q_3^2 = M_3^2$, $q_4^2 = M_4^2$. 
The Mandelstam invariants are\footnote{We use Mandelstam variables written with capital letters to refer 
to the physical process. Later, we will use Mandelstam variables for families of integrals; those we will write 
with small letters.} 
\be
S = (q_1+q_2)^2 = (q_3 + q_4)^2,\;\;\; T = (q_1 - q_3)^2 = (q_2 - q_4)^2,\;\;\; U = (q_1 - q_4)^2 = (q_2 - q_3)^2;
\label{eq_man}
\ee
they satisfy  the standard constraint $S + T + U = M_3^2 + M_4^2$.   The physical values of these kinematic variables are 
$M_3^2 > 0, M_4^2 > 0$,   $S > (M_3 + M_4)^2$, $ T < 0$ and $U < 0$.  Further constraints on these variables can be derived 
by considering the center-of-mass frame of colliding partons and  expressing  the transverse 
momentum of each of the vector bosons $\vec q_\perp$ through $T$ and $U$ variables. We find
\be
\vec q_\perp^{~~2} = \frac{( T U - M_3^2 M_4^2)}{S}.
\ee
In addition, the square of the three-momentum of each of the vector bosons in the center-of-mass frame reads 
\be
\vec q\;^2 = \frac{S^2 - 2 S (M_3^2 + M_4^2) + (M_3^2 - M_4^2)^2}{4 S} \,.
\label{eqp}
\ee
The constraints on $T$ and $U$ for given $S, M_3^2, M_4^2$ follow from the obvious inequalities 
\be
 0 \le \vec q_\perp^{~~2} \le \vec q\;^2.
\ee

In general, the complete 
kinematics of the process is defined by four variables that we take to be $S$, $T$, $M_3^2$ and 
$M_4^2$.  However, the dependence on one of these variables is redundant, 
since any Feynman integral can be written as a function 
of three dimensionless ratios of these variables and an overall factor that 
is fully fixed by the mass dimension of an integral.  For all planar integrals we choose the following parametrization 
\be
\frac{S}{M_3^2} = (1+x) ( 1 + x y),\;\;\; \frac{T}{M_3^2} = -xz,\;\;\; \frac{M_4^2}{M_3^2}  = x^2 y.
\label{eq_xyz}
\ee
This parametrization is motivated  by the appearance of a complicated square root 
in expressions for master integrals\footnote{These square roots 
are proportional to a relative three-momentum  of the vector bosons, c.f. Eq.~(\ref{eqp}).} 
that  becomes a simple rational function when expressed in these variables 
\be
\sqrt{ S^2 - 2 S (M_3^2 + M_4^2) + (M_3^2 - M_4^2)^2}  = M_3^2 x (1-y).
\ee
As we will see in the next Section, 
once we rationalize  the square root, the solution of a system of differential equations 
is easily achieved using Goncharov polylogarithms. 
We note that in terms of the variables $x,y,z$, the physical region corresponds to 
\be
x > 0,\;\;\; y > 0,\;\;\;\; y < z < 1. 
\ee

\begin{figure}[tbp]
  \centering
  \includegraphics[width=0.5\textwidth]{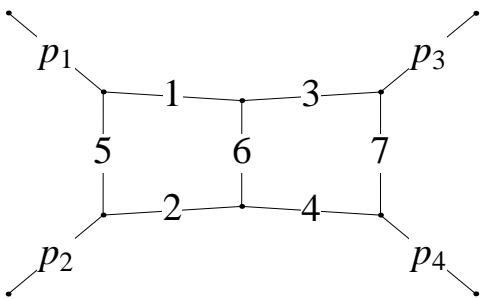}
  \caption{Double box graph. The numbering of the internal lines corresponds to the notation used in Eqs.~(\ref{eq_g}), (\ref{defpropagators}). The ingoing external momenta satisfy $\sum_i {p_i^\mu}=0$. 
  Different choices of on-shell conditions for them define the three planar integral families considered in the main text.}
  \label{figuredoublebox}
\end{figure}

All planar two-loop diagrams that are required for the production 
of two off-shell vector bosons
can be   described by a 
single meta-graph shown in Figure \ref{figuredoublebox}.  Three mappings, that define three 
distinct families of integrals, need to be considered:
\begin{enumerate} 
\item family P12 :  $ p_1 = -q_3,\; p_2 = -q_4,\;  p_3 = q_1,\;  p_4 = q_2$;
\item family P13: $ p_1 = -q_3,\;  p_2 = q_1,\;  p_3 = -q_4,\;  p_4 = q_2$;
\item family P23:  $ p_1 =  q_2,\ p_2 = -q_4,\; p_3 = -q_3,\; p_4 = q_1$.
\end{enumerate} 

For each of these families, we define a set of integrals that is closed under the application of 
integration-by-parts  identities. Specifically, 
\be 
%G(\{a_i\}) = 
G_{a_1,\ldots,a_9}=
\int \frac{{\rm d}^D k_1}{ i \pi^{D/2} } 
\frac{{\rm d}^D k_2}{ i \pi^{D/2} }
\frac{1}{[1]^{a_1} [2]^{a_2} [3]^{a_3} [4]^{a_4} [5]^{a_5} [6]^{a_6} [7]^{a_7} [8]^{a_8} [9]^{a_9}},
\label{eq_g}
\ee
and 
\be
\begin{split} \label{defpropagators}
& [1] = -k_1^2,\;\;\; [2] = -(k_1+p_1+p_2)^2,\;\;\; [3] = -k_2^2,\; \;\; \\
& [4] = -(k_2+p_1+p_2)^2,\;
[5] = -(k_1+p_1)^2, \;\;\; [6] = -(k_1 -k_2)^2, \\
& [7] = -(k_2-p_3)^2,\;\;\;\; [8] = -(k_2 + p_1)^2,\;\;\;\;\; [9] = -(k_1-p_3)^2.
\end{split} 
\ee
Here, the exponents can take any integer values,  with the restriction that $a_8 \le 0$ and $a_9\le 0$.
These factors are  used to represent irreducible numerators. For each of the three families, 
integration-by-parts identities can be used to express all the integrals of that type to a minimal 
set of (master) integrals. 
Our choice of master integrals can be found  in Section~\ref{sec:masters}. 
These master integrals satisfy differential equations in the external 
kinematic variables. In the next Section we discuss how such systems of equations can be solved.

\section{Differential equations} 
\label{sec:diffeqs}

In this Section we discuss how the master integrals can be calculated. 
To this end, we derive  systems of differential equations for each of the above families. 
This is a relatively standard procedure, see e.g. \cite{kotikov,remiddi} and we do not discuss 
it further. When deriving differential equations we performed a reduction to master integrals
using {\tt FIRE} \cite{Smirnov:2008iw,Smirnov:2013dia}.
We choose all master integrals to be dimensionless, such that they depend only
on the three variables $x,y,z$, and obtain 
\be
\partial_\xi \vec f = \epsilon {A}_\xi \vec{f},
\label{eq_f}
\ee
where $\xi = x,y$ or $z$ and $\vec{f}$ is a vector of master integrals. 
The matrices ${\tilde A}_\xi$ contain simple rational functions.
They satisfy the integrability conditions
\be
\left (  \partial_\xi \partial _\eta - \partial_\eta \partial_\xi \right ) \vec f =0 \;\;\;\;
\Rightarrow\;\;\;\;\; 
\partial_{\xi} A_{\eta} - \partial_{\eta} A_{\xi} = 0\,,\qquad [A_{\eta} , A_{\xi} ] = 0\,,
\ee
for $\xi,\eta \in \{x,y,z\}$.
The structure of the equations can be further clarified by writing them in the
combined form
\be \label{DEdifferentialform}
d\, \vec{f}(x,y,z;\eps) = \epsilon\,  d \, \tilde{A}(x,y,z) \, \vec{f}(x,y,z;\eps) \,,
\ee
where the differential $d$ acts on $x,y$ and $z$.
%, of the form  
%$1/(x_\xi - a)$ that, taken together, form  an alphabet.
For our choice of master integrals (see Section~\ref{sec:masters}), 
the matrix $\tilde{A}$ can be written in the following way
\begin{align}\label{Atilde}
\tilde{A}  = \sum_{i=1}^{15}  \tilde{A}_{\alpha_i} \, \log(\alpha_{i} ) \,,
\end{align}
where the $ \tilde{A}_{\alpha_i}$ are {\it constant} matrices, and 
the arguments of the logarithms $\alpha_{i}$, called {\it letters}, are simple
functions of $x,y,z$. We find 
\bea\label{alphabet}
& \alpha = \{ x, y, z, 1 + x, 1 - y, 1 - z,  1 + x y, z-y,
1 + y(1 + x) - z,  x y + z, \nonumber \\
&
 1 + x(1 +  y - z), 1 + x z,  1+y-z, z + x (z - y) + x y z ,   z-y + y z + x y z \}.
\label{eq_alphabet}
\eea
We call Eq.~(\ref{alphabet}) the {\it alphabet} relevant to the functions $\vec{f}$.
For example, in case of family~P12,  the first twelve of these letters are required.
Eq.~(\ref{DEdifferentialform}) makes it manifest that the analytic solution, to all orders in the $\eps$ expansion, can be written in terms of multiple polylogarithms defined by the alphabet (\ref{alphabet}).
In general, the solution to Eq.~(\ref{DEdifferentialform}) can be written in the elegant form
\be\label{solution_chen}
 \vec{f}(x,y,z;\eps)  = {\mathbb P} e^{\eps \int_{\mathcal{C}} d \tilde{A} } \vec{f}_0(\eps)  \,,
\ee
where $ {\mathbb P} $ refers to path ordering of the matrix exponential, and the
integrals are Chen iterated integrals \cite{Chen:1977oja} along the contour $\mathcal{C}$ in the space of kinematical variables $x,y,z$. The vector 
$ \vec{f}_0(\eps)$ represents the boundary value at the base point of the contour $\mathcal{C}$.
Eq.~(\ref{solution_chen}) is to be understood as a series expansion for small $\eps$.
The homotopy invariance of (\ref{solution_chen}) allows for many equivalent representations
of the same functions, corresponding to different choices and parametrizations 
of $\mathcal{C}$.\footnote{For a recent example in the context of Bhabha scattering, 
see Ref.~\cite{Henn:2013woa}.}
For this reason Eq.~(\ref{solution_chen}) is probably the most compact and invariant representation
of the functions $\vec{f}$. However, for practical applications, we find it convenient to make a specific choice of the integration  contour.

Indeed, the linearity of the alphabet $(\ref{alphabet})$ allows us to write a simple representation of $\vec{f}$ in terms of multiple polylogarithms. This can be thought of as a specific choice 
of the contour $\mathcal{C}$. Another 
way to arrive at such a solution is to integrate Eqs.~(\ref{eq_f}) 
over one variable at a time.
In the next Section, we will discuss this in more detail. 

Note that singular points of the differential equations (\ref{DEdifferentialform}) can be read off from
the alphabet (\ref{alphabet}). They correspond to special kinematic points  such as 
singular limits, threshold or pseudo-threshold configurations of the multivalued functions $\vec{f}$.
A useful feature of the differential equations is that they allow one to easily determine 
the behavior of $\vec{f}$ close to singular points, and this is helpful in determining 
the boundary conditions \cite{Henn:2013nsa}. A practical example of how this is done 
can be found in  Section~\ref{sec:boundary}.

Finally, we wish to point out that the letters in Eq.~(\ref{eq_alphabet}) all have a 
definite sign in the physical regions. This means that all iterated integrals needed 
for calculating $\vec{f}$ can be written in a manifestly real way, and imaginary parts appear 
only through explicit factors of $i$. The latter come from the boundary 
conditions in the physical region.  

\section{Solution in terms of multiple polylogarithms} 
\label{sec:solution}

The vector of master integrals $\vec f$ can be expanded in powers of $\epsilon$, 
\be
\vec f = \sum \limits_{i=0}^{4} \vec f^{(i)} \epsilon^i +  {\mathcal{O}}(\eps^5).
\ee
To construct a solution of the differential equation, we need to iteratively solve Eq.~(\ref{eq_f}) order-by-order 
in dimensional-regularization parameter $\epsilon$.  Suppose the solution is constructed up to $i = n-1$. The set of differential 
equations for $\vec f^{(n)}$ is then 
\be
\partial_x \vec f^{(n)} = A_x \vec{f}^{(n-1)}, \;\;\; \partial_y  \vec{f}^{(n)} =  A_y \vec{f}^{(n-1)},
\;\;\; \partial_z  \vec{f}^{(n)} =  A_z \vec{f}^{(n-1)}.
\label{eq_de}
\ee
To find $\vec f^{(n)}$, we integrate the first equation over $x$; this determines 
the solution up to a function of $y,z$
\be
\vec{f}^{(n)}(x,y,z) = \vec{h}^{(n)}(y,z) + 
\int \limits_{0}^{x} {\rm d} \bar x A_x(\bar x, y,  z)  \vec{f}^{(n-1)}(\bar x, y,  z).
\label{eqx}
\ee
It follows from Eqs.~(\ref{DEdifferentialform}),(\ref{Atilde}), and (\ref{alphabet})
that the integration kernels appearing on the right-hand side of Eq.~(\ref{eqx}) 
only contain terms of the form $d \bar{x}/(\bar{x} - a)$, for some $a$'s.
Therefore, the integration over $\bar{x}$ can be performed systematically provided that 
$\vec f^{(n-1)}$ is  written in terms of Goncharov polylogarithms 
\be
G(a_n,a_{n-1},....a_1,t) = \int \limits_{0}^{t} \frac{{\rm d} t_n }{ t_n -a_n }
G(a_{n-1},....a_1,t_n). 
\label{eq_gn}
\ee
For the simplicity of integration, 
it is important to keep the same order of integration, e.g. always start with $x$, for all the integrals
that contribute to the vector $\vec f$.  If this is not done consistently -- so that integration variables 
also appear in indices of Goncharov polylogarithms in addition to their  arguments --
one has  to use various  identities between Goncharov polylogarithm 
to remedy this situation and enable the 
integration  as in Eq.~(\ref{eq_gn}). 
Substituting the  solution in Eq.~(\ref{eqx}) into the second term in Eq.~(\ref{eq_de}), we find the  
differential equation for the function $\vec{h}^{(n)}(y,z)$
\be
\partial_y \vec{h}^{(n)}(y,z) = B_y \vec{h}^{(n-1)}(y,z),
\ee 
where $B_y$ is a  matrix related to the original matrix $\tilde A$ in a non-trivial way.
Note, however,  that this equation can {\it only} depend on the elements of the alphabet that are independent of $x$; this provides a non-trivial 
check of the consistency of reconstructed solutions. Integrating this equation over $y$, we find 
\be
\vec{h}^{(n)}(y,z) = \vec{g}^{(n)}(z) + 
\int \limits_{0}^{y} {\rm d} \bar y B_y(\bar y,  z)  \vec{h}^{(n-1)}(\bar y, z),
\label{eqy}
\ee
where $\vec{g}^{(n)}(z)$ is an arbitrary function of a single variable $z$. Substituting  Eq.~(\ref{eqx}) with $\vec{h}^{(n)}(y,z)$ from 
Eq.~(\ref{eqy}) into the third equation in Eq.~(\ref{eq_de}), we find a differential equation for $\vec{g}^{(n)}(z)$ that is independent 
of $y$ and $x$ 
\be
\partial_y \vec{h}^{(n)}(z) = C_z \vec{g}^{(n-1)}(z). 
\ee
The solution to this equation 
\be
\vec{g}^{(n)}(z) = \vec{e}^{(n)} + \int \limits_{0}^{z} {\rm d} \bar z C_z(\bar z)  \vec{g}^{(n-1)}(\bar z),
\label{eqy2}
\ee
is determined up to a constant of integration $\vec{e}^{(n)}$. This  constant of integration has to be determined from 
the boundary conditions that we will discuss presently. Once $\vec f^{(n)}$ 
is found, we employ the same strategy to obtain 
$\vec f^{(n+1)}$.

\section{Boundary conditions in the physical region} 
\label{sec:boundary}

It is common practice (see e.g. Refs.~\cite{Gehrmann:2002zr,tancredi})
that a solution to differential equations is first constructed in an 
unphysical region, where the solution 
is real and unique, and then properly continued into the physical region. 
We have found it difficult to follow this approach here.  
The reason has to do with the mapping from the kinematic variables $S,T,U$ and masses $M_3^2, M_4^2$, 
where the 
analytic continuation is simple, to the $x,y,z$ variables. It is the non-linear nature of this mapping that  
makes  it difficult to perform the {\it proper} analytic continuation once the result is written in 
$x,y,z$ variables.  Because of that, we decided to perform computations directly in the physical region. 
Note that an analysis of master integrals for $q \bar q \to VV$ reported 
recently in Ref.~\cite{tancredi} arrives at a similar 
conclusion:  all, {\it but one},  of the integrals described in that reference 
are obtained using analytic continuation, while the remaining integral 
is computed directly in the physical region since the analytic continuation 
becomes too cumbersome.  We, however, decided in favor of  a unified 
approach for computing all the integrals for planar graphs.  

To understand how solutions in the physical region are constructed, we note that 
a Goncharov polylogarithm may  develop an imaginary part  when its  argument is larger than at least 
one of the indices.  Inspecting the alphabet in Eq.~(\ref{eq_alphabet}), it is easy to realize that in the 
kinematic region of interest, every entry in the alphabet is sign-definite. Therefore, upon 
integrating over $x$, $y$ and  $z$ from zero to their actual values, we can explicitly construct a real-valued 
solution, thereby by-passing all the subtleties related to analytic continuation of Goncharov polylogarithms. 
However, since in the physical region Feynman integrals do have imaginary parts, we should be able to get them 
in our approach as well and it is clear that, in case one has a sign-definite 
alphabet,   imaginary parts can only appear through the boundary conditions. 

To determine boundary conditions, we consider the limit $x \to 0$, $z \to 1$ and $y \to 1$.  Physically, this 
limit corresponds to the production of two vector bosons, one with the mass $p^2 = M_3^2$ 
and the other with the mass 
$M_3^2 x^2$.  The total energy squared of the collision is $M_3^2 (1+x)^2$, 
which implies that the two vector bosons are {\it at  rest }
in the center-of-mass frame of the colliding 
partons. For two families, P12 and P13,  the only singularities 
that are developed in this limit, are related to the mass of the lightest  of the 
two vector bosons; for them, $y$ and $z$ can be set to one and the limit of small 
$x$-values needs to be approached carefully.  A typical behavior of an integral in that 
limit  is  $f \sim f_{a} x^{-n_a \ep} $, where $n_a$ is some integer. 
Unfortunately, for some 
integrals required for the family~P23, the limit $z \to 1, y \to 1$ is also not smooth due to the appearances 
of the so-called double-parton scattering singularities \cite{Gaunt:2011xd}.  For such integrals, 
a typical asymptotic in the limit $x\to 0, y \to 1, z \to 1$ reads 
\be
f \sim f_a x^{-n_1 \ep} + f_b x^{-n_2 \ep} \left [ ( z-y)(1-z) \right ]^{-n_3 \ep}, 
\ee
where $n_{1,2,3}$ are integers. Our goal is to compute constants the $f_{a,b}$ to the relevant order in $\epsilon$ 
and then use them to construct solutions of differential equations as explained in the previous 
Section.

%*****
There are at least two ways to compute asymptotics in the required limits. One option is to simply 
take the limit $z \to 1, y \to 1, x\to 0$ 
in an integrand of a relevant Feynman integral. Since most of the integrals diverge in 
at least  one of these limits, 
we need to resort to asymptotic expansions to evaluate them. 
To this end, one can use the strategy of expansion by 
regions \cite{Beneke:1997zp,Smirnov:2002pj} (for a recent review see Chapter~9 of Ref.~\cite{Smirnov:2012gma})
and its implementation in  an open computer code {\tt asy.m} \cite{Pak:2010pt,Jantzen:2012mw}
which is now included into {\tt FIESTA} \cite{Smirnov:2013eza}.
To apply this code to a given Feynman integral, one has to specify the  propagators, their 
powers  and the limit of interest,   by identifying the small parameter  in the problem.
As an output one  obtains  contributions of regions relevant for  the given limit, 
in terms of Feynman-parametric integrals. Such  integrals are further evaluated 
by the method of Mellin--Barnes representation~\cite{Smirnov:1999gc,Tausk:1999vh,Smirnov:2012gma}.
In fact, for some of the master integrals of family P23, we considered two limits, 
 $x \to 0$ and $z,y \to 1$.
When we evaluated asymptotics in the second limit, we used parametric integrals
obtained after taking the first limit as an input for  the second limit, also using
the code {\tt asy.m}.

An alternative, and in some cases simpler,  way to 
get the boundary conditions for complicated integrals, is provided by the differential equations. 
To illustrate  it, we consider a differential equation 
in the $z$-variable for the box integral $g_{17}$ of the family P12. The definition of the integral 
can be found in the next Section.   Writing the differential 
 equation in the limit $z \to 1, y \to 1$, we find 
\be
\partial_z f_{17}^{P12} = \ep \left (\frac{1}{z-1} + \frac{1}{z-y}  \right ) 
\left ( -\frac{3}{2} f_{1}^{P12} + \frac{1}{2} f_2^{P12} + f_3^{P12} - f_6^{P12} + f_{17}^{P12} \right ) +..., 
\label{eq5p2}
\ee
where ellipses stand for less singular terms. 
In $z \to 1, y \to 1$ limit, all the integrals in the family $P_{12}$ must have finite limits. 
The consistency of this requirement 
with Eq.~(\ref{eq5p2}) leads to a relation between different integrals
\be
\lim_{z,y \to 1} f_{17}^{P12} - \frac{3}{2} f_{1}^{P12} + \frac{1}{2} f_2^{P12} + f_3^{P12} - f_6^{P12}  = 0.
\label{eq_g17}
\ee
As can be seen from Section~\ref{sec:masters}, where all master integrals are defined, 
the integrals $f_{1,2,3}^{P12}$ are the two-loop two-point functions and $f_6^{P12}$ is a relatively simple 
three-point function, whose $y \to 1, z \to 1, x \to 0$ limits are straightforward to obtain.
We find 
\be
f_1^{P12} \sim  -x^{-2\ep},\;\;\;  f_2^{P12}  \sim  -e^{2i\pi \epsilon} x^{-4\ep},\;\;\; 
f_3^{P12} \sim  -e^{2i\pi\ep},
\;\;\; f_{6}^{P12} \sim  - e^{2i\pi\ep}.
\ee
We then read off the limit of  the integral $f_{17}^{P12}$ from Eq.~(\ref{eq_g17})  implies 
\be
f_{17}^{P12} \sim  \frac{1}{2} e^{2i \pi \ep} x^{-4\ep} - \frac{3}{2} x^{-2\ep}.
\ee
%******************************
Finally, we note that the 
boundary conditions in the physical region for all the master integrals
are reported in Section~\ref{sec:masters}.  To make sure that the boundary conditions 
are correct, we  have often used both  strategies described above  to evaluate them.
An  agreement between these independent computations is a non-trivial check of the 
correctness of the boundary conditions.

\section{Analytic continuation}
\label{sec:analytic}
 
In the previous Section, we described how we determined the boundary behavior of the integrals directly in the physical region, thereby avoiding the necessity of any analytic continuation.
As we pointed out, the analytic continuation is not obvious to perform in the $x,y,z$ variables.
The problem is that the change of variables Eq.~(\ref{eq_xyz})  is non-linear. Therefore, our insistence 
on writing results  in terms of Goncharov polylogarithms makes the analytic structure 
 of the solution less obvious.

Here, we wish to show how the analytic continuation can be easily done in the language of 
Chen iterated integrals, in terms of 
the original variables, $S, T, M_{a}^2, M_{b}^2$.  We will take the integral family P23 as an
example. This will also be a useful check of our results, since the boundary behavior for this integral
family is particularly complicated in the physical region. 

The integrals of family P23 depend on the variables $s,t,p_2^2, p_3^2$. 
We can start from a non-physical region with $s<0, t<0, p_2^2 <0, p_2^2<0$. 
The physical region is then reached by analytically continuing to $p_{2}^2 >0, p_{3}^2>0$, 
keeping in mind the Feynman $i0$ prescription. Note that such an analytic 
continuation is possible, since the integrals 
in the P23 family do not have discontinuities in the Mandelstam variable $u$, so that the incorrect 
$i0$ prescription for the Mandelstam variable 
$u$,  induced by the analytic continuation of $p_{2,3}^2$,  is not relevant. 

We will  discuss a single-parameter 
slice of the functions, which is obtained by fixing two Mandelstam variables and 
varying the remaining two. Specifically, we choose 
\begin{align}\label{param-r}
s= t= -1 \,,\quad p_{2}^2 = p_{3}^2 =  - \frac{2 r}{1+r^2}  \,.
\end{align}
A nice feature of this parametrization is that the alphabet (\ref{alphabet}) needed to describe the functions
becomes simply
\begin{align}\label{alphabet-r}
\alpha \longrightarrow \left\{ r , 1-r ,1+r, 1+r^2 \right\} \,.
\end{align}
The boundary constants in the non-physical region $ r> 0$  are easily fixed. In fact, they can be obtained
from the requirement that no branch cuts should start in that region.
In the present case, the potential singularity at $r=1$, cf. Eq.~(\ref{alphabet-r}), must be spurious. 
Experience shows that such conditions usually allow one to determine all boundary constants without calculations \cite{Henn:2013tua,Henn:2013nsa}. The same is true here. 
For the basis choice $\vec{g}^{P23}$ made in Section~\ref{sec:masters}, 
one easily sees that the boundary values at $r=1$ are given by 
\begin{align}
  \vec{g}^{P23}|_{r=1} = 
\{ b_1, b_1, b_1, b_1, 0, 0, 0, 0, b_2, 0, b_2, 0, b_2, b_2,
0, 0, 0, 0, 0, 0, 
0, 0, 0, 0, 0, 0, 0, 0\} \,.
\end{align}
Here $b_1$ and $b_2$ are just the explicit values of trivial bubble-type integrals.
They are given by
\begin{align}
b_{1} =&-
   \Gamma^3(1-\eps) \Gamma(1 + 2 \eps)/
 \Gamma(1 - 3 \eps) \,,\\
b_{2} =&  \Gamma^4(1 - \eps)  \Gamma^2(
  1 + \eps)/\Gamma^2(1 - 2 \eps)\,.
\end{align}
Taking into account that
\begin{align}
\ln[ \Gamma(1+\eps) ]  = - \gamma_{\rm E} \eps + \sum_{k\ge 2} (-1)^k  \zeta_{k} \frac{\eps^k }{k}  \,,
\end{align}
we see that after multiplying with $e^{2  \gamma_{\rm E} \eps }$, the $\eps$ expansion
of these functions has uniform weight.
This, together with the differential equations (\ref{DEdifferentialform}), shows that the solution has
uniform weight in the $\eps$ expansion, to all orders in $\eps$.

Let us now discuss the analytic continuation in $r$ to negative values of $r$.
The Feynman prescription implies that $r$ should have a small negative imaginary part.
The alphabet in Eq.~(\ref{alphabet-r}) 
indicates that poles in the complex $r$ plane are located at $-1,0,1,i,-i$, and at infinity. 
As we discussed earlier, the pole at $r=1$ is  spurious. 
There are branch cuts along the negative real axis, starting at $r=0$, 
and possibly along the imaginary axis starting from $r=\pm i$.

It is now clear how to analytically continue to negative values of $r$. We can choose a path below
the negative real axis, but with $\Im (r)>-1$, thereby avoiding branch cuts. Then we simply evaluate the Chen iterated path integral along this contour. We have done so for a path consisting of two segments, the first along the real axis from $r=1$ to $r=1/2$, 
and the second along the semi-circle $r= \frac{1}{2} e^{-i \pi t}$, with $t \in [0,1]$.
In this way, we numerically verified the values for $\vec{g}^{P23}$ obtained in the physical region at $r=-1/2$. 
In terms of the $x,y,z$ variables of Eq.~(\ref{eq_xyz}), this point corresponds to $x=2, y= 1/4, z=5/8$.

Given the simplicity of the alphabet (\ref{alphabet-r}) arising from the parametrization (\ref{param-r}), it is also possible to perform the analytic continuation in a more algebraic way. 
Indeed, the terms that require analytic continuation are the ones that develop logarithmic singularities as $r \to 0$. In the present case, functions corresponding to the alphabet (\ref{alphabet-r}) can be written as Goncharov
polylogarithms with indices $0,\pm 1, \pm i$. 
The terms with logarithmic divergences are the ones with $0$'s at the rightmost entry.  
This behavior can be made manifest by using shuffle relations for iterated integrals,
e.g. 
\begin{align}
G(1,0;r) = G(0;r) G(1;r) - G(0,1;r) \,,
\end{align} 
and so on, where we explicitly see $G(0;r) = \log r$.
The logarithmic terms are then analytically continued according to $\log r \to \log(-r) - i \pi$.
In this way, one arrives at a representation valid for $r<0$.

In summary, the formulation of Eq.~(\ref{solution_chen}) in terms of iterated path integrals has
many conceptional advantages; here we exploited its manifest homotopy invariance in order to perform
the analytic continuation. On the other hand, if one first fixes an integration contour, in order, 
for example,  to obtain an expression in terms of Goncharov polylogarithms, one looses much of this flexibility.

\section{Master integrals} 
\label{sec:masters}

For each family of integrals, the Mandelstam variables are given by $s = (p_1+p_2)^2 = (p_3+p_4)^2$, 
$t = (p_1+p_3)^2 = (p_2+p_4)^2$, $u = (p_2+p_3)^2 = (p_1+p_3)^2$. Their relation to the physical Mandelstam 
variables $S,T,U$ and the ensuing parametrization in terms of variables $x,y,z$ can be read off using 
the $q \to p$ mapping just before Eq.~(\ref{eq_g}) and Eqs.~(\ref{eq_man}), (\ref{eq_xyz}).

When choosing the master integrals we followed the strategy proposed in Ref.~\cite{jhenn}
to find master integrals having uniform weight. As guiding principles for  finding such integrals
we analyzed generalized unitarity cuts, as well as explicit (Feynman) parameter representations 
of the integrals.  
Technically this is very similar to the analysis of certain three-loop massless integrals studied in 
Refs. \cite{Henn:2013tua,Henn:2013nsa}. In fact, some of the two-loop integrals 
with two off-shell legs are contained in those three-loop integrals as subintegrals. 
For more detailed explanations and examples, see Section 2 of Ref.~\cite{Henn:2013tua}.

Below we present the master integrals, and the boundary conditions in the physical region 
that we used to evaluate them.  For convenience, we re-scale and renormalize the master
integrals. In particular, for the families P12 and P13 we choose master integrals to be 
 $f_i^{\rm P12,P13} = N_0 (p_1^2)^{2 \eps}  \, e^{2 \gamma_{\rm E} \eps} \, g_{i}^{\rm P12,P13}$,
while for  the family P23, we choose master integrals as 
 $f_i^{\rm P23} =  N_0 (p_3^2)^{2 \eps}  e^{2 \gamma_{\rm E} \eps} \, g_{i}^{\rm P23}$. The 
normalization constant $N_0$ is 
\be
N_0 = 1 + \frac{\pi^2}{6} \ep^2 + \frac{32 \zeta_3 }{3} \ep^3 + \frac{67 \pi^4 \ep^4}{360}.
\ee
Furthermore, to present the master integrals and the results for the limits, we use the following notation 
\be
\begin{split} 
& N_1 = 1 + i \pi \ep - \frac{2 \pi^2 \ep^2 }{3}
        - \left ( i \frac{\pi^3}{3} - 2 \zeta_3 \right )\ep^3
        + \left (\frac{\pi^4}{10} + 2 i \pi \zeta_3 \right ) \ep^4,
\\ 
& N_2 = 1  + 6 \ep^3 \zeta_3 + \frac{\ep^4 \pi^4}{10},\;\;\;
N_3 = 1 - i\ep\pi - \frac{\pi^2 \ep^2}{6} - \left (\frac{i\pi^3}{6} + 14\zeta_3 \right )\ep^3. 
\\
& 
R_{12}=\sqrt{p_1^2 + (p_2^2 - s)^2 - 2 p_1^2 (p_2^2 + s)},
\;\;\;R_{13}=\sqrt{p_1^2 + (p_3^2 - t)^2 - 2 p_1^2 (p_3^2 + t)},
\\
& R_{23}=\sqrt{(s + t)^2-4 p_2^2 p_3^2 }. 
\end{split} 
\ee
The pictures below are intended to give a general idea of how the corresponding master integrals look like, but obviously do not show doubled propagators or numerators and prefactors. Also, in some cases we chose linear combinations of integrals as master integrals, and in those cases only one representative figure is given.

\begingroup

\allowdisplaybreaks

The master integrals and their boundary asymptotic behaviour at the point $x \to 0, y \to 1, z \to 1$ 
for the family P12 read 
\begin{small}
 \begin{align}
\picturepage{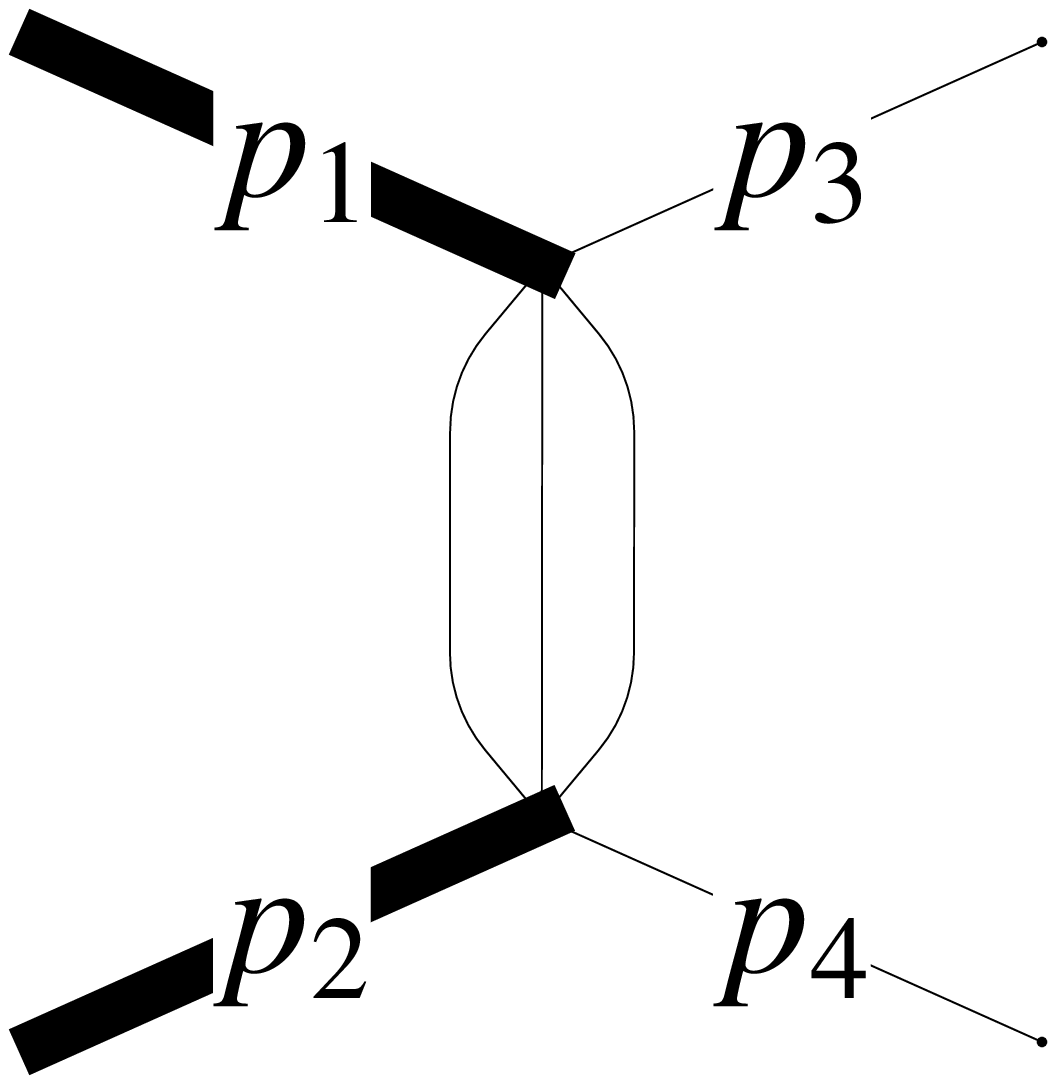}
{  \bea     
  g^{\rm P12}_1 &=& \eps^2 \; t \;G_{0, 0, 0, 0, 1, 2, 2, 0, 0} 
\,,~~~~~~~~~~~~~~~~~~~~~~~~~~~~~~~~~~~~~~~~~~~~
  \\
     f^{\rm P12}_{1} &\sim &  -x^{-2\ep}\;, \nn 
\eea} \nn \\
\picturepage{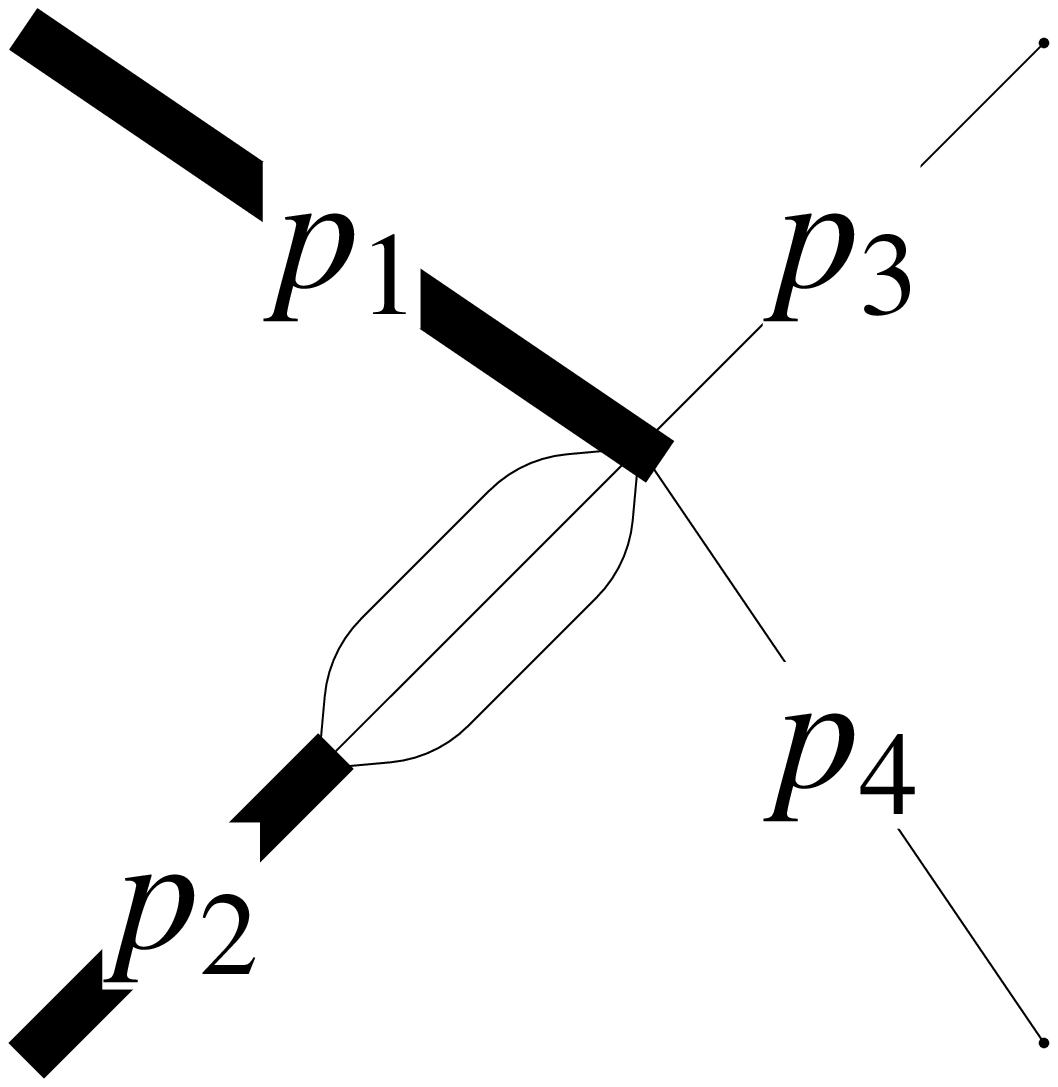}
{  \bea       g^{\rm P12}_2 &=&  \eps^2 \; p_2^2 \;  G_{0, 0, 0, 1, 2, 2, 0, 0, 0}
\,,~~~~~~~~~~~~~~~~~~~~~~~~~~~~~~~~~~~~~~~~~~~~
\\
f^{\rm P12}_{2} &\sim&  -e^{2\pi i \ep} \; x^{-4\ep}\;,   \nn
\eea} \nn \\
\picturepage{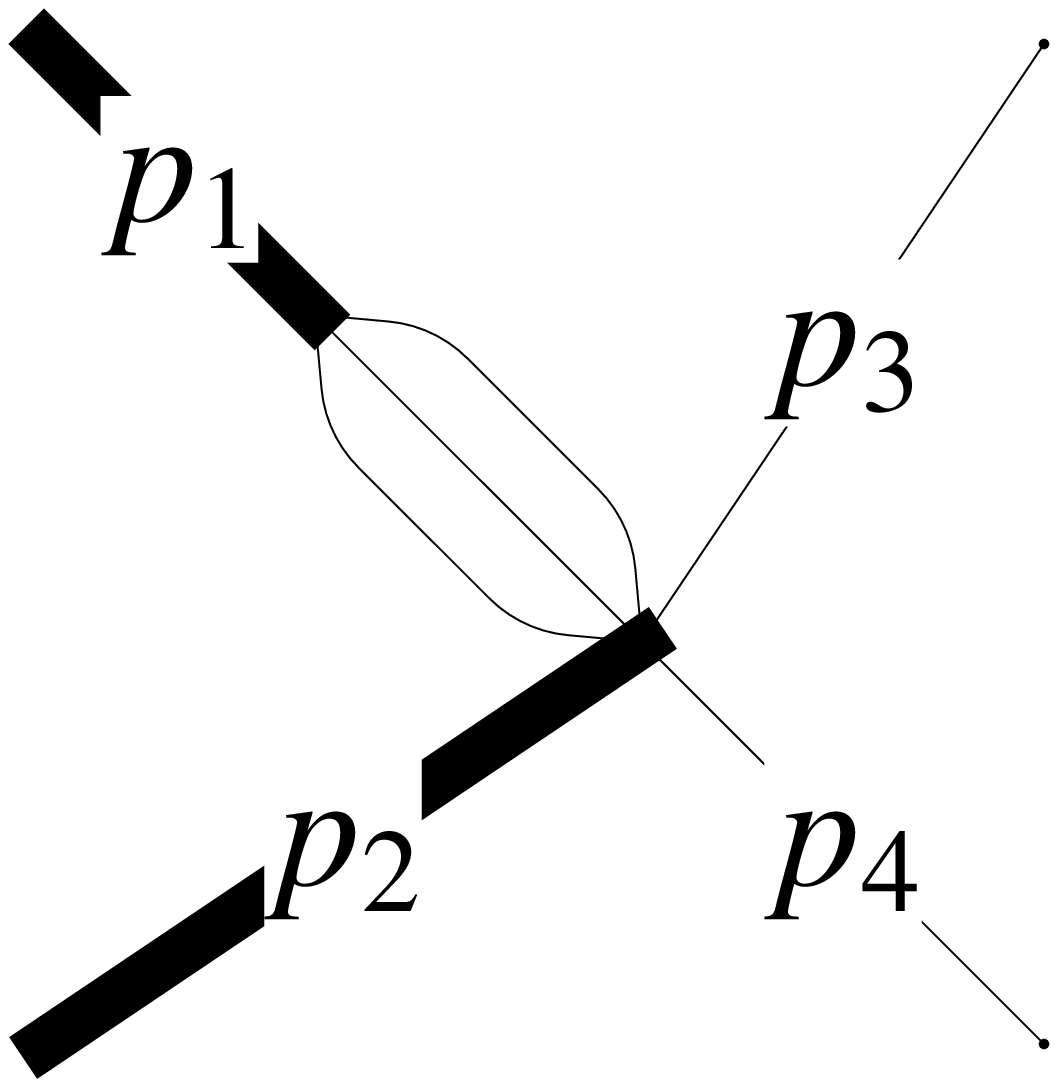}
{  \bea      g^{\rm P12}_3 &=&   \eps^2 \; p_1^2 \;  G_{0, 0, 1, 0, 2, 2, 0, 0, 0} 
\,,~~~~~~~~~~~~~~~~~~~~~~~~~~~~~~~~~~~~~~~~~~~~
 \\
    f^{\rm P12}_{3} &\sim&  -e^{2\pi i \ep}\;, \nn
\eea} \nn \\
\picturepage{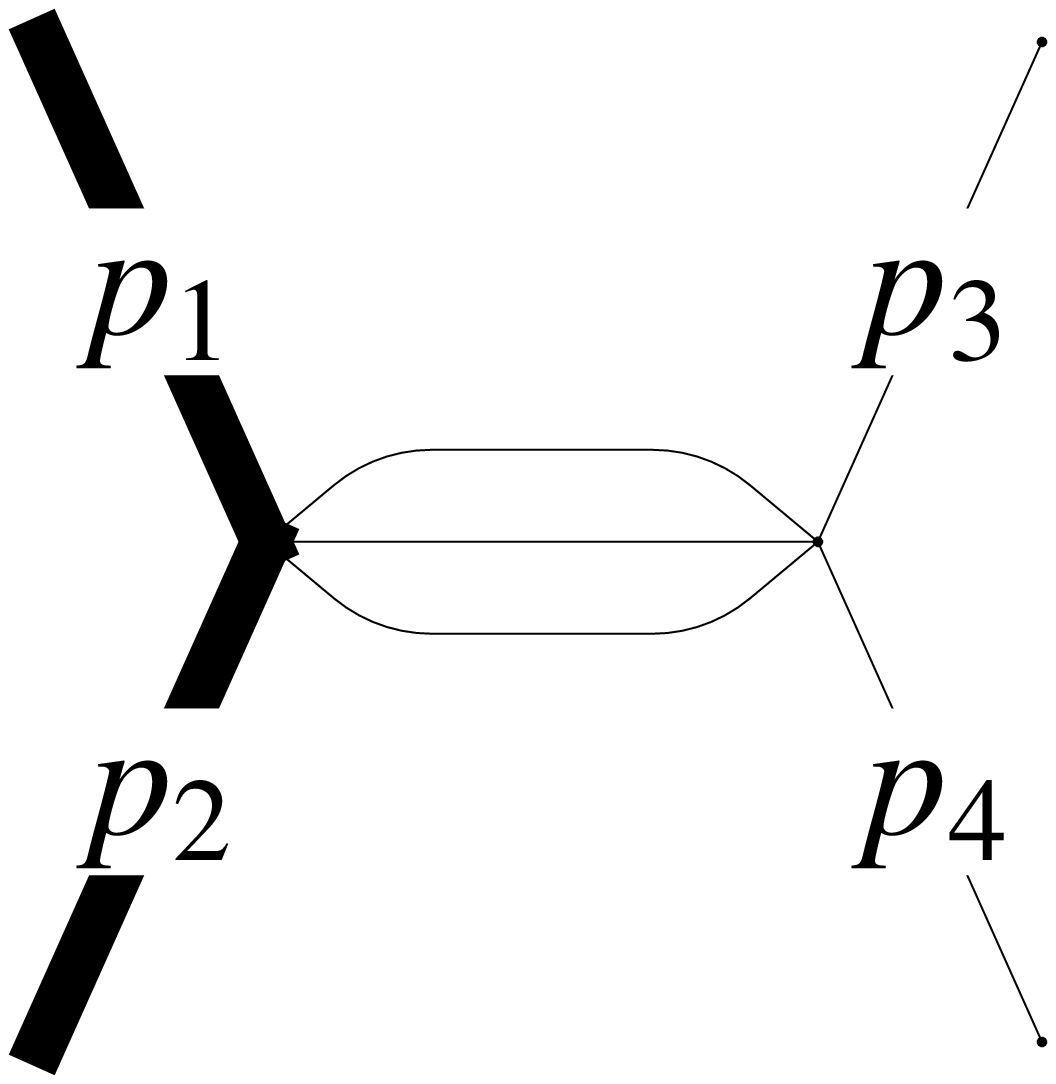}
{  \bea    g^{\rm P12}_{4} &=&  \eps^2 \; s \; G_{0, 1, 2, 0, 0, 2, 0, 0, 0}
\,,~~~~~~~~~~~~~~~~~~~~~~~~~~~~~~~~~~~~~~~~~~~~~~   
 \\
    f^{\rm P12}_{4} &\sim& -e^{2\pi i \ep}\;, \nn
\eea} \nn \\
\picturepage{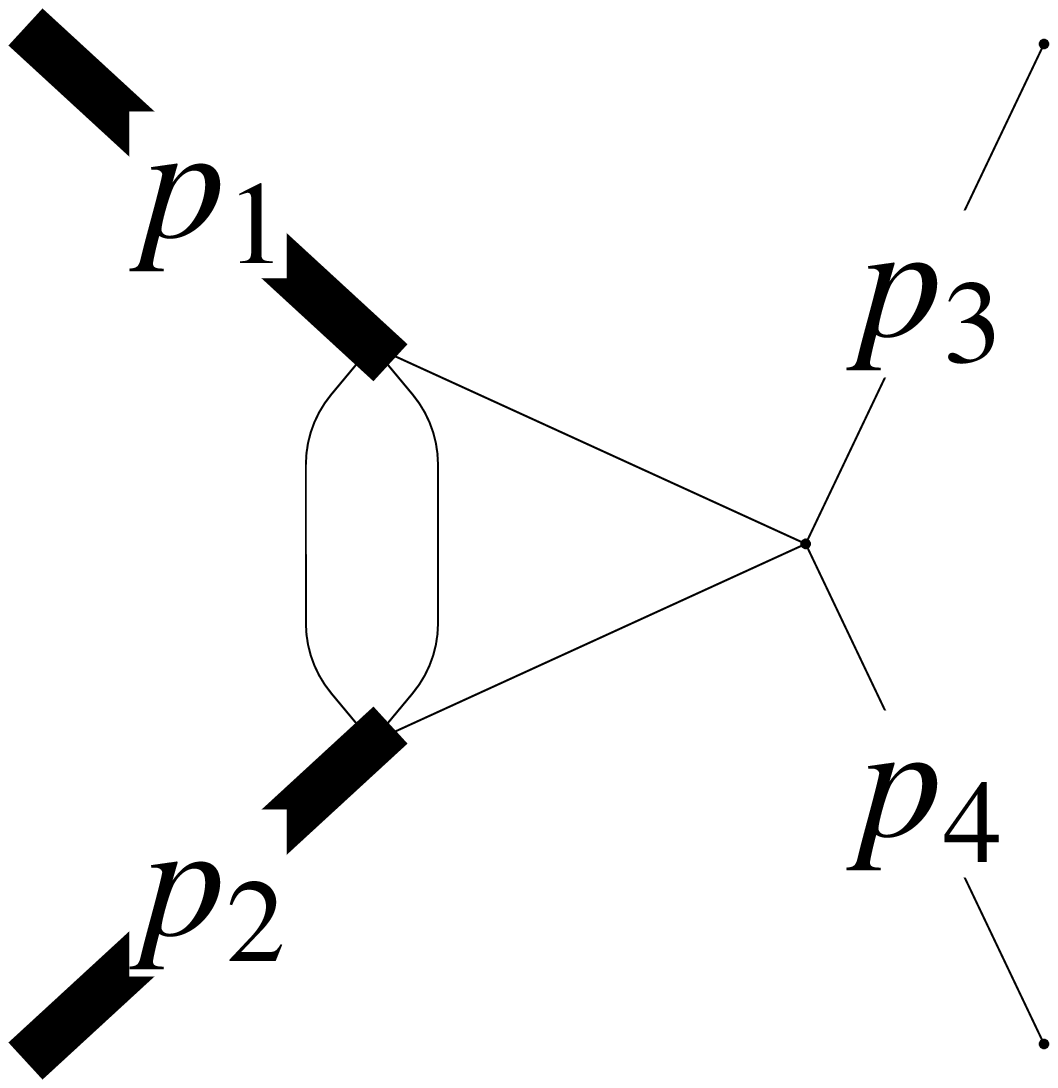}
{  \bea      g^{\rm P12}_5 &=&  \eps^3 \; R_{12} \; G_{0, 0, 1, 1, 1, 2, 0, 0, 0}
\,,~~~~~~~~~~~~~~~~~~~~~~~~~~~~~~~~~~~~~~~~~~~
 \\
    f^{\rm P12}_{5} &\sim& 0\;,  \nn
\eea} \nn \\
\picturepage{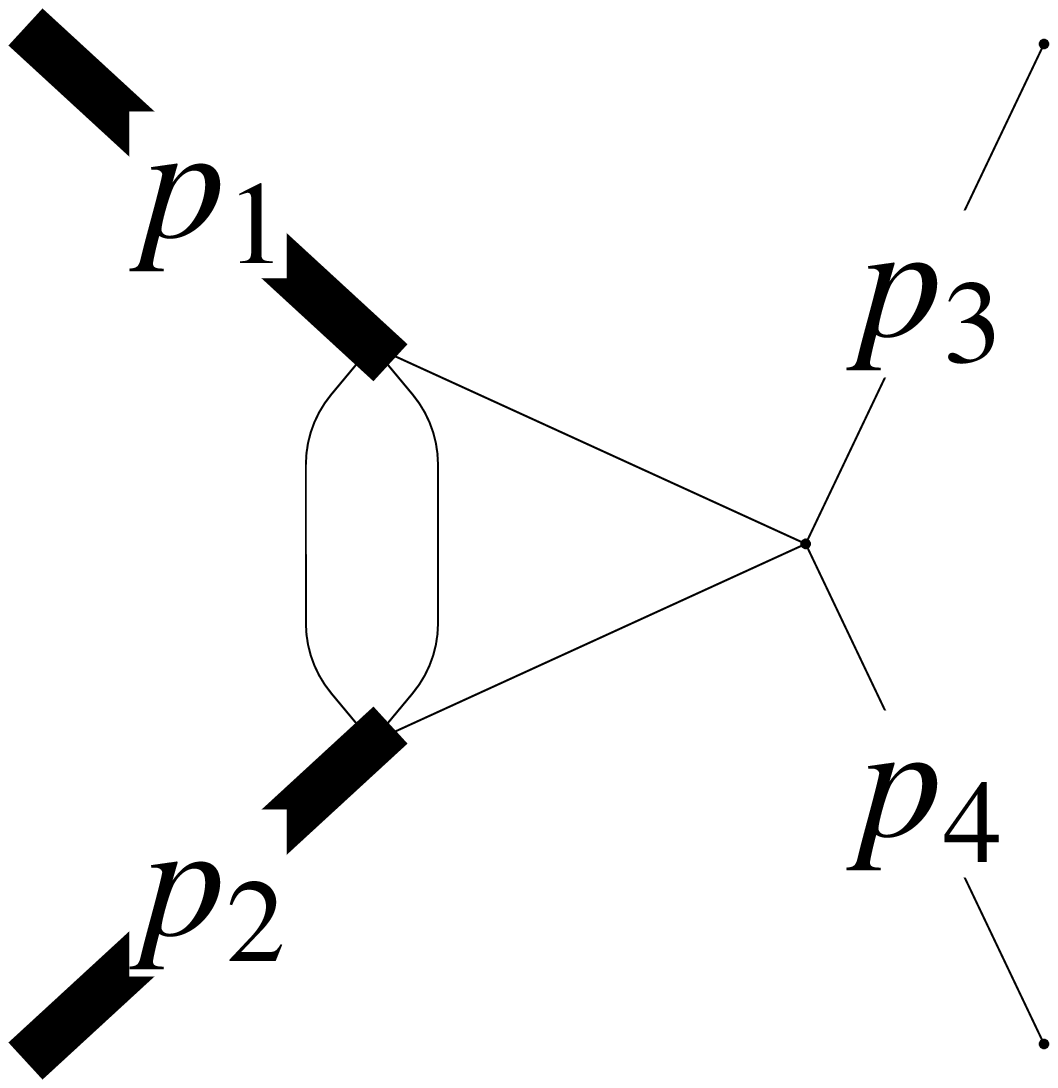}
{  \bea      ~~~g^{\rm P12}_6 &=&   \eps^2 \Big [-\frac{1}{2} \eps (p_1^2  - p_2^2  - s)
  G_{0, 0, 1, 1, 1, 2, 0, 0, 0} 
+  s G_{0, 0, 2, 1, 1, 2, 0, -1, 0}   \Big ]  \,,
 \\
    f^{\rm P12}_{6} &\sim&  -e^{2\pi i \ep}, \nn
\eea} \nn   \\
\picturepage{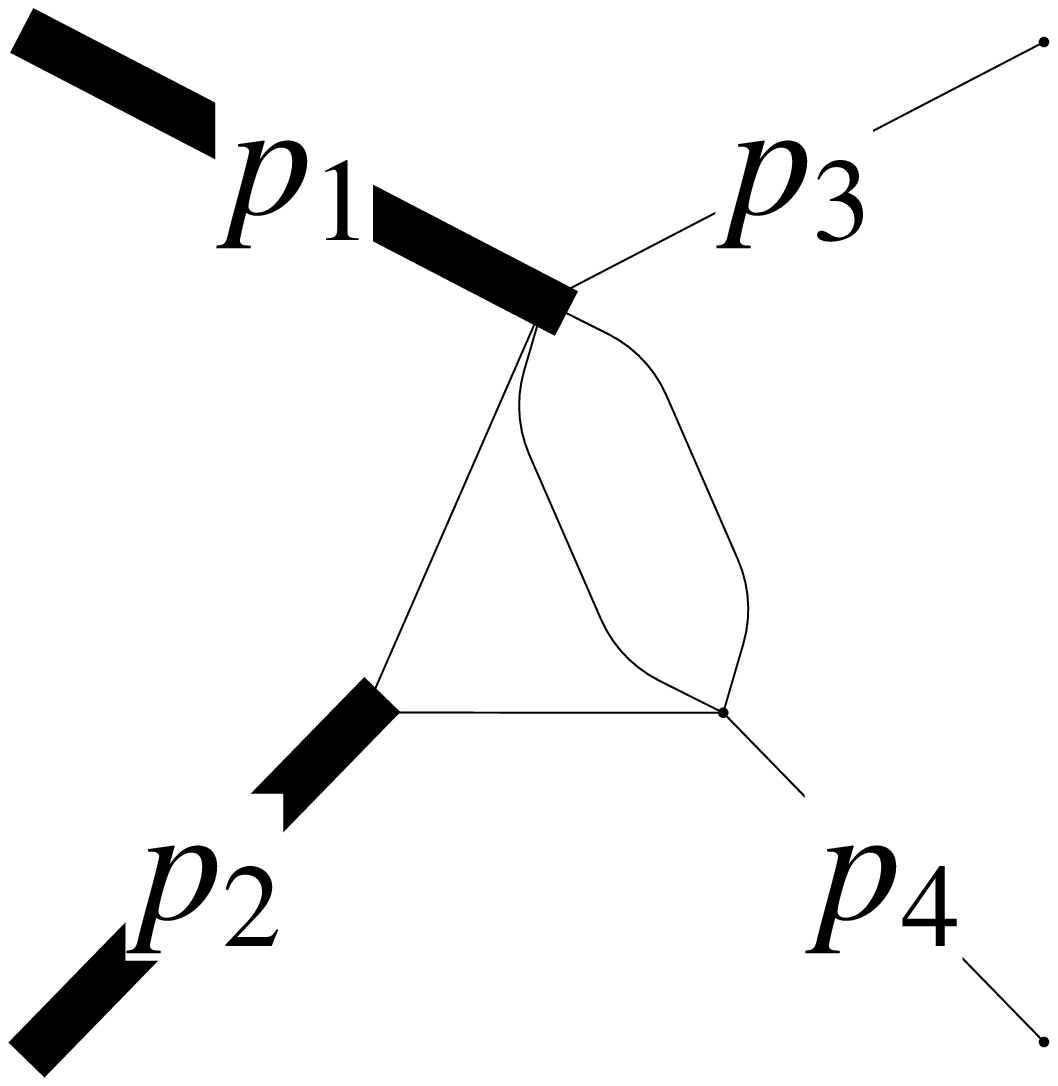}
{  \bea      g^{\rm P12}_7 &=&    
\eps^3 \; (p_2^2  - t) \; G_{0, 1, 0, 0, 1, 1, 2, 0, 0}
\,,~~~~~~~~~~~~~~~~~~~~~~~~~~~~~~~~~~~~~~~
 \\
    f^{\rm P12}_{7} &\sim& -\frac{x^{-2\ep}  }{2} 
+ \frac{x^{-3\ep}}{2}  N_1, \nn
\eea} \nn \\
%
%\end{align}
%\end{small}
%%%%%%%%%%%%%%%%
%\begin{small}
%\begin{align}
%
\picturepage{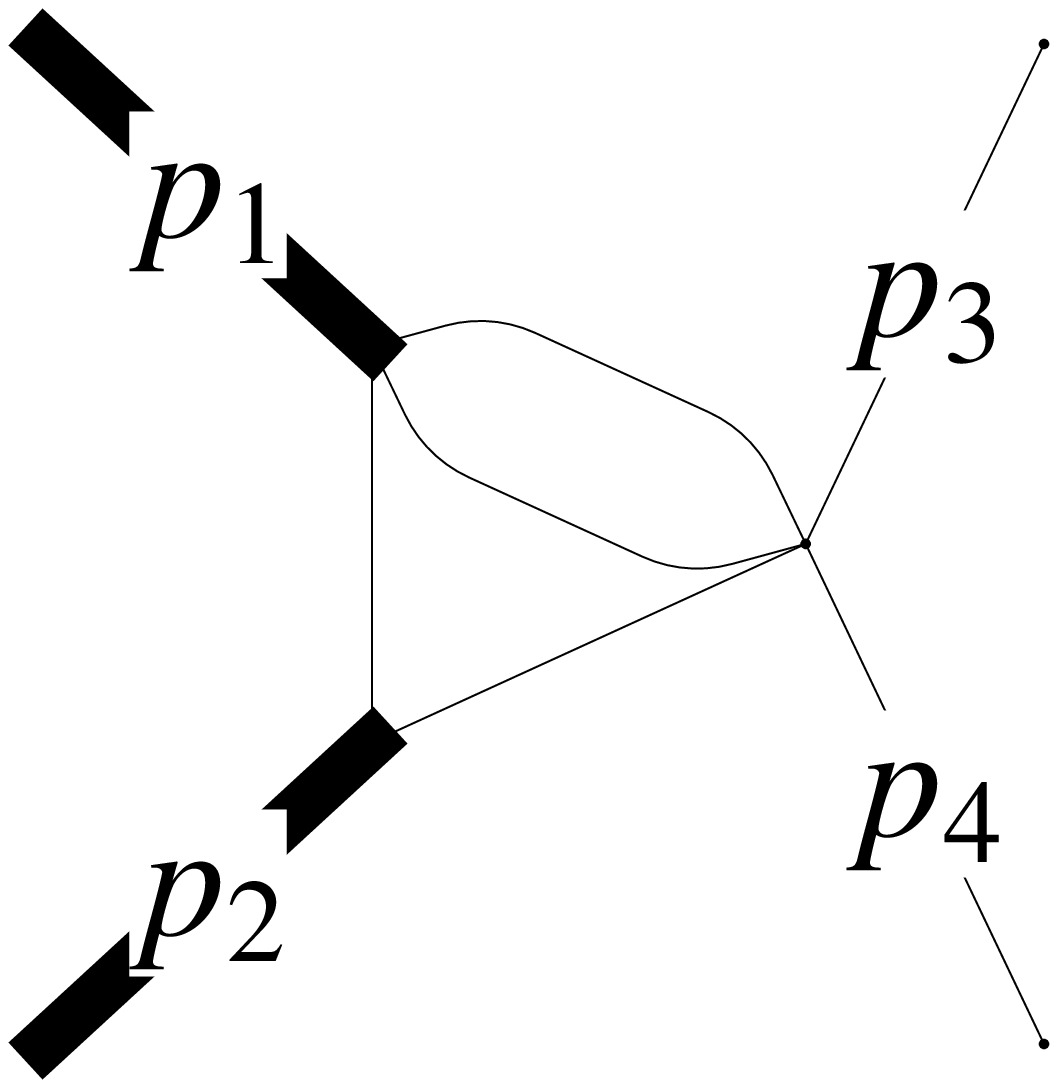}
{  \bea      g^{\rm P12}_8 &=&  \eps^3 \; R_{12} \; G_{0, 1, 1, 0, 1, 2, 0, 0, 0}
\,,~~~~~~~~~~~~~~~~~~~~~~~~~~~~~~~~~~~~~~~~~~~~~~~~~~~~~
 \\
    f^{\rm P12}_{8} &\sim& 0, \nn
\eea} \nn \\
\picturepage{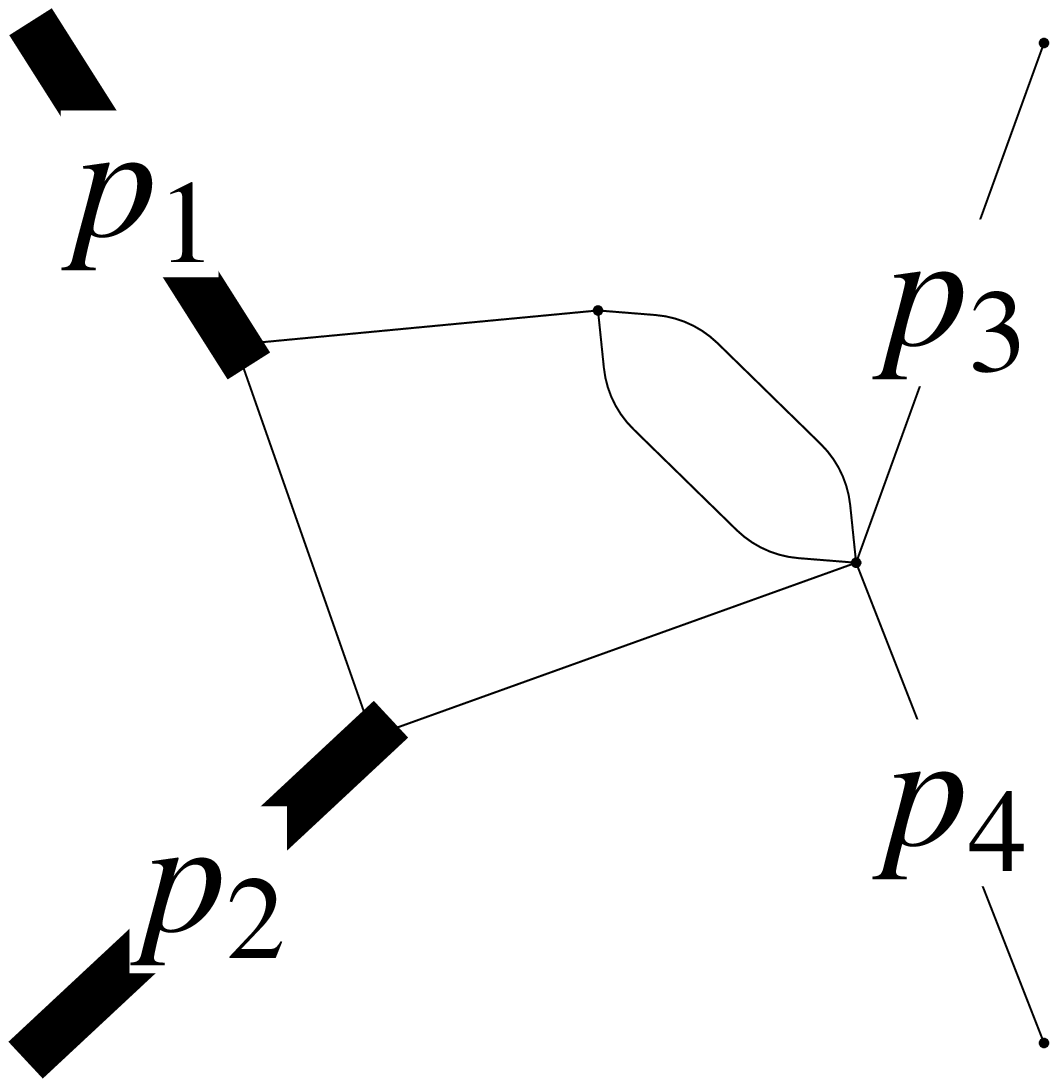}
{  \bea      g^{\rm P12}_9 &=&  \eps^2 \Big [ \frac{3}{2} \eps (p_1^2  - p_2^2  + s) G_{0, 1, 1, 0, 1, 2, 0, 0, 0} 
+  (1 + \eps) p_1^2  s G_{1, 1, 1, 0, 1, 2, 0, 0, 0} \Big ]   \,,
\\
    f^{\rm P12}_{9} &\sim&  \frac{3 e^{2i\pi \ep} }{2}  - x^{-2\ep} e^{2i\pi \ep} N_2,  \nn
\eea} \nn \\
\picturepage{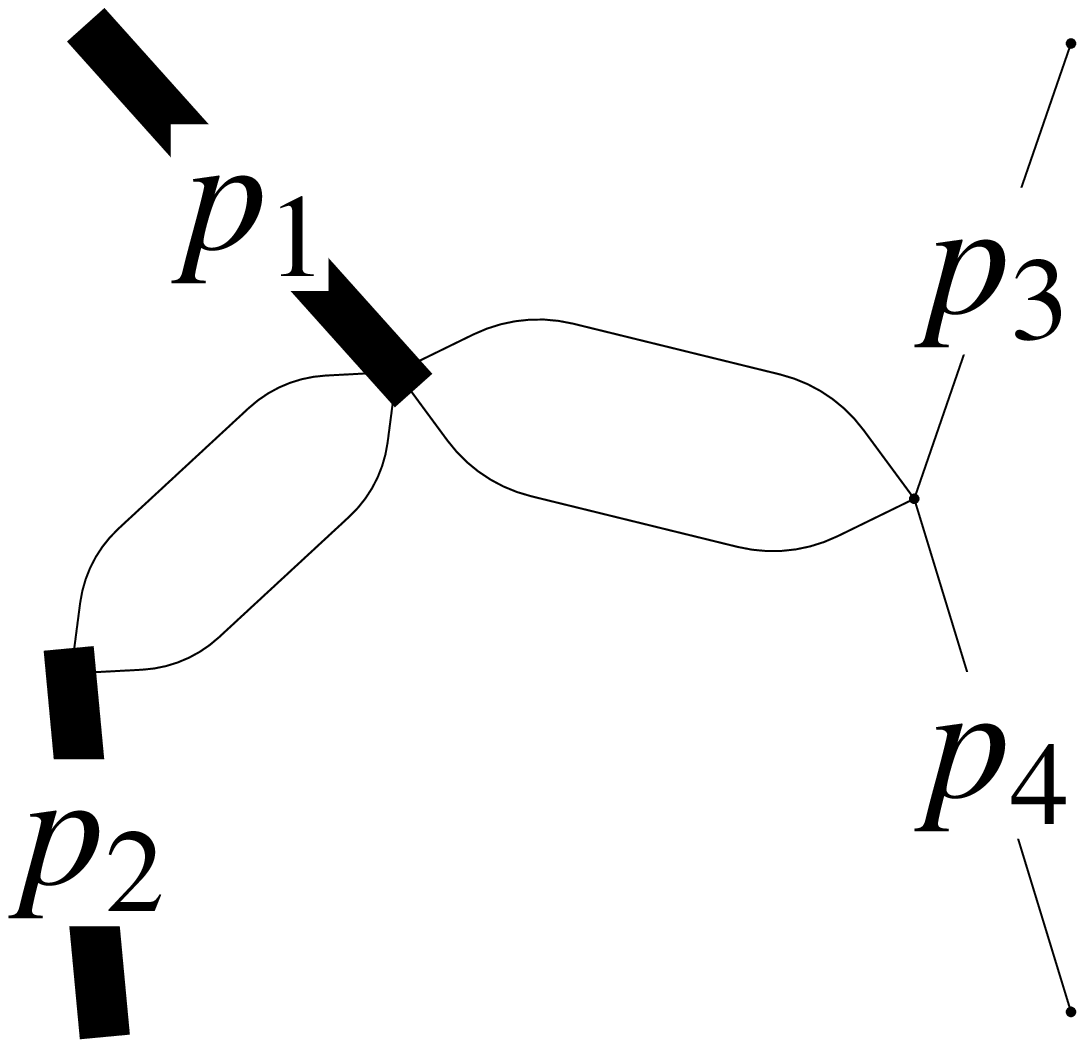}
{  \bea      g^{\rm P12}_{10} &=&    \eps^2 p_2^2  s G_{0, 1, 1, 2, 2, 0, 0, 0, 0} 
\,,~~~~~~~~~~~~~~~~~~~~~~~~~~~~~~~~~~~~~~~~~~~~~~~~~~~~~~   
 \\
    f^{\rm P12}_{10} &\sim&  x^{-2\ep} e^{2i\pi \ep} N_2,  \nn
\eea} \nn \\
\picturepage{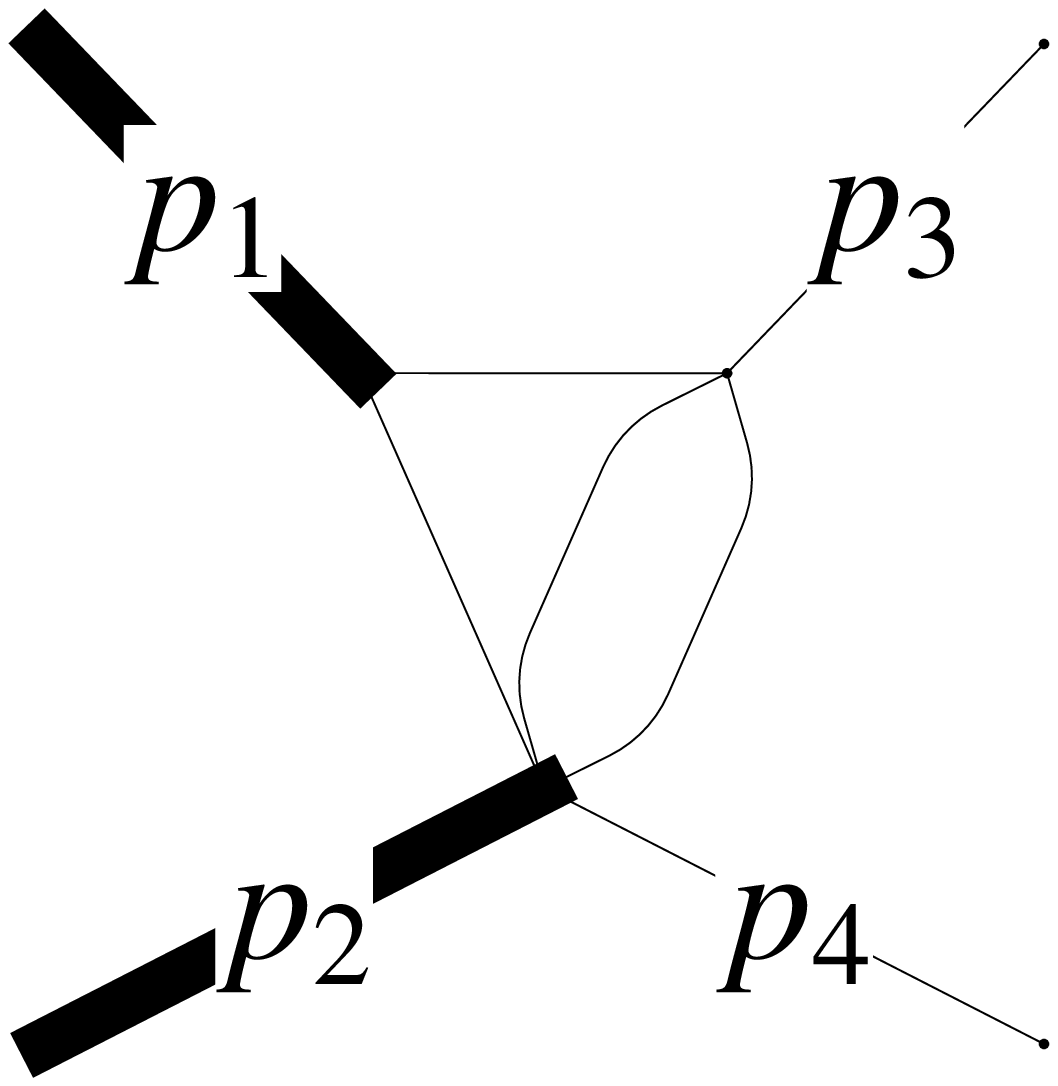}
{  \bea      g^{\rm P12}_{11} &=&   \eps^3 (p_1^2  - t) G_{1, 0, 0, 0, 1, 1, 2, 0, 0}  
\,,~~~~~~~~~~~~~~~~~~~~~~~~~~~~~~~~~~~~~~~~~~~~~~~~
 \\
    f^{\rm P12}_{11} &\sim&  -\frac{x^{-2\ep}}{4} + 
    e^{2i\pi \ep} \left ( \frac{1}{4} +  \frac{\pi^2 \ep^2}{12} + \frac{\zeta_3 \ep^3 }{2} 
+  \frac{\pi^4 \ep^4}{40}    \right ),
\nn
\eea} \nn \\
\picturepage{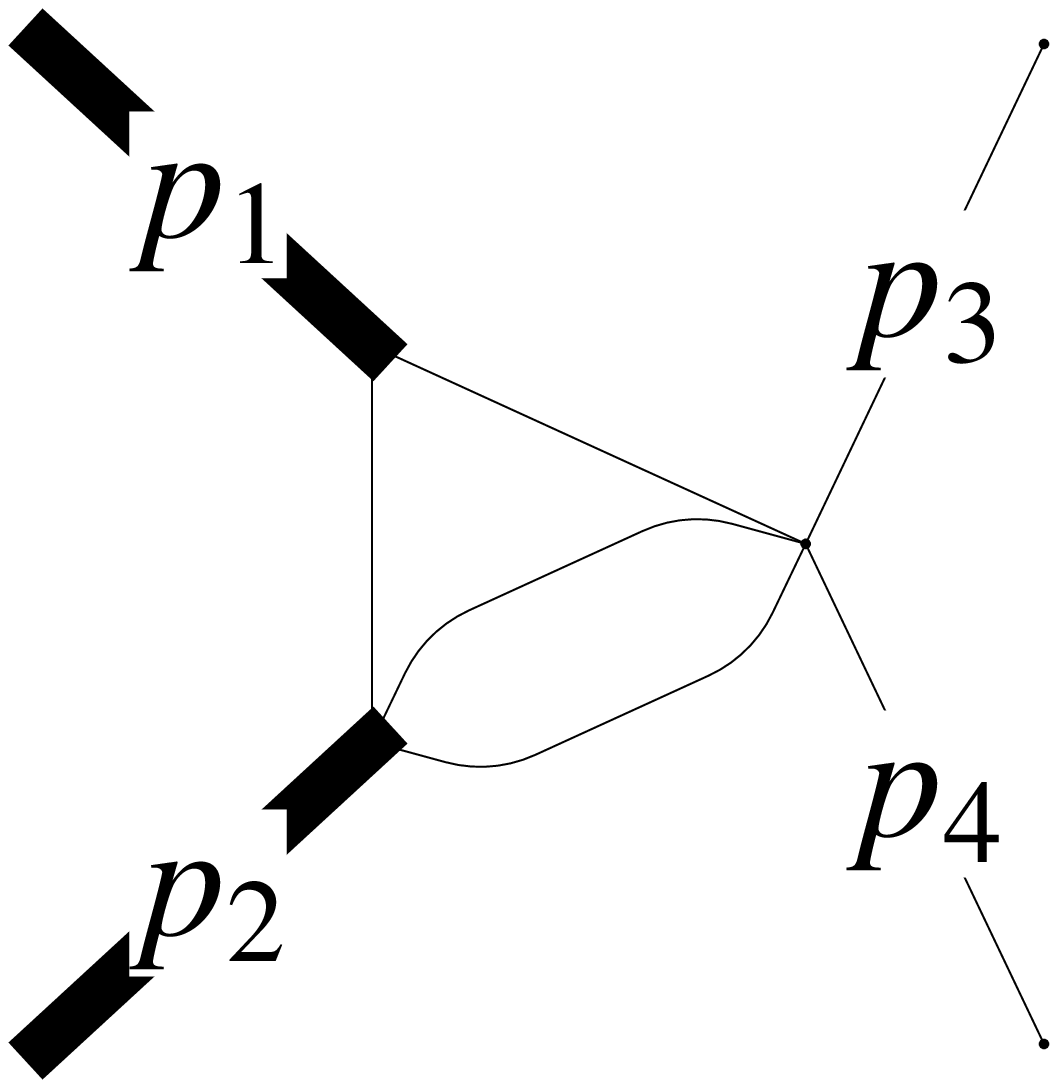}
{  \bea      g^{\rm P12}_{12} &=& \eps^3 \; R_{12} \; G_{1, 0, 0, 1, 1, 2, 0, 0, 0}
\,,~~~~~~~~~~~~~~~~~~~~~~~~~~~~~~~~~~~~~~~~~~~~~~~~~~~~~~
\\
    f^{\rm P12}_{12} &\sim& 0\;, \nn
\eea} \nn \\
\picturepage{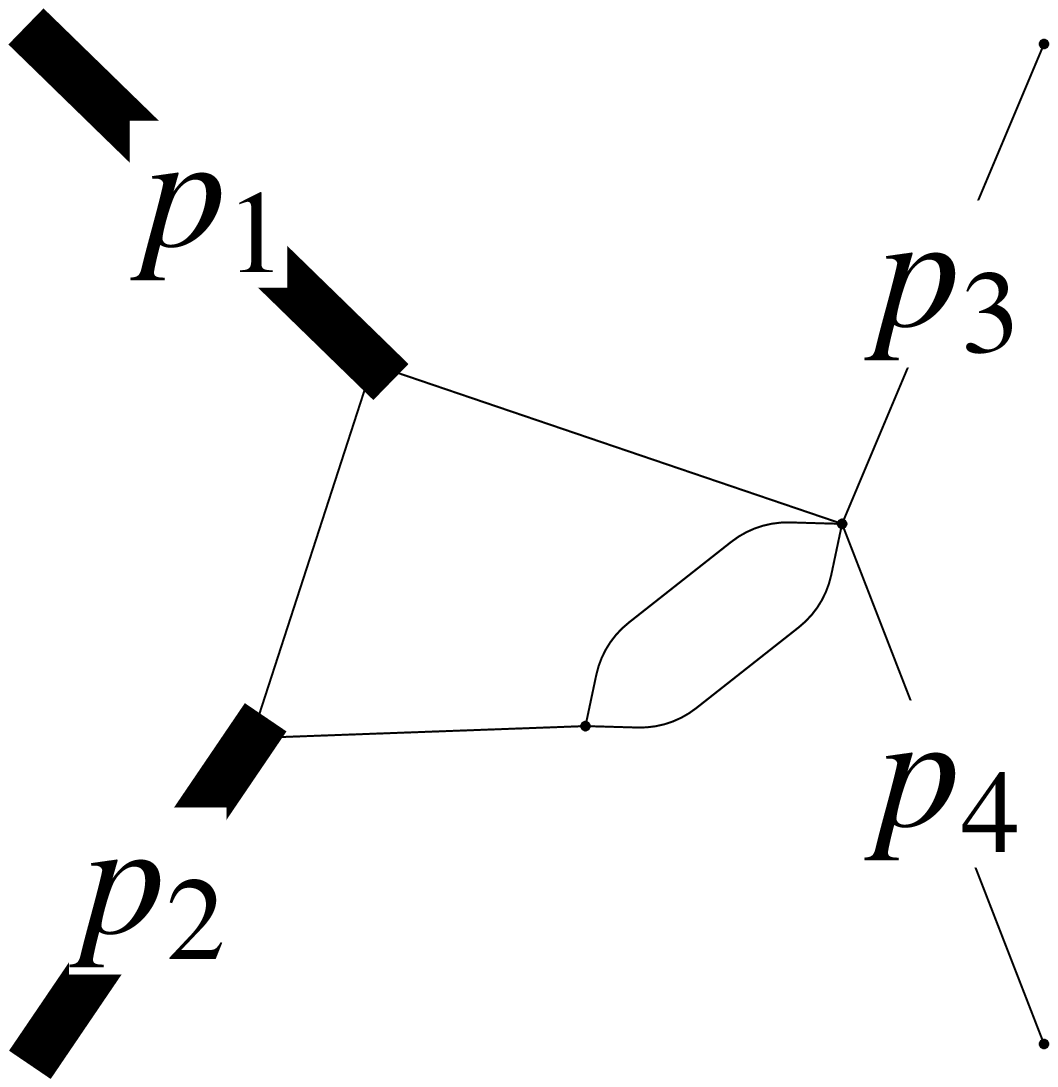}
{  \bea      g^{\rm P12}_{13} &=&  \eps^2 \Big [ \frac{3}{2} \eps (s - p_1^2  + p_2^2  ) 
 G_{1, 0, 0, 1, 1, 2, 0, 0, 0}  + (1 + \eps) p_2^2  s G_{1, 1, 0, 1, 1, 2, 0, 0, 0} \Big ] \;,
\\
    f^{\rm P12}_{13} &\sim&  \frac{e^{2i\pi \ep}}{2} x^{-4\ep},  \nn
\eea} \nn \\
\picturepage{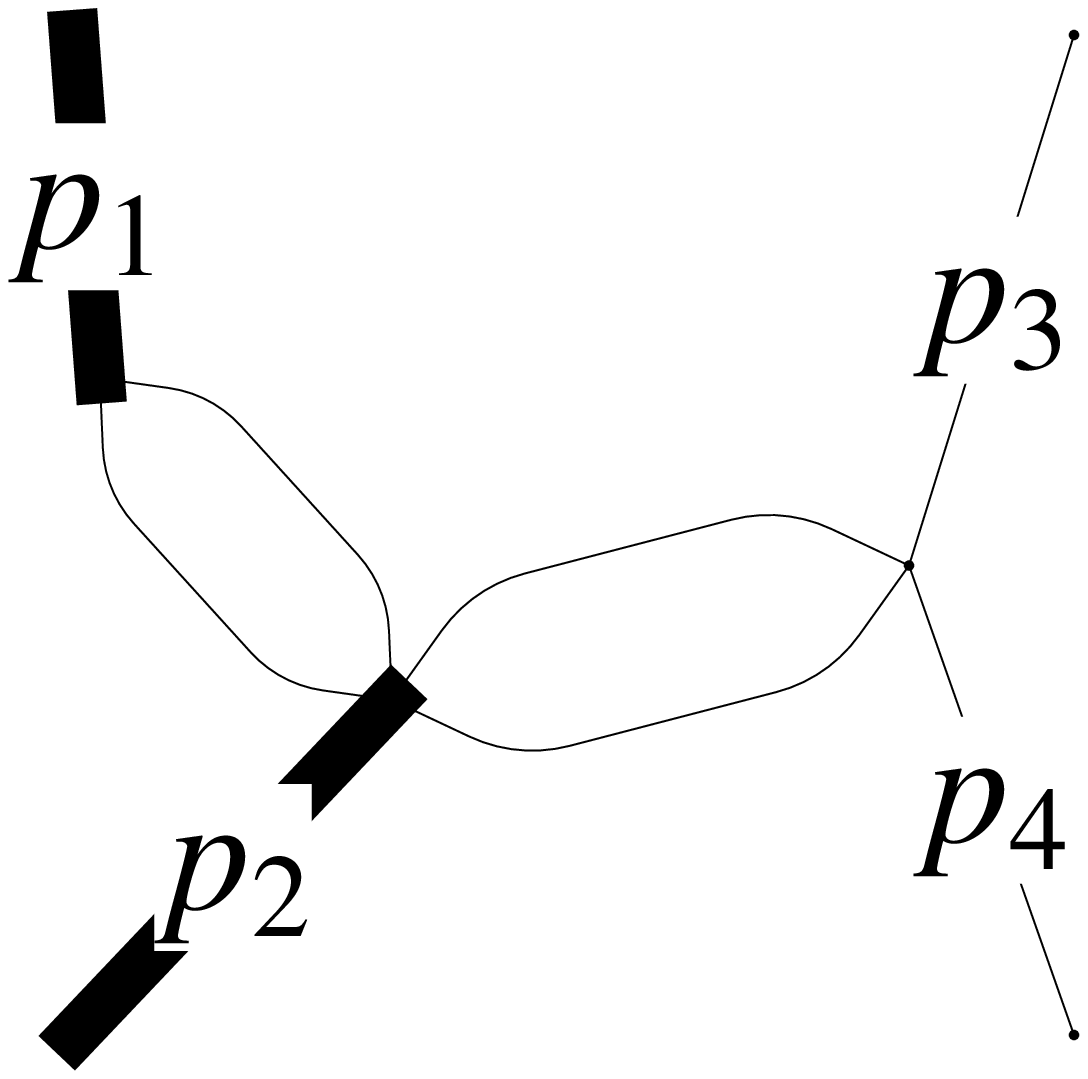}
{  \bea      g^{\rm P12}_{14} &=&    \eps^2 \; p_1^2  \; s \; G_{1, 0, 1, 2, 2, 0, 0, 0, 0} \,,
\,,~~~~~~~~~~~~~~~~~~~~~~~~~~~~~~~~~~~~~~~~~~~~~~~~~~~~ 
\\
    f^{\rm P12}_{14} &\sim&  e^{2i\pi \ep} N_2, \nn
\eea} \nn \\
\picturepage{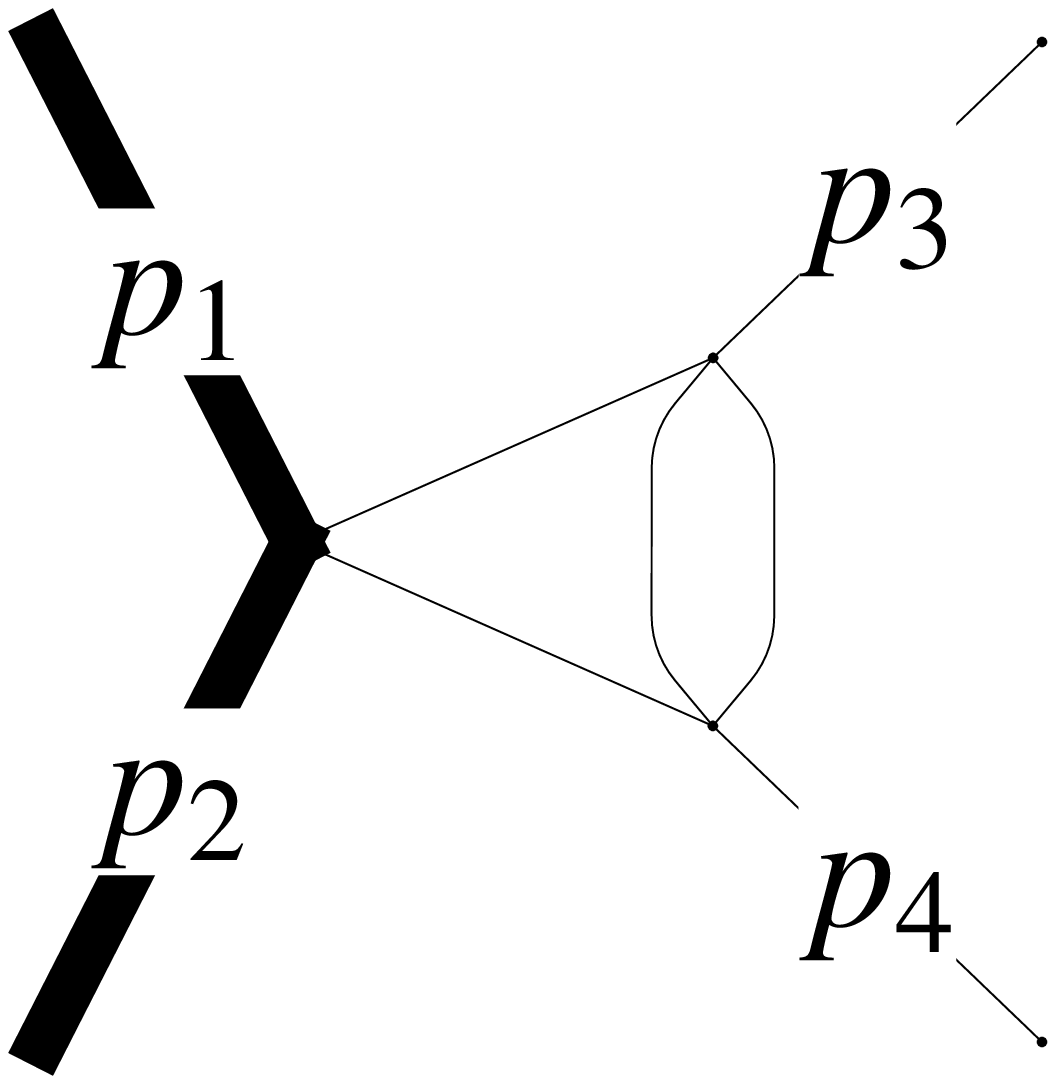}
{  \bea      g^{\rm P12}_{15} &=& \eps^3 \; s \; G_{1, 1, 0, 0, 0, 1, 2, 0, 0}   
\,,~~~~~~~~~~~~~~~~~~~~~~~~~~~~~~~~~~~~~~~~~~~~~~~~~~~~~~~~~ 
\\
    f^{\rm P12}_{15} &\sim&     e^{2i\pi \ep} \left ( \frac{1}{4} + \frac{\pi^2 \ep^2}{12} + \frac{\zeta_3\ep^3}{2}
 + \frac{\pi^4\ep^4}{40} \right ), 
\nn
\eea} \nn \\
\picturepage{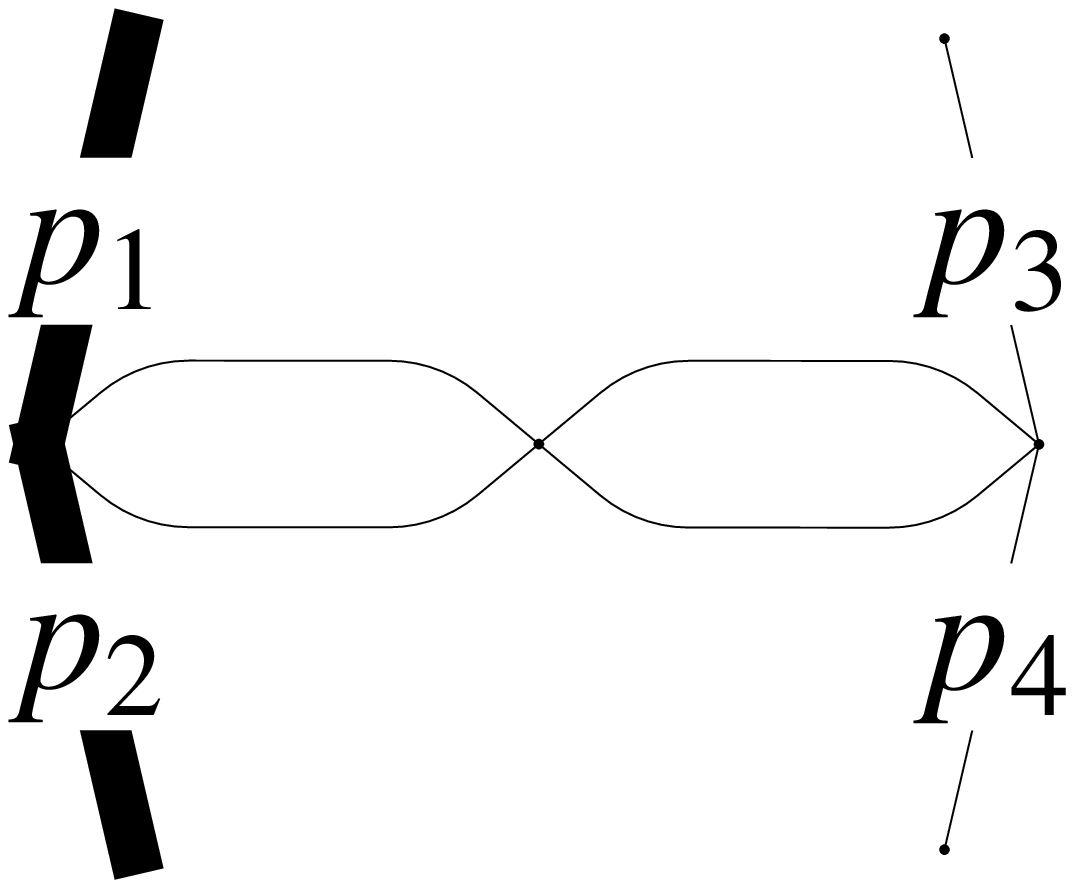}
{  \bea      g^{\rm P12}_{16} &=&  \eps^2 \; s^2 \; G_{1, 2, 1, 2, 0, 0, 0, 0, 0}  
\,,~~~~~~~~~~~~~~~~~~~~~~~~~~~~~~~~~~~~~~~~~~~~~~~~~~~~~~~~
 \\
    f^{\rm P12}_{16} &\sim& e^{2i\pi \ep} N_2, \nn
\eea} \nn \\
%   
%   
%   
%   
%\end{align}
%\end{small}
%%%%%%%%%%%%%%%%%%%
%
%\begin{small}
%\begin{align}
\picturepage{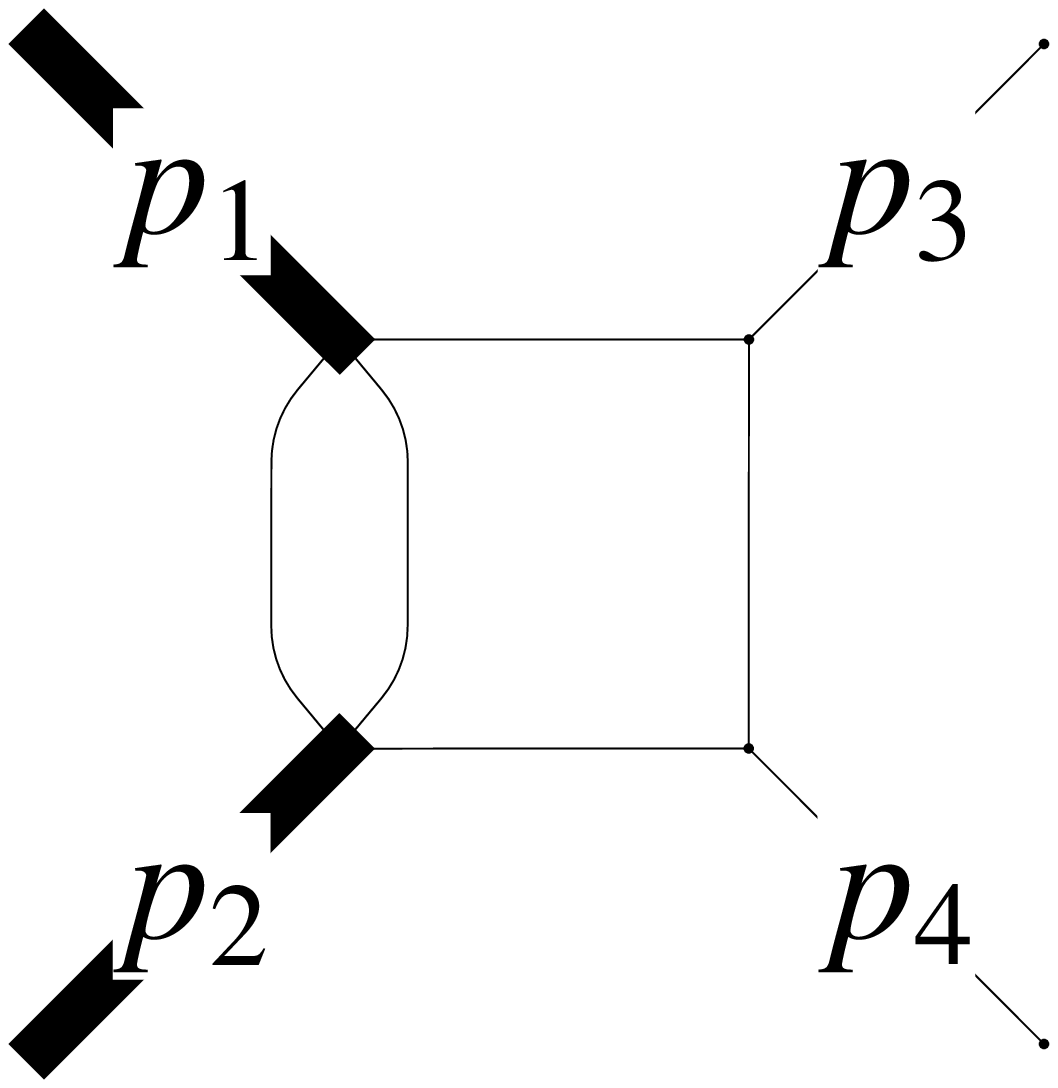}
{  \bea      g^{\rm P12}_{17} &=&   \eps^3 s t G_{0, 0, 1, 1, 1, 2, 1, 0, 0} 
\,,~~~~~~~~~~~~~~~~~~~~~~~~~~~~~~~~~~~~~~~~~~~~~~~~~~~~~~
\\
    f^{\rm P12}_{17} &\sim&   \frac{e^{2i\pi \ep} x^{-4\ep} }{2} - \frac{3x^{-2\ep}}{2},  \nn
\eea} \nn \\
\picturepage{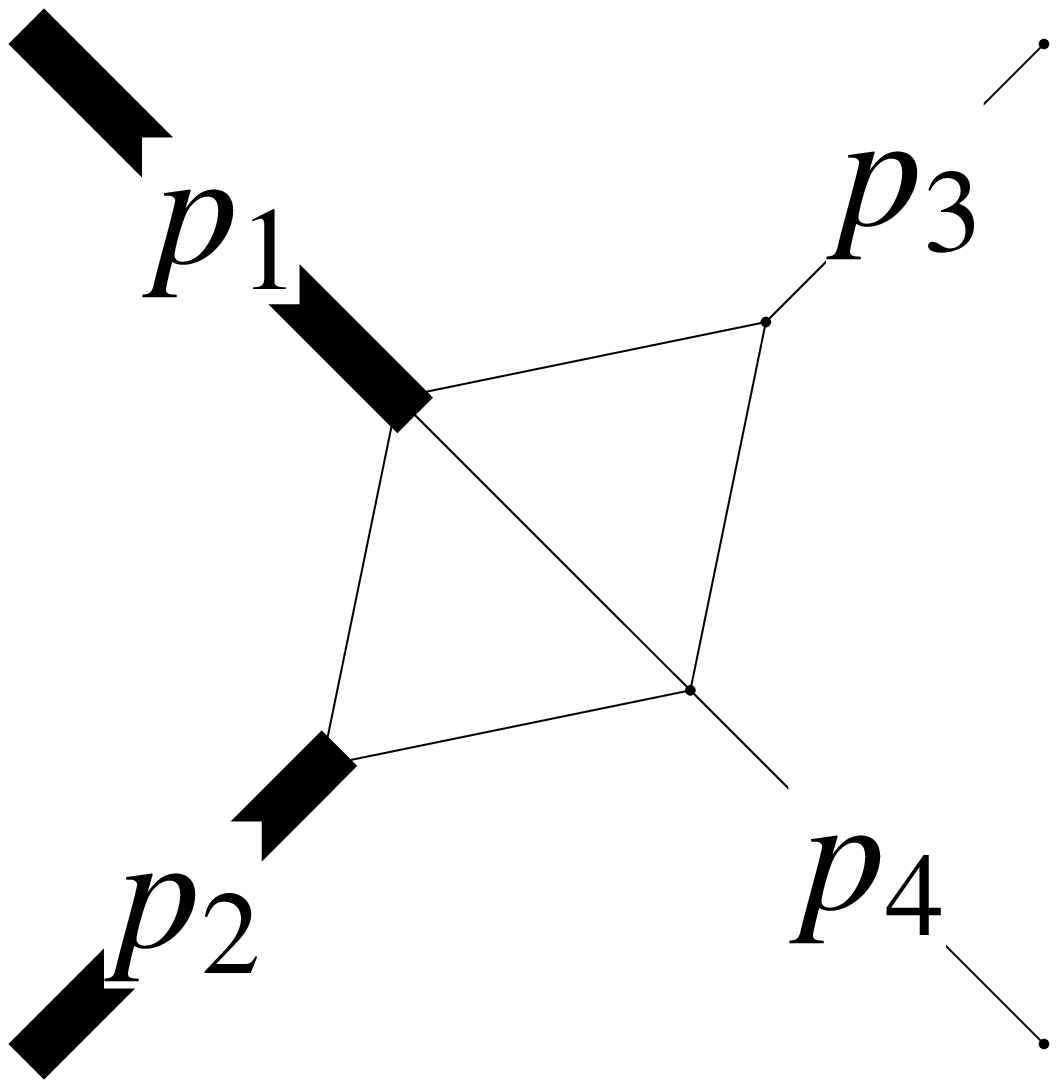}
{  \bea      g^{\rm P12}_{18} &=& \eps^4 (p_1^2  - s - t) G_{0, 1, 1, 0, 1, 1, 1, 0, 0}  
\,,~~~~~~~~~~~~~~~~~~~~~~~~~~~~~~~~~~~~~~~~~~
\\
    f^{\rm P12}_{18} &\sim& 0, \nn
\eea} \nn \\
\picturepage{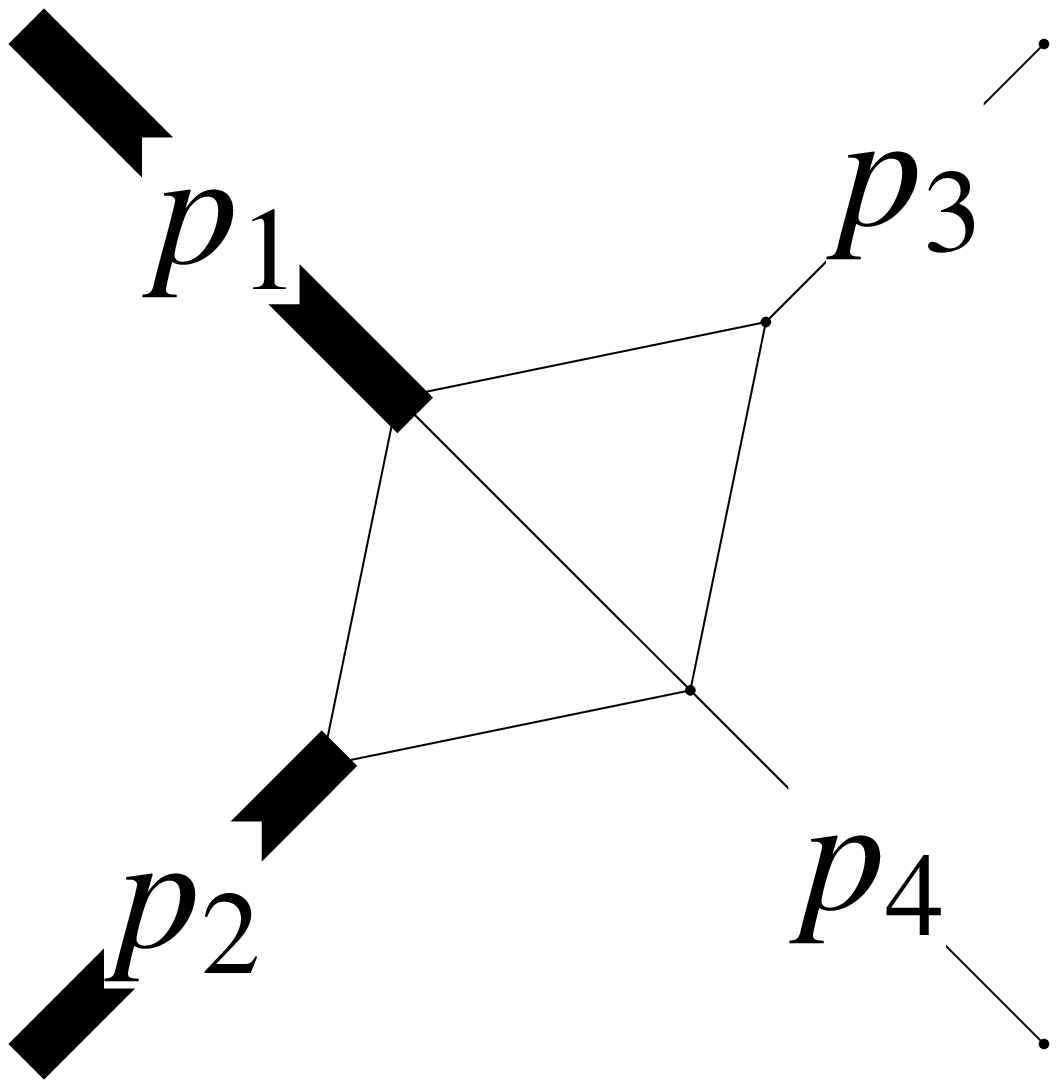}
{  \bea      g^{\rm P12}_{19} &=&  \eps^3 s t G_{0, 1, 1, 0, 1, 2, 1, 0, 0}  
\,,~~~~~~~~~~~~~~~~~~~~~~~~~~~~~~~~~~~~~~~~~~~~~~~~~~~~~~
\\
    f^{\rm P12}_{19} &\sim& -\frac{3 x^{-2\ep} }{2}   
      + x^{-3\ep} N_1,  \nn
\eea} \nn \\
\picturepage{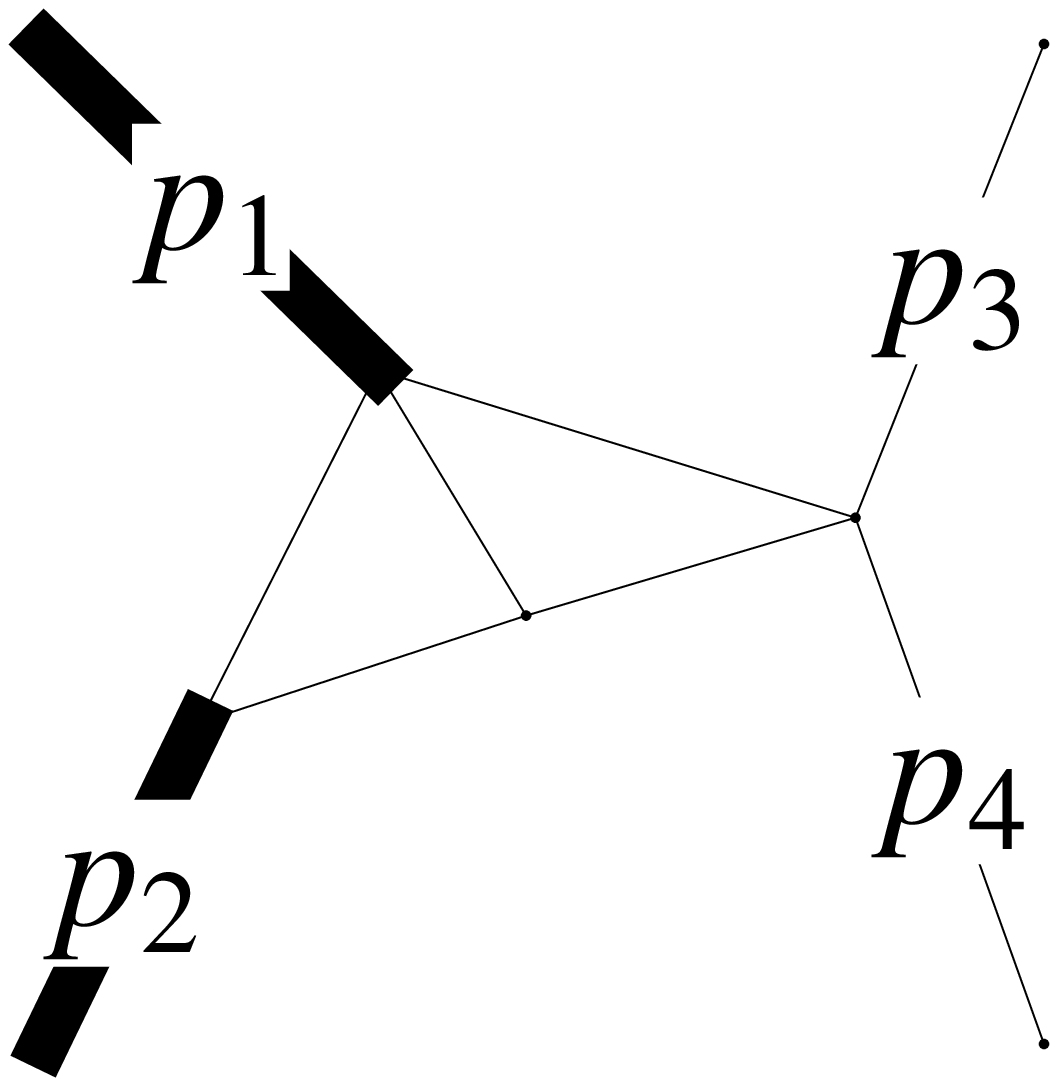}
{  \bea      g^{\rm P12}_{20} &=&  \eps^4 R_{12} G_{0, 1, 1, 1, 1, 1, 0, 0, 0} 
\,,~~~~~~~~~~~~~~~~~~~~~~~~~~~~~~~~~~~~~~~~~~~~~~~~~~~~~
\\
    f^{\rm P12}_{20} &\sim& 0, \nn
\eea} \nn \\
\picturepage{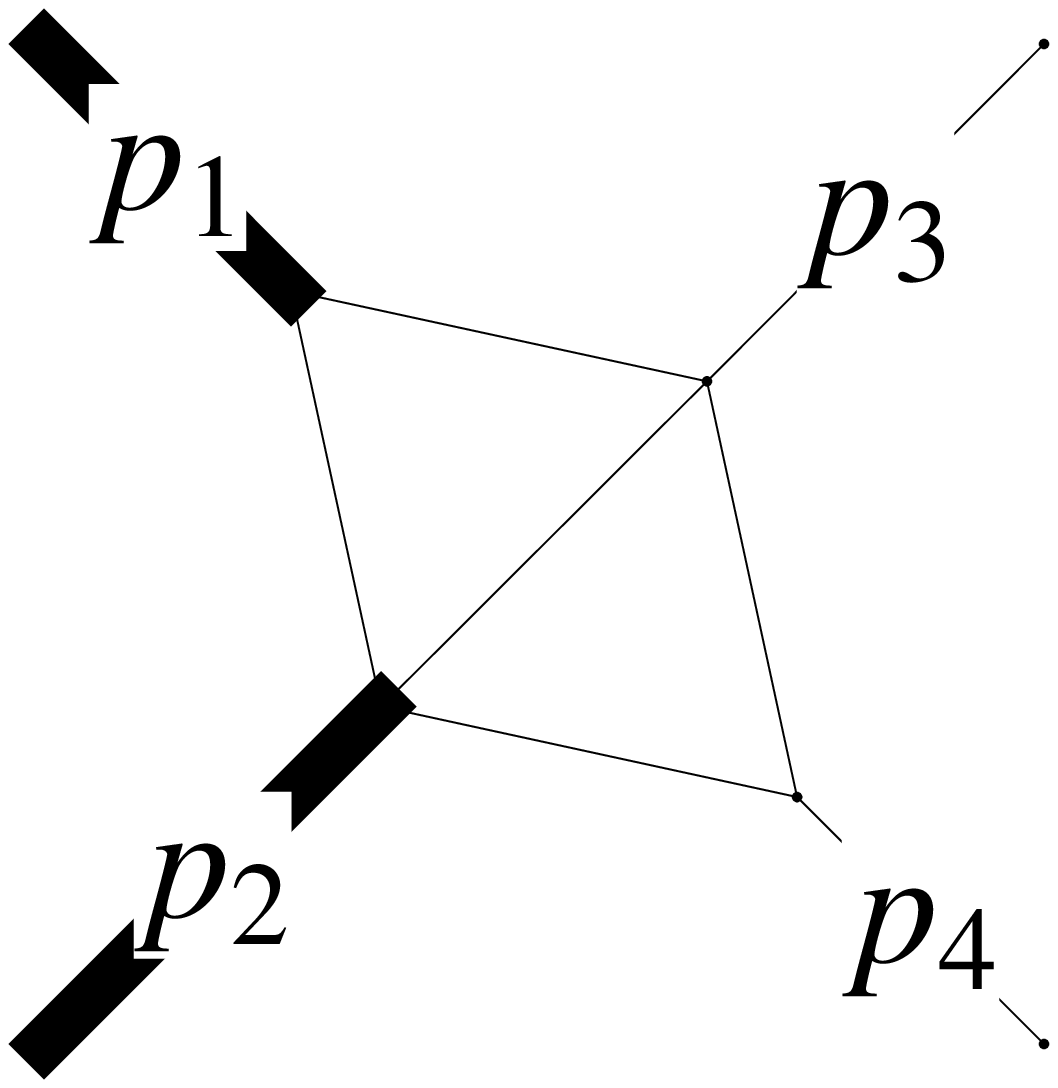}
{  \bea      g^{\rm P12}_{21} &=&   -\eps^4 (p_2^2  - s - t) G_{1, 0, 0, 1, 1, 1, 1, 0, 0} 
\,,~~~~~~~~~~~~~~~~~~~~~~~~~~~~~~~~~~~~~~~~
\\
    f^{\rm P12}_{21} &\sim&   -e^{2i\pi \ep} \left ( \frac{\pi^2 \ep^2}{12} + \frac{\zeta_3 \ep^3}{2}
+ \frac{\pi^4 \ep^4}{40}  \right ) 
- \frac{ x^{-2\ep} }{4} 
\nn \\
& + &     \frac{x^{-2\ep}}{4}
 \left (1 + \frac{\pi^2 \ep^2 }{3} + 14\zeta_3\ep^3 + \frac{2 \pi^4\ep^4 }{3} \right ), \nn
\eea} \nn \\
\picturepage{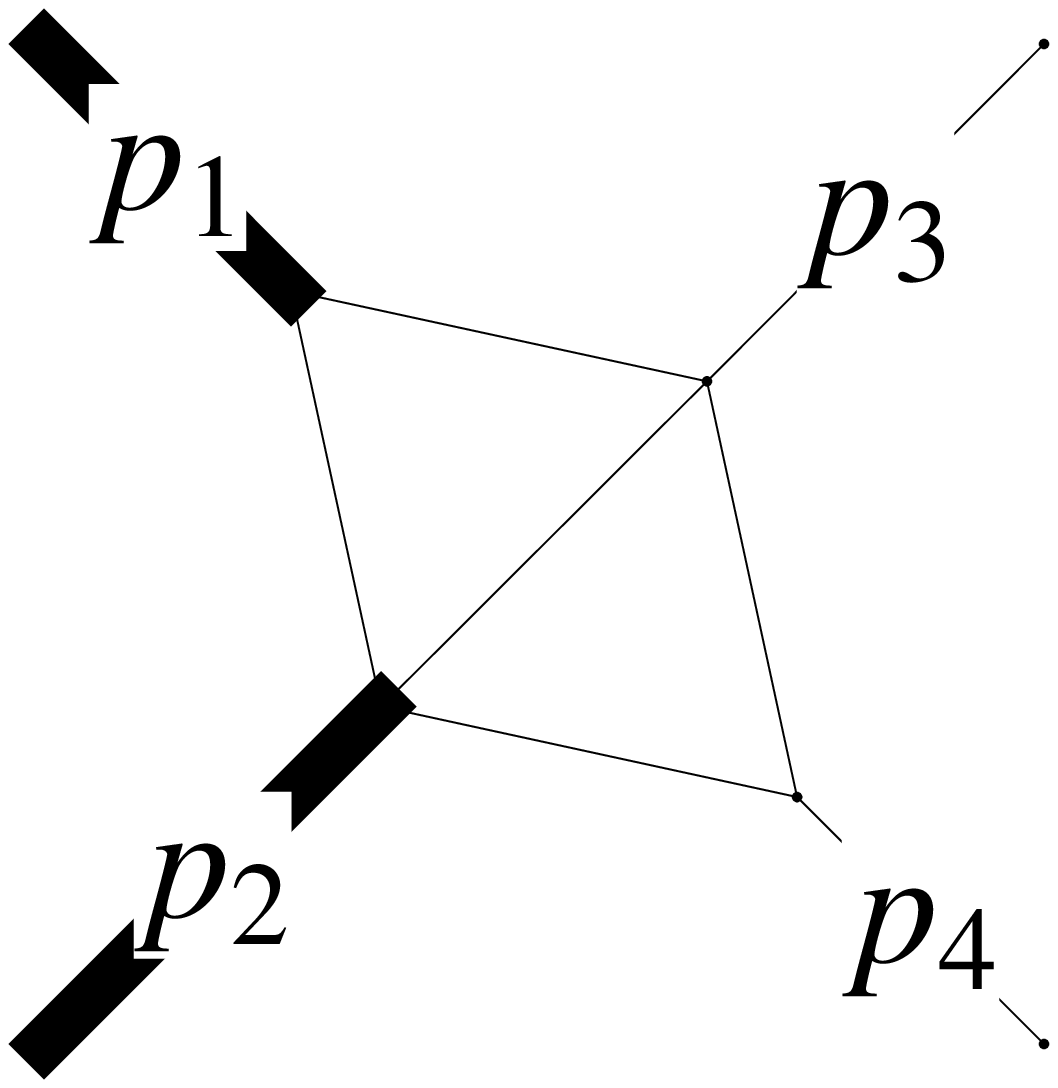}
{  \bea      g^{\rm P12}_{22} &=&\eps^3 \; s \; t \; G_{1, 0, 0, 1, 1, 2, 1, 0, 0}\; ,
\,,~~~~~~~~~~~~~~~~~~~~~~~~~~~~~~~~~~~~~~~~~~~~~~~~~~
 \\
    f^{\rm P12}_{22} &\sim&     \frac{e^{2i\pi \ep}  x^{-4 \ep} }{2} 
- x^{-2 \ep}
 \left (1 - \frac{\pi^2 \ep^2 }{3} - 7\zeta_3\ep^3 - \frac{ \pi^4 \ep^4}{3}  \right ), \nn
\eea} \nn \\
\picturepage{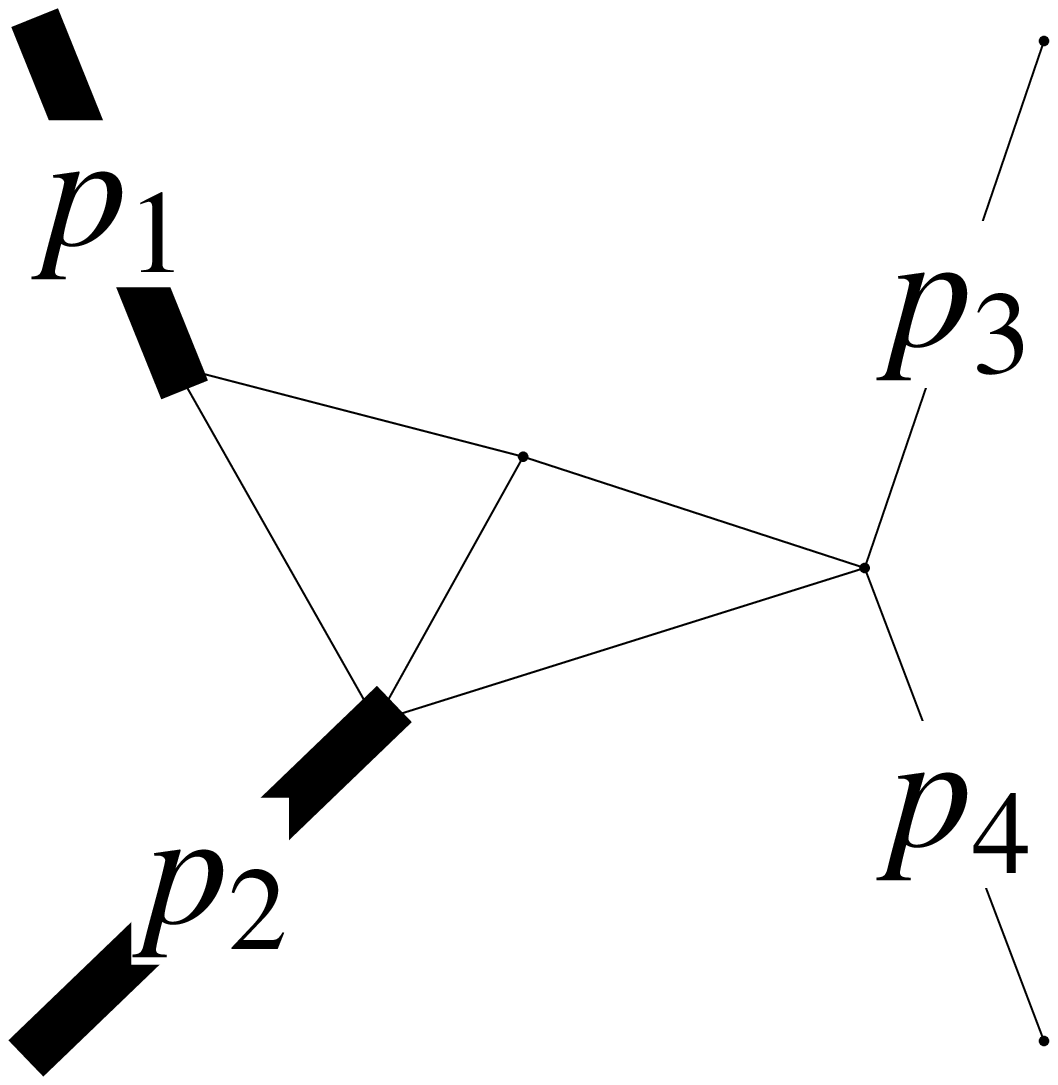}
{  \bea      g^{\rm P12}_{23} &=&  \eps^4 \; R_{12} \; G_{1, 0, 1, 1, 1, 1, 0, 0, 0}  
\,,~~~~~~~~~~~~~~~~~~~~~~~~~~~~~~~~~~~~~~~~~~~~~~~~~~
 \\
    f^{\rm P12}_{23} &\sim& 0, \nn
\eea} \nn \\
\picturepage{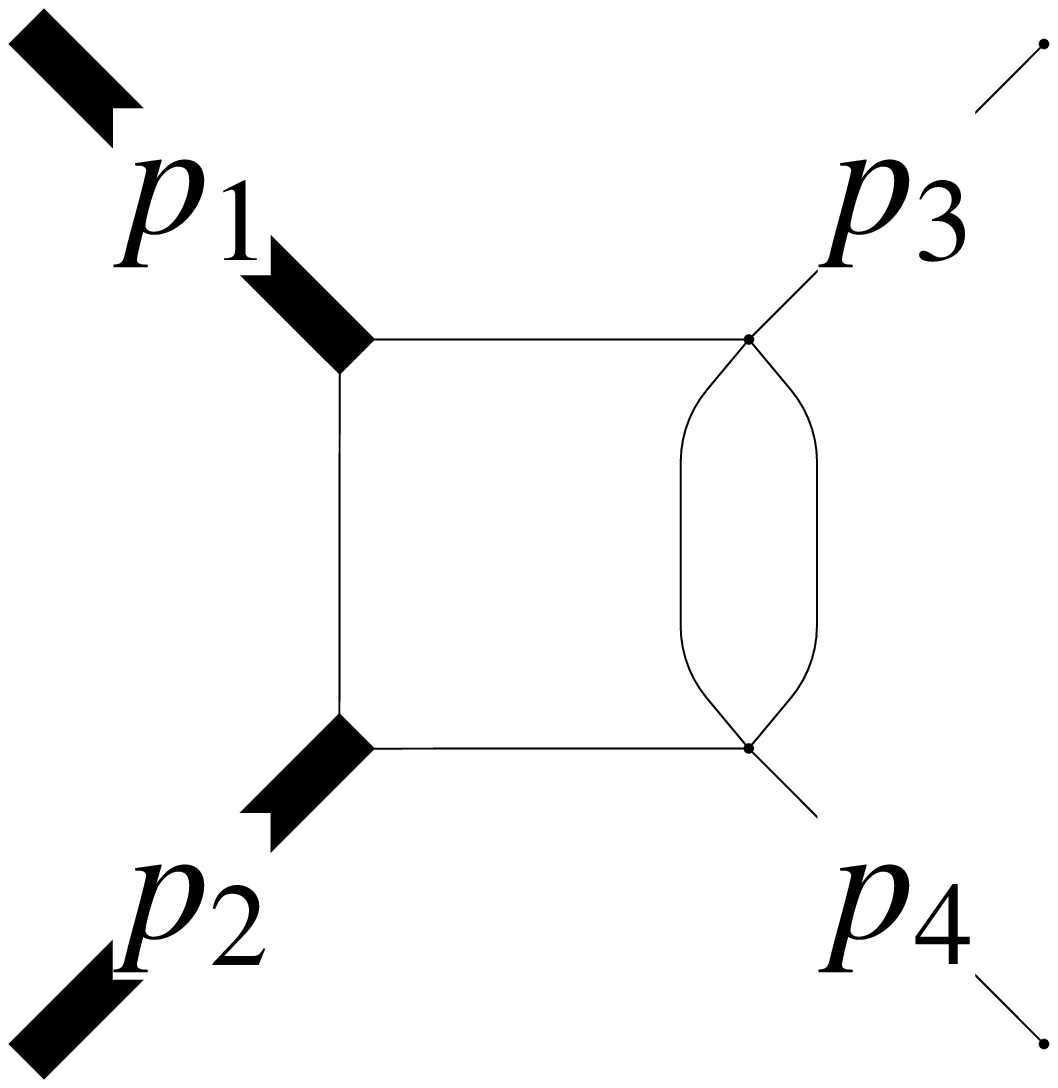}
{  \bea      g^{\rm P12}_{24} &=& \eps^3 \; s \; t \; G_{1, 1, 0, 0, 1, 1, 2, 0, 0}   
\,,~~~~~~~~~~~~~~~~~~~~~~~~~~~~~~~~~~~~~~~~~~~~~~~~~~~~
\\
    f^{\rm P12}_{24} &\sim&  -\frac{3 x^{-2\ep}}{4} + 
       \frac{x^{-3 \ep}}{2}N_1, \nn
\eea} \nn \\
%\end{align} 
%\end{small}
%
%\begin{small}
%\begin{align}
\picturepage{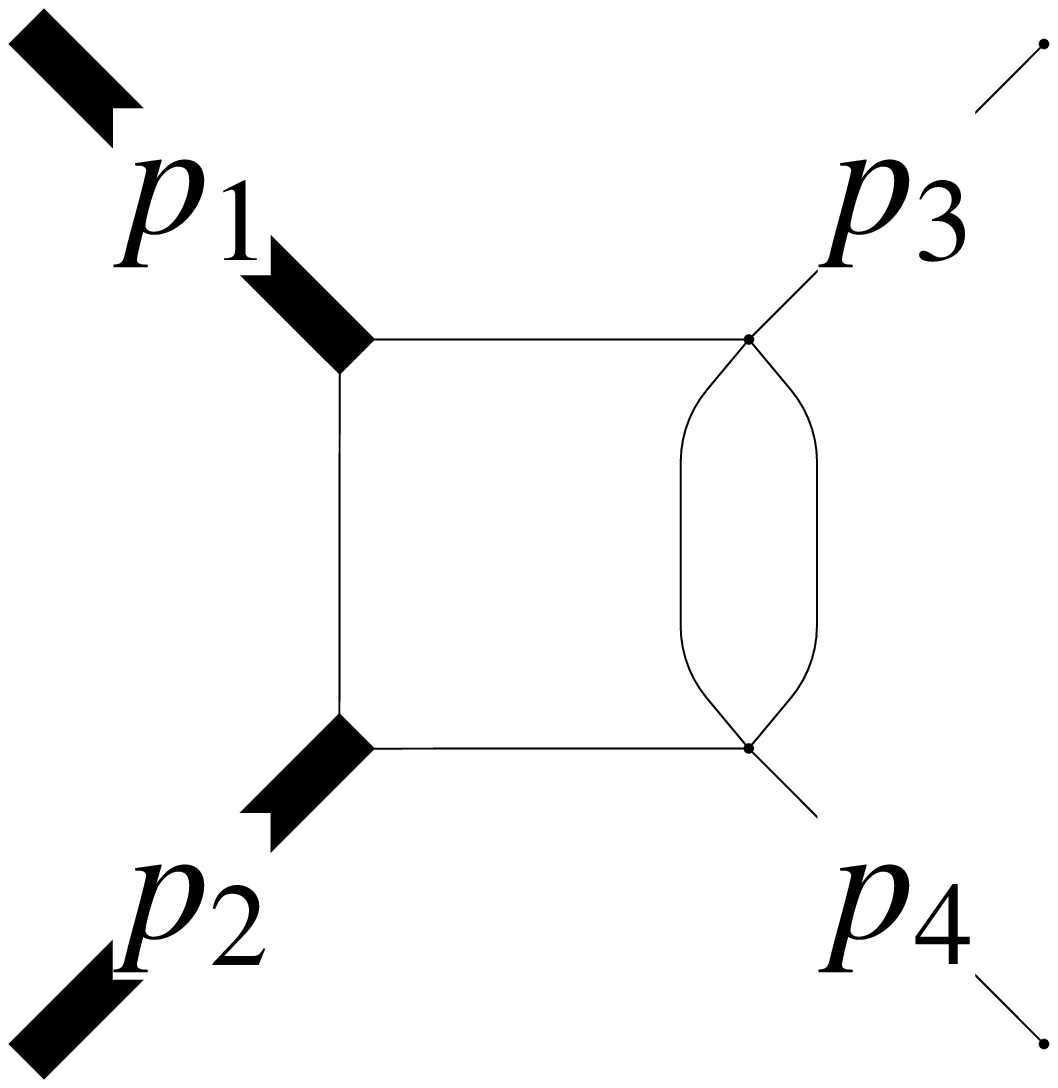}
{  \bea      g^{\rm P12}_{25} &=&  \eps^3 R_{12} G_{1, 1, 0, 0, 1, 1, 2, 0, -1}  
\,,~~~~~~~~~~~~~~~~~~~~~~~~~~~~~~~~~~~~~~~~~~
  \\
    f^{\rm P12}_{25} &\sim&  0, \nn
\eea} \nn \\
\picturepage{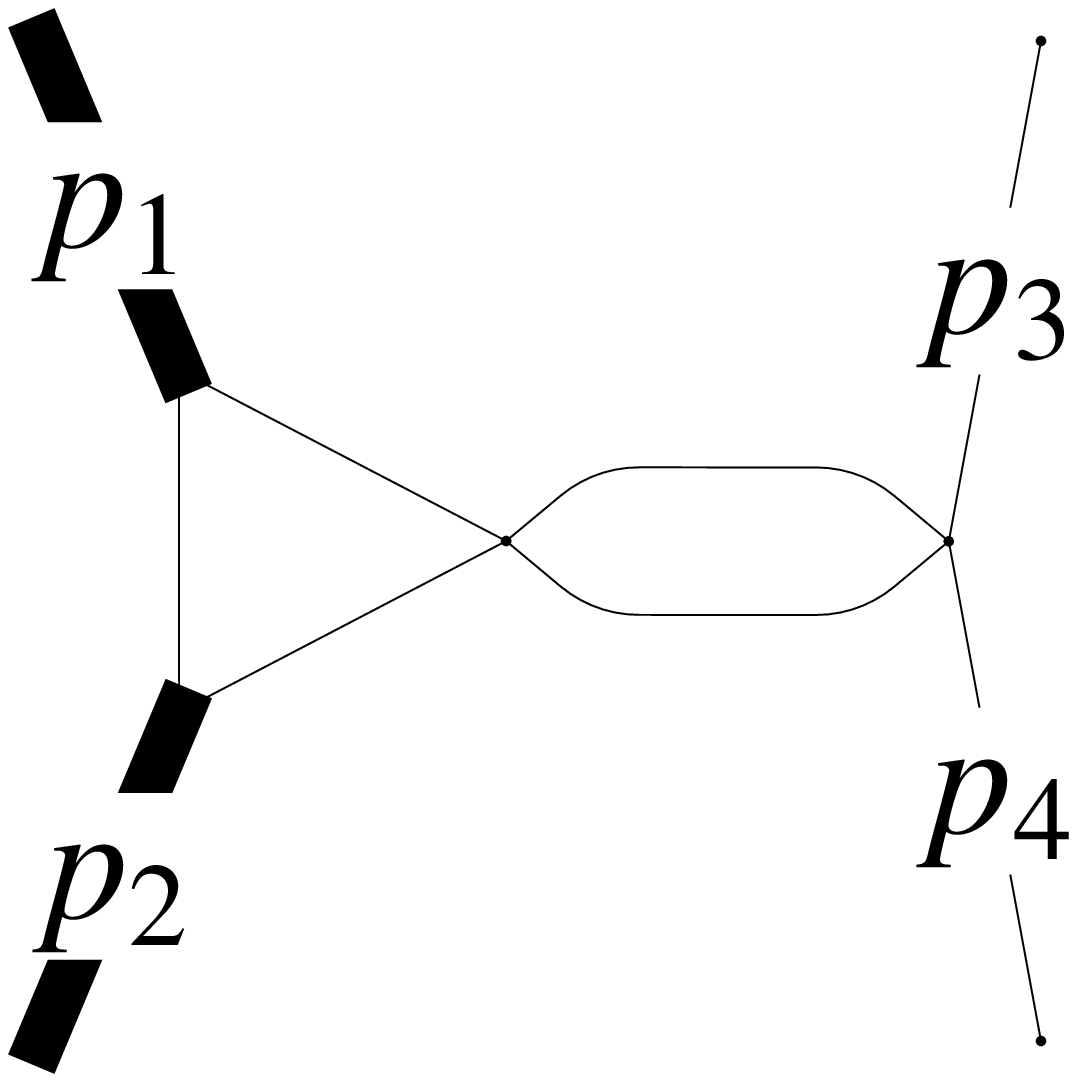}
{  \bea      g^{\rm P12}_{26} &=&    \eps^3 R_{12} s G_{1, 1, 1, 2, 1, 0, 0, 0, 0} 
\,,~~~~~~~~~~~~~~~~~~~~~~~~~~~~~~~~~~~~~~~~~~~~
 \\
    f^{\rm P12}_{26} &\sim& 0, \nn
\eea} \nn \\
\picturepage{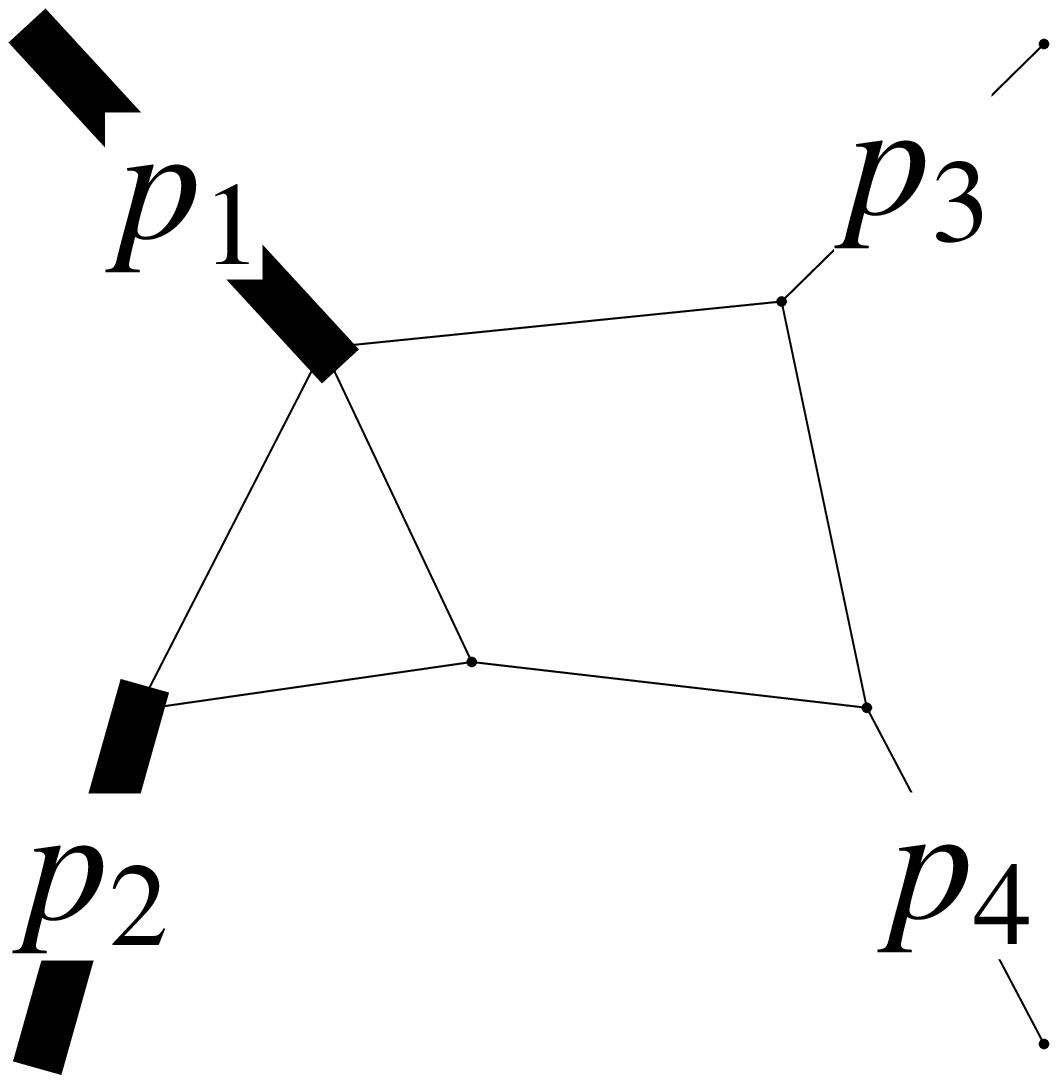}
{  \bea      g^{\rm P12}_{27} &=& -\eps^4 s (p_2^2  - t) G_{0, 1, 1, 1, 1, 1, 1, 0, 0} 
\,,~~~~~~~~~~~~~~~~~~~~~~~~~~~~~~~~~~~~~~
  \\
    f^{\rm P12}_{27} &\sim& \frac{e^{2i\pi \ep} x^{-4\ep} }{4} 
            + \frac{3 x^{-2 \ep}}{4}  - x^{-3 \ep} N_1, \nn
\eea} \nn \\
\picturepage{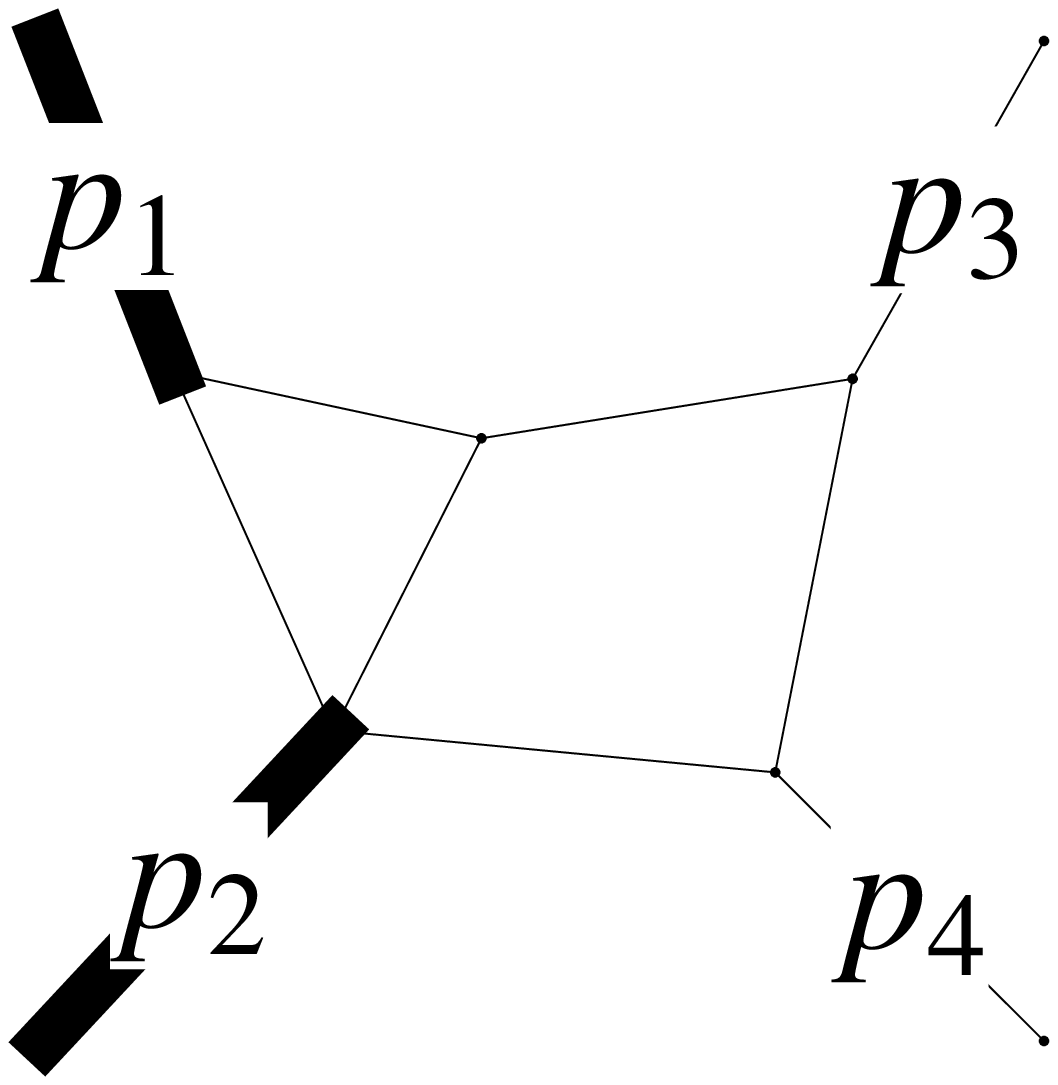}
{  \bea      g^{\rm P12}_{28} &=& -\eps^4 s (p_1^2  - t) G_{1, 0, 1, 1, 1, 1, 1, 0, 0} 
\,,~~~~~~~~~~~~~~~~~~~~~~~~~~~~~~~~~~~~~~~
 \\
    f^{\rm P12}_{28} &\sim&   -\frac{e^{2i\pi \ep}}{4} \left ( 1 + \ep^2\pi^2 +  30\zeta_3 \ep^3 
  + \frac{7\ep^4 \pi^4}{10} \right ) 
\nn \\
   & +&  
     \frac{x^{-2\ep}}{4} \left ( 1 + \frac{\pi^2 \ep^2}{3} + 14 \zeta_3 \ep^3 + \frac{2 \pi^4 \ep^4}{3} 
          \right ), \nn
\eea} \nn \\
\picturepage{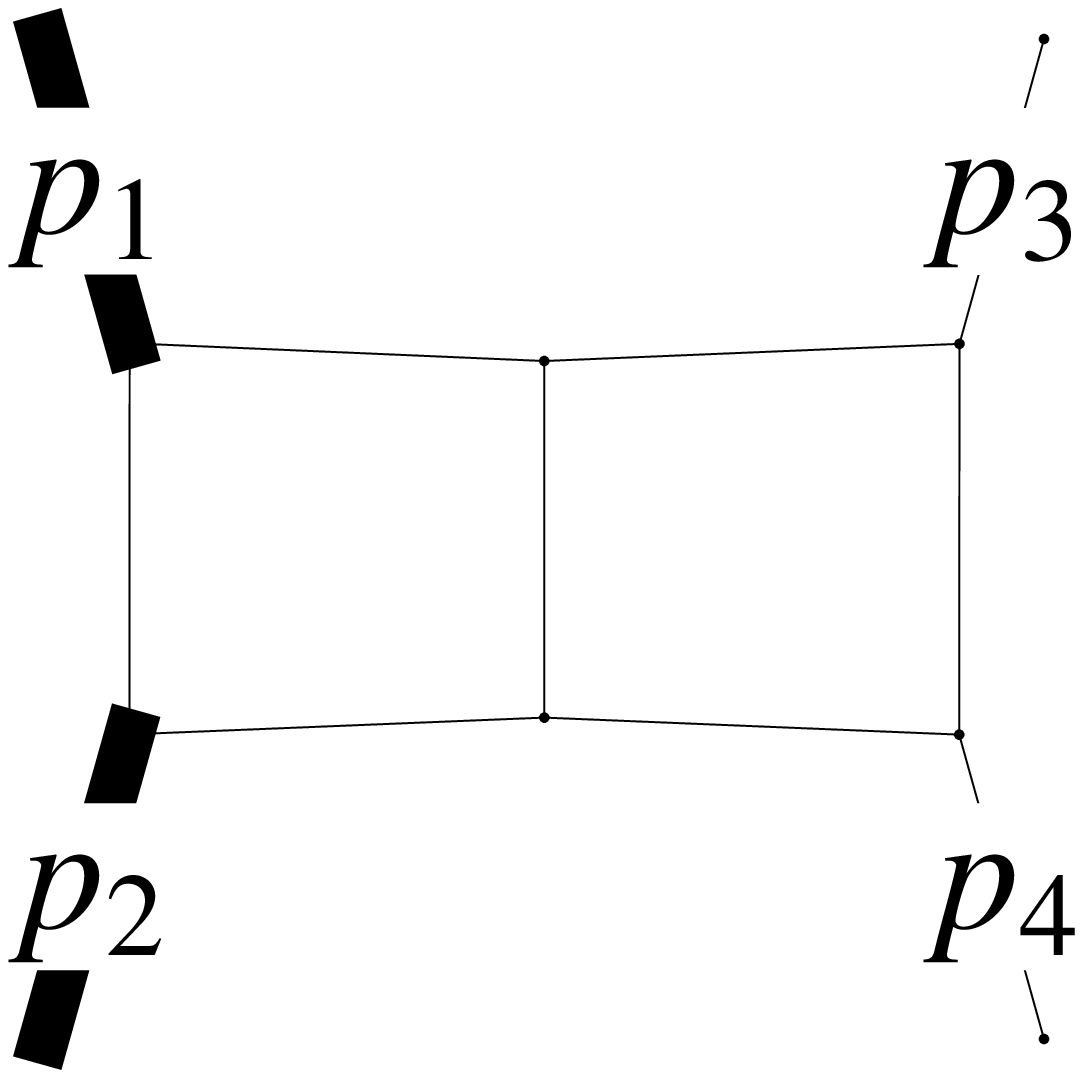}
{  \bea      g^{\rm P12}_{29} &=& \eps^4 \; s^2 \; t \; G_{1, 1, 1, 1, 1, 1, 1, 0, 0}    \,,
\,,~~~~~~~~~~~~~
 \\
    f^{\rm P12}_{29} &\sim&   -\frac{e^{2i\pi \ep} x^{-4\ep}}{4} + 
     x^{-3\ep} N_1
      - \frac{x^{-2\ep}}{2} \left ( 2 + \frac{\pi^2 \ep^2}{6} + 7 \zeta_3 \ep^3 
        + \frac{\pi^4 \ep^4}{3} \right ),  \nn
\eea} \nn \\
\picturepage{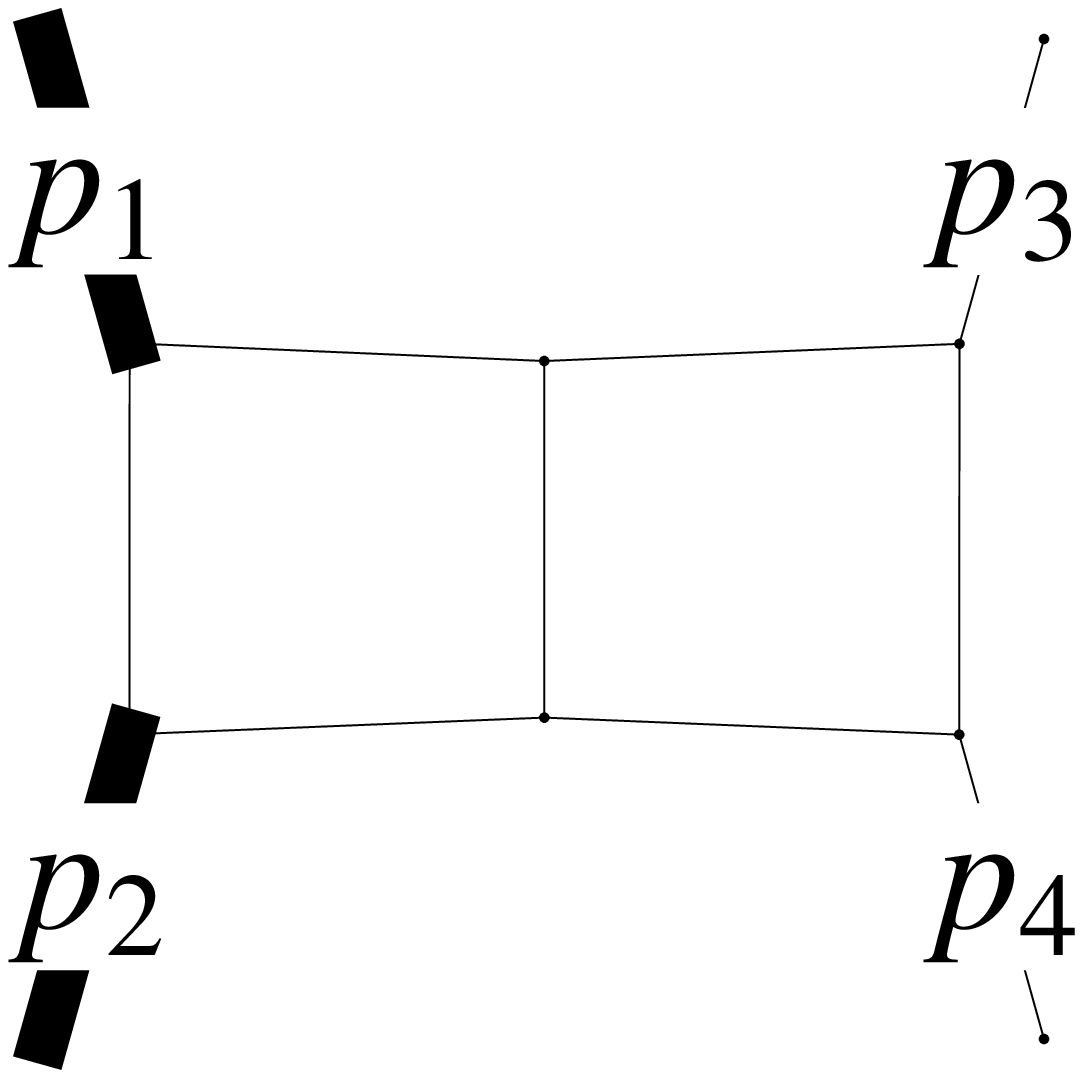}
{  \bea       g^{\rm P12}_{30} &=& \eps^2 \Big[-\frac{1}{2} \eps \; p_1^2  s G_{0, 1, 1, 0, 1, 2, 1, 0, 0} - 
    \frac{1}{2} \eps \; p_2^2  s G_{1, 0, 0, 1, 1, 2, 1, 0, 0}     
\,,~~~~~~~~~~
 \\
 &&+ \eps (p_1^2  + p_2^2 ) s G_{1, 1, 0, 0, 1, 1, 2, 0, 0} + 
    \eps^2 s^2 G_{1, 1, 1, 1, 1, 1, 1, -1, 0}\Big]\,, \nn \\    
      f^{\rm P12}_{30} &\sim&   \frac{3}{4} x^{-2\ep} -
    \frac{x^{-3\ep}}{2} N_1, \nonumber
    \eea  } \nn \\
\picturepage{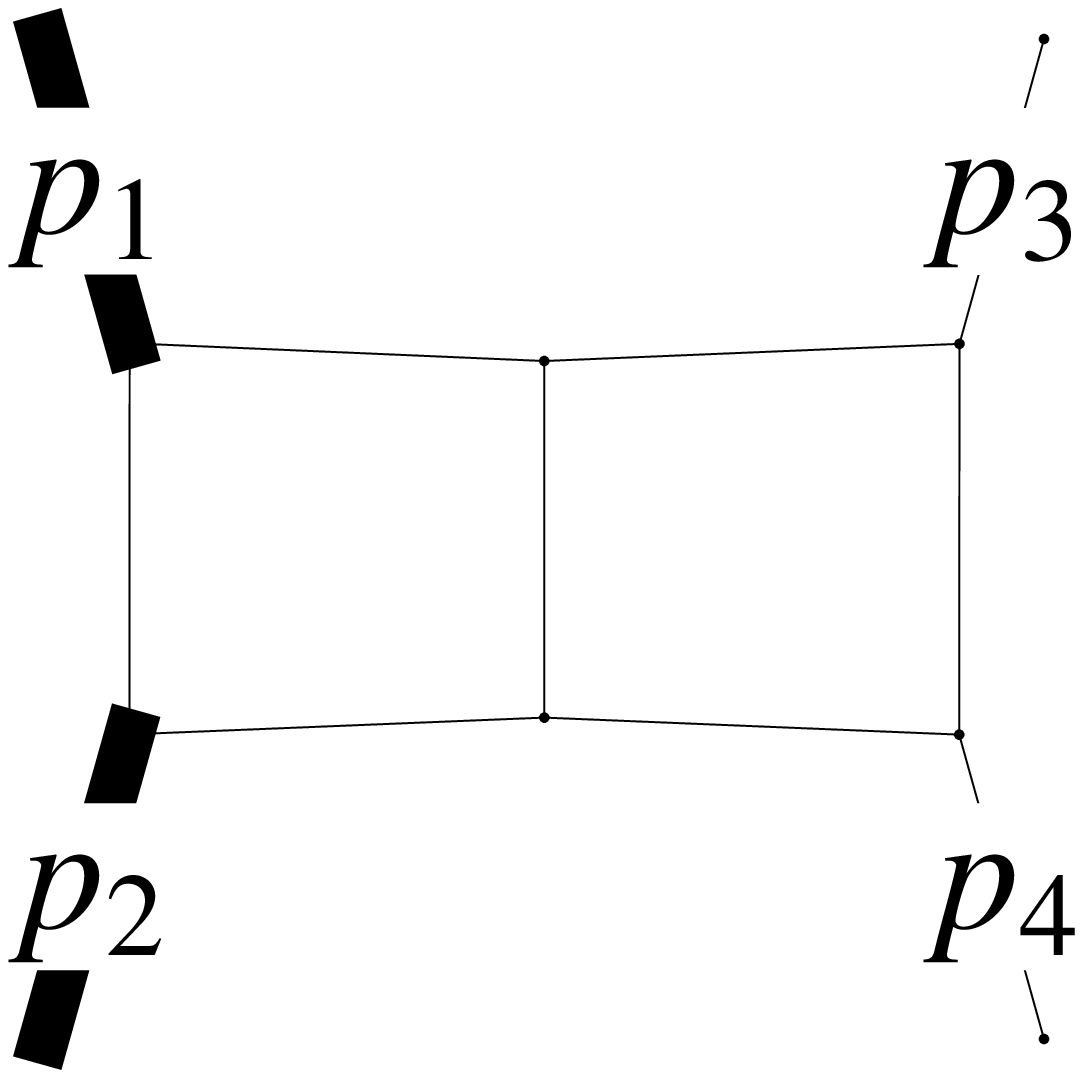}
{  \bea       
    g_{31}^{\rm P12} & =&\eps^4 R_{12} s G_{1, 1, 1, 1, 1, 1, 1, 0, -1}
\,,~~~~~~~~~~~~~~~~~~~~~~~~~~~~~~~~~~~~~~~~~~~~~~
\\
f^{\rm P12}_{31} &= & 0.     \nn
\eea  } \nn 
 \end{align}
\end{small}
%
%
%%%%%%%%%%%%%%%%%%%%%%%%%%%%%   P13
%\newpage 

The master integrals for the family P13 and their limits in the kinematic point $x \to 0,~ y \to 1,~ 
z \to 1$  read 
\begin{small}
\begin{align}
\picturepage{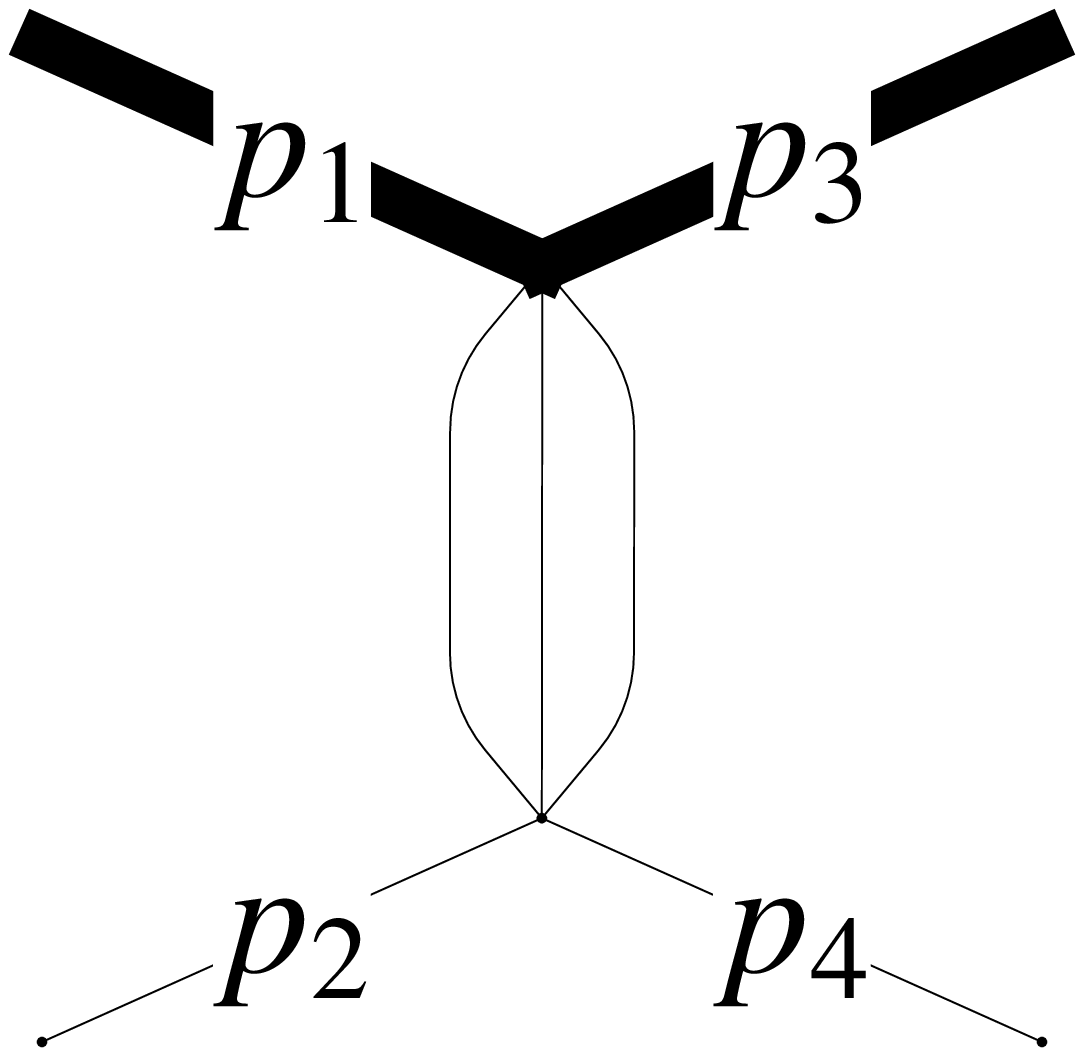}
{  \bea      g^{\rm P13}_1 &=& \eps^2 t G_{0, 0, 0, 0, 1, 2, 2, 0, 0} 
\,,~~~~~~~~~~~~~~~~~~~~~~~~~~~~~~~~~~~~~~~~~~~~~
\\
    f^{\rm P13}_{1} &\sim&  -e^{2i\pi \ep},\;\;\;   \nn
\eea} \nn \\
\picturepage{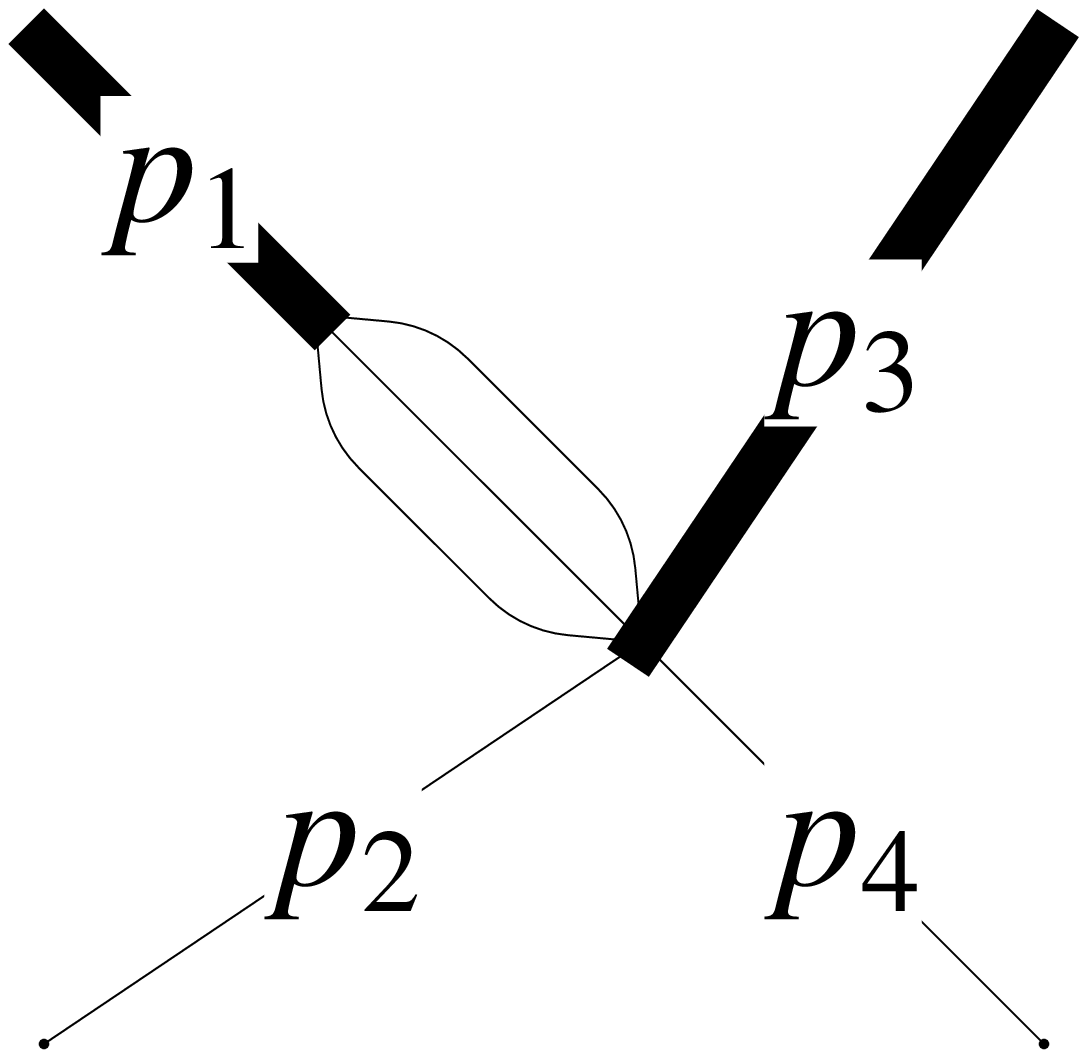}
{  \bea      g^{\rm P13}_{2} &=&  \eps^2 p_1^2  G_{0, 0, 1, 0, 2, 2, 0, 0, 0}  
\,,~~~~~~~~~~~~~~~~~~~~~~~~~~~~~~~~~~~~~~~~~~~~~~
 \\
    f^{\rm P13}_{2} &\sim& -e^{2i\pi \ep}, \;\;\;\  \nn
\eea} \nn  
\\
\picturepage{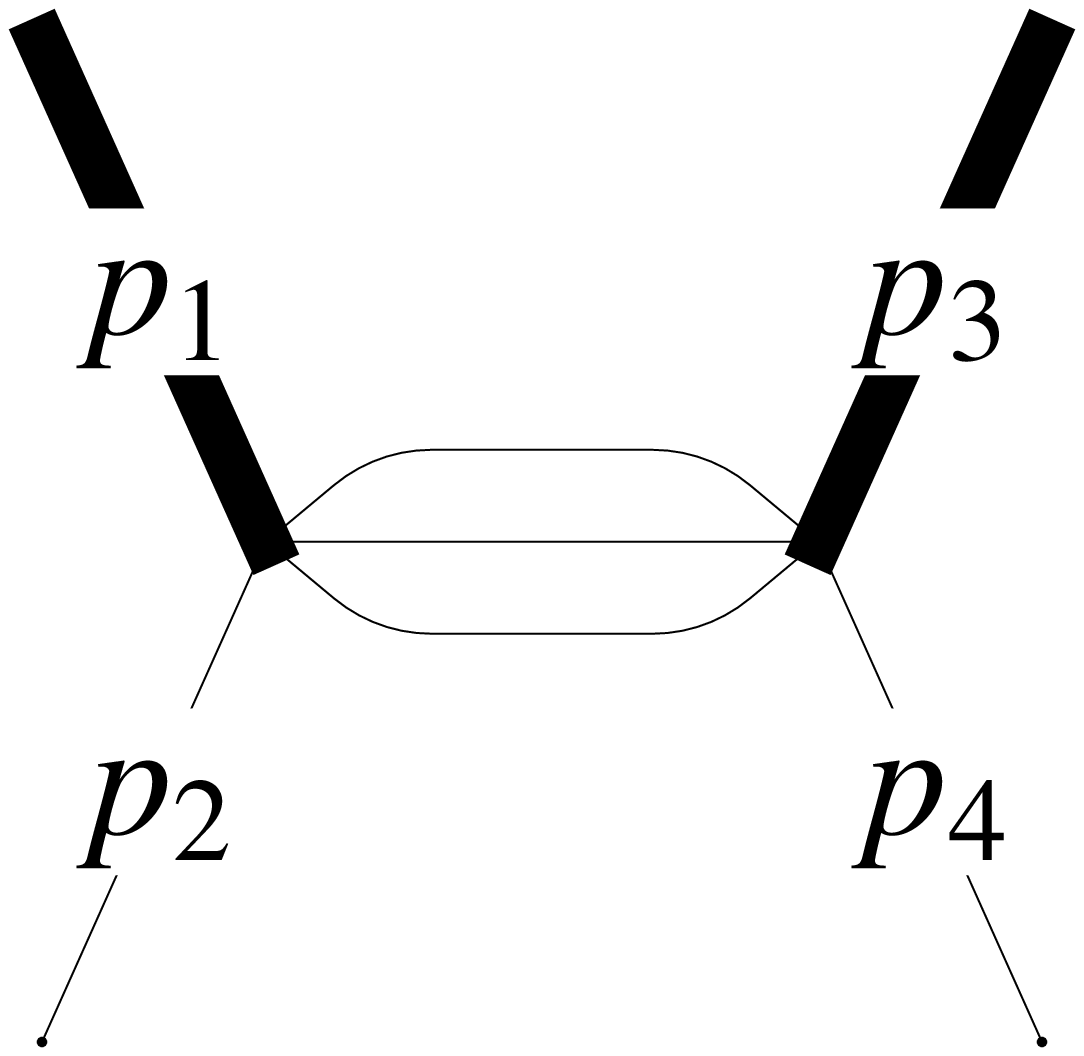}
{  \bea      g^{\rm P13}_{3} &=&  \eps^2 s G_{0, 2, 2, 0, 0, 1, 0, 0, 0}
\,,~~~~~~~~~~~~~~~~~~~~~~~~~~~~~~~~~~~~~~~~~~~~~~~~ 
\\
    f^{\rm P13}_{3} &\sim& -x^{-2\ep},  \nn
\eea} \nn 
\\
\picturepage{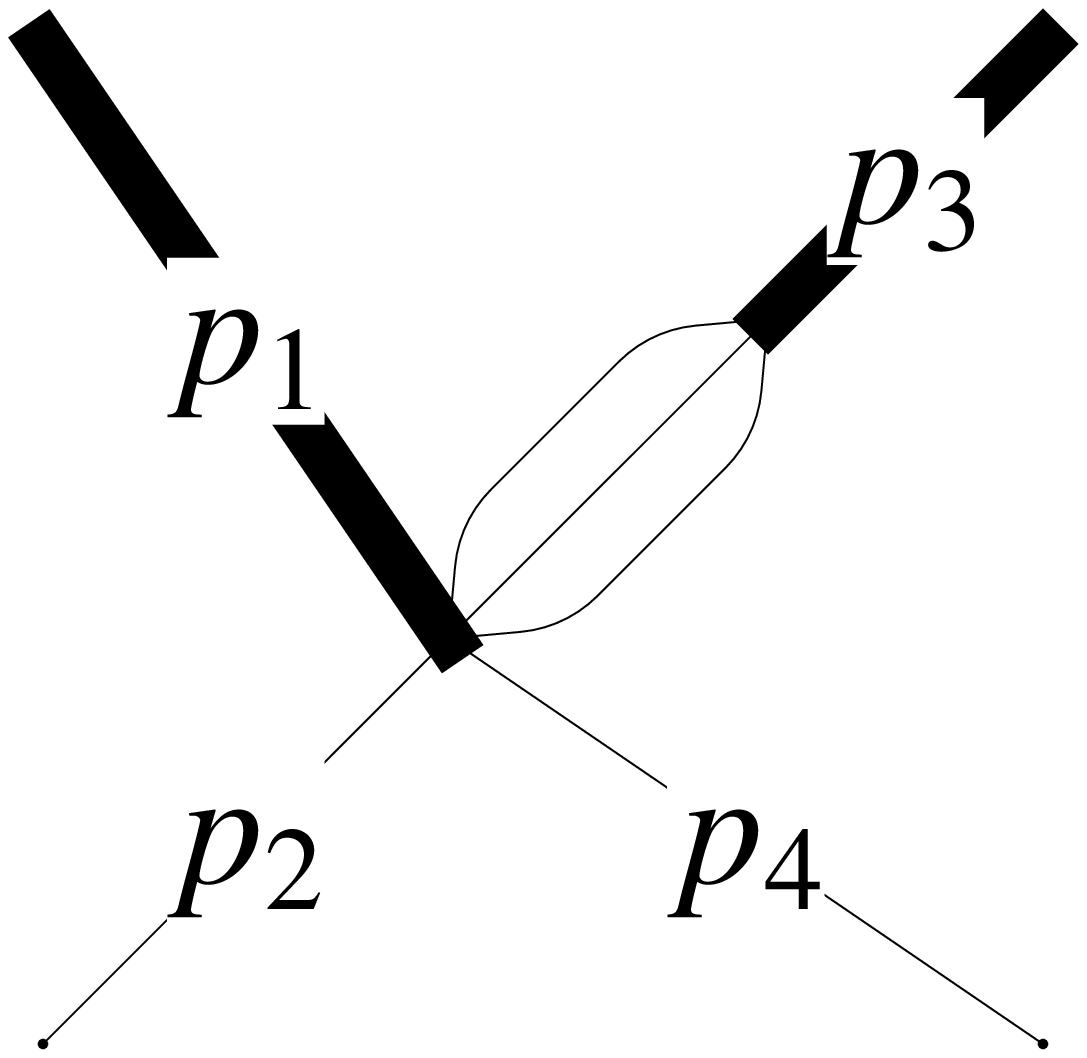}
{  \bea      g^{\rm P13}_{4} &=&  \eps^2 p_3^2  G_{1, 0, 0, 0, 0, 2, 2, 0, 0}
\,,~~~~~~~~~~~~~~~~~~~~~~~~~~~~~~~~~~~~~~~~~~~~~~~
 \\
    f^{\rm P13}_{4} &\sim& -e^{2i\pi \ep} x^{-4\ep},\nn
\eea} \nn 
\\
\picturepage{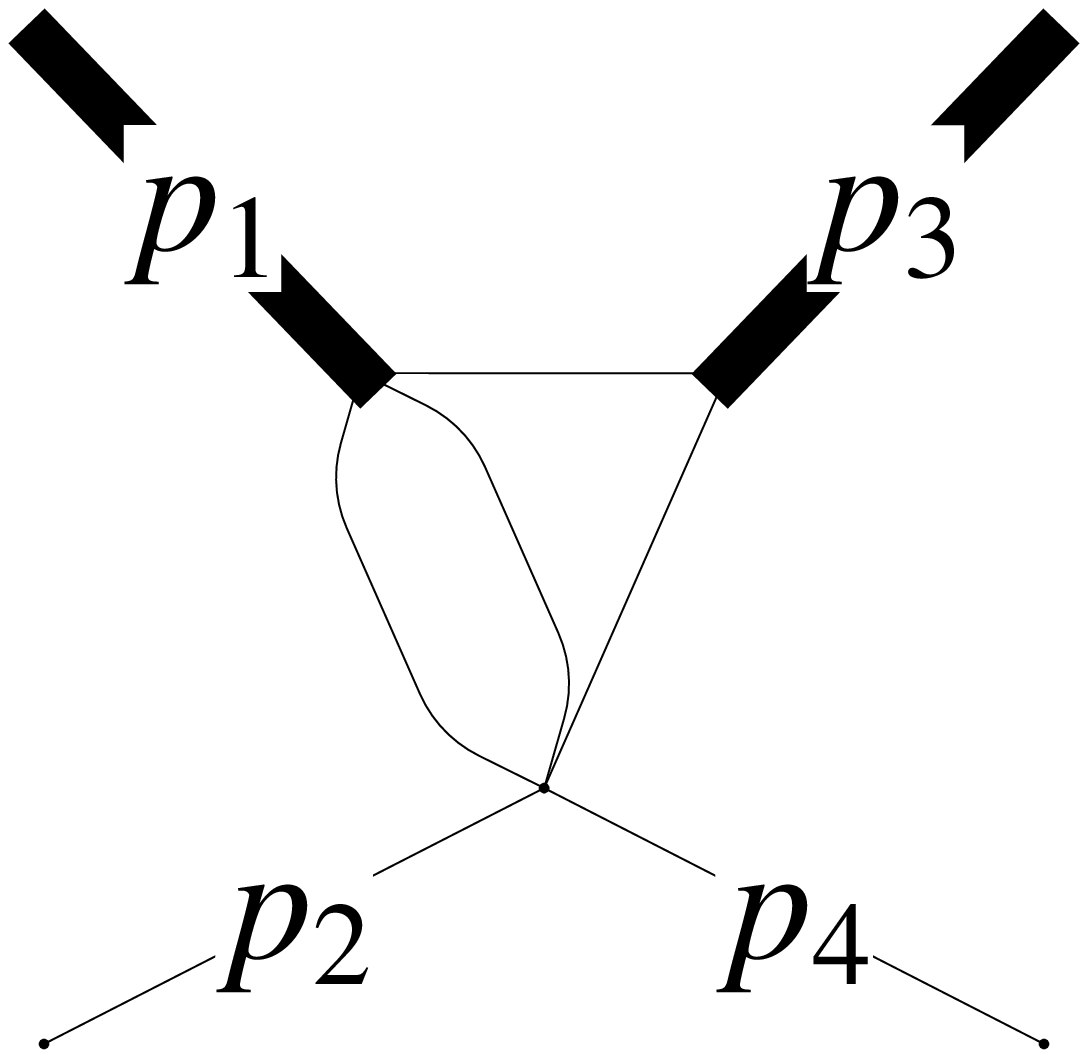}
{  \bea      g^{\rm P13}_{5} &=&  \eps^3 R_{13} G_{0, 0, 1, 0, 1, 2, 1, 0, 0}
\,,~~~~~~~~~~~~~~~~~~~~~~~~~~~~~~~~~~~~~~~~~~~~~~
 \\
    f^{\rm P13}_{5} &\sim& 0\;, \nn
\eea} \nn 
\\
\picturepage{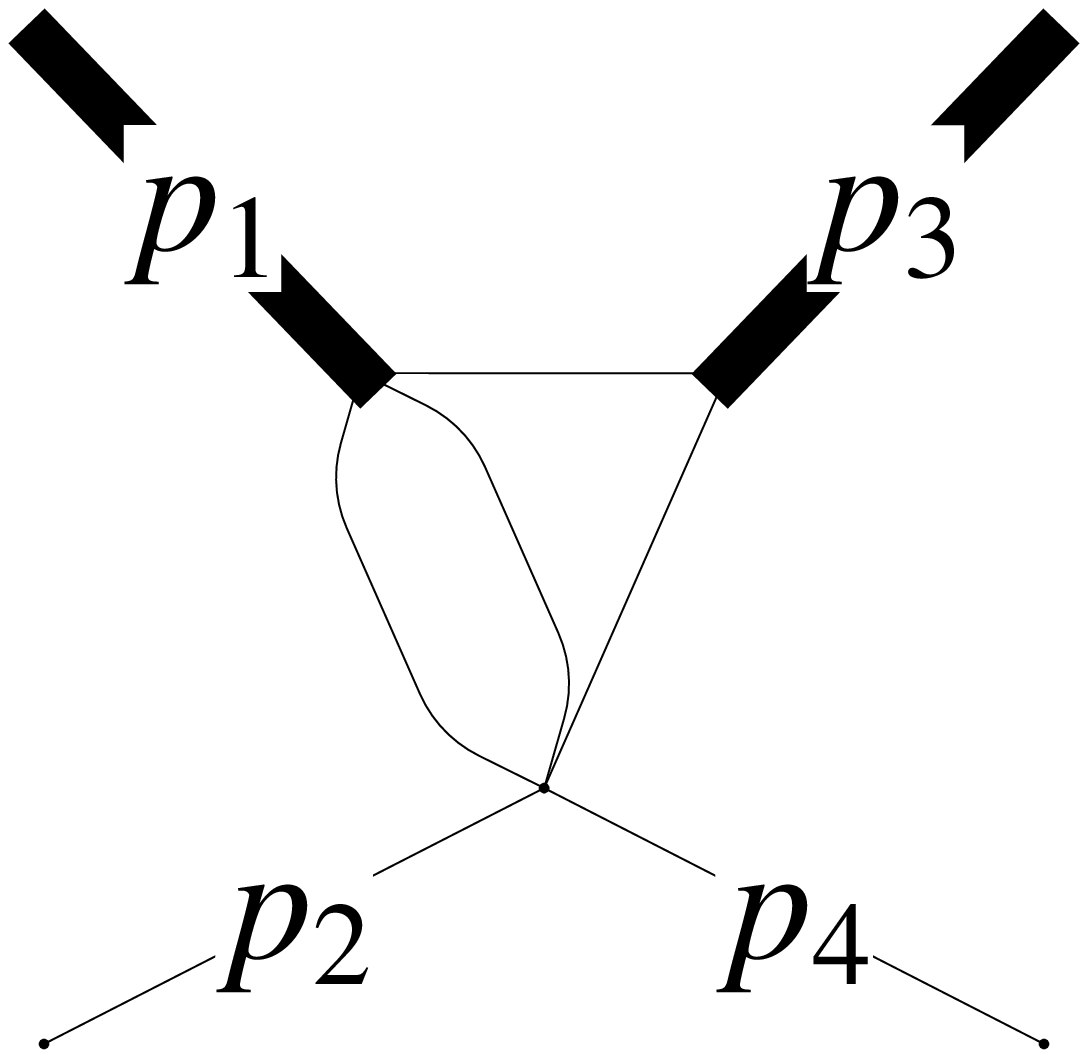}
{  \bea      g^{\rm P13}_{6} &=&  \eps^2 \Big[ \frac{1}{2} \eps (t-p_1^2  + p_3^2 ) G_{0, 0, 1, 0, 1, 2, 1, 0, 0} + 
    p_3^2  G_{0, 0, 2, 0, 1, 2, 1, -1, 0}\Big]\,
\;,
 \\
    f^{\rm P13}_{6} &\sim& -e^{2i\pi \ep} x^{-2\ep} \left (1 + 6 \zeta_3\ep^3 + \frac{\pi^4 \ep^4}{10} \right ),  \nn
\eea} 
\nn  \\
\picturepage{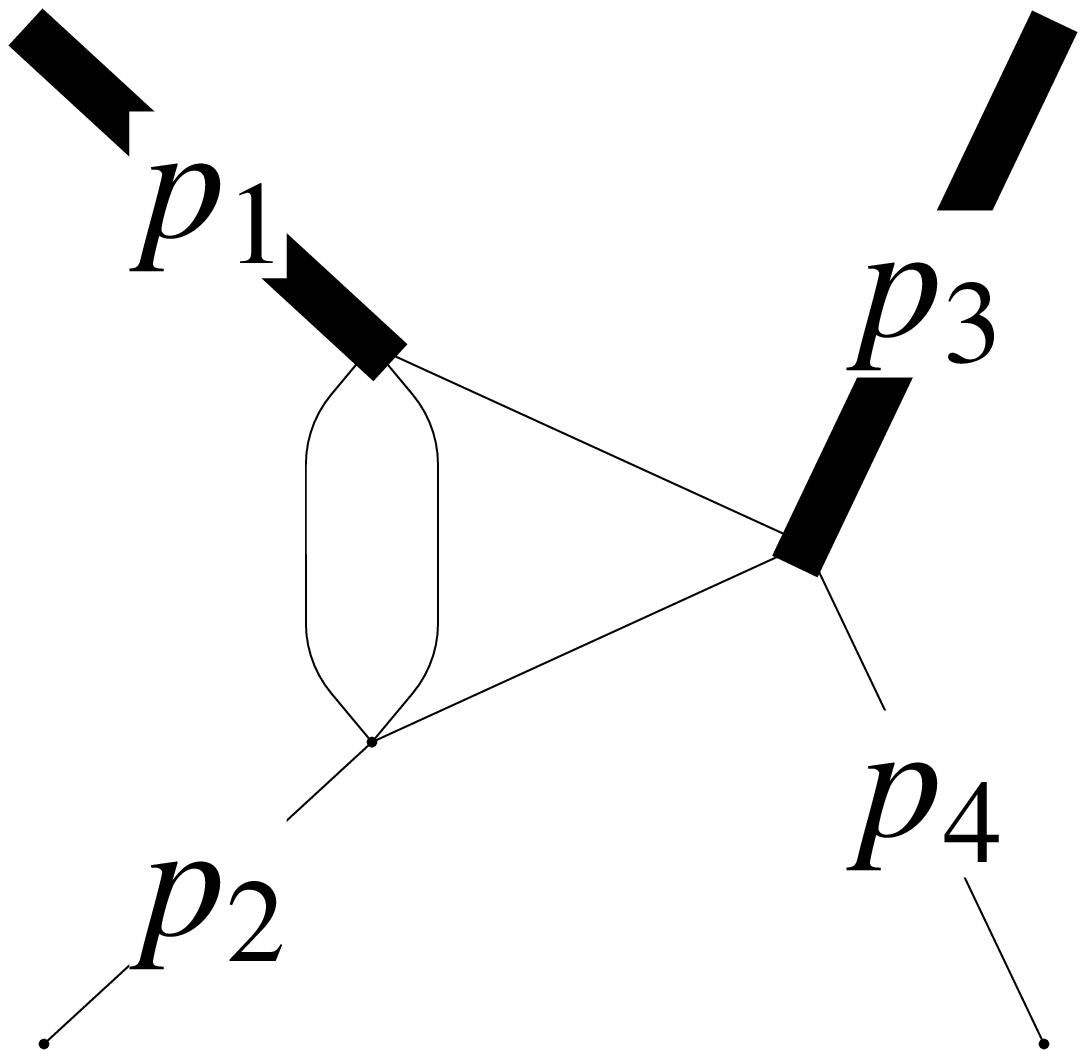}
{  \bea      g^{\rm P13}_{7} &=&  \eps^3 (p_1^2  - s) G_{0, 0, 1, 1, 1, 2, 0, 0, 0}\,, 
\;\;\;~~~~~~~~~~~~~~~~~~~~~~~~~~~~~~~~~~~~~~
 \\
    f^{\rm P13}_{7} &\sim&  \frac{e^{2i\pi \ep}}{2} -\frac{x^{-\ep}}{2}  N_1, \nn
\eea} \nn 
\\
\picturepage{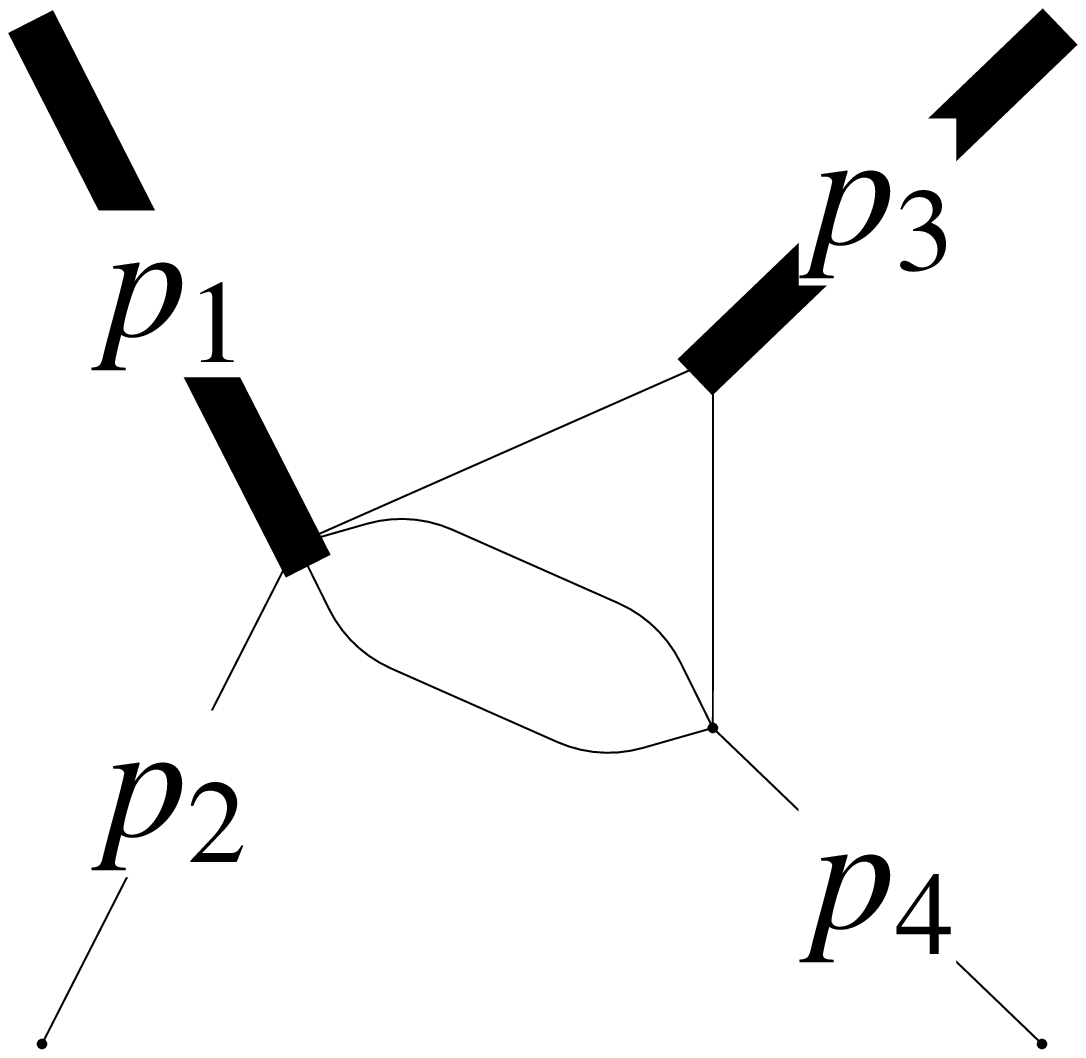}
{  \bea      g^{\rm P13}_{8} &=&  \eps^3 (p_3^2  - s) G_{0, 1, 1, 0, 0, 2, 1, 0, 0}
\,,~~~~~~~~~~~~~~~~~~~~~~~~~~~~~~~~~~~~~~~~~
 \\
    f^{\rm P13}_{8} &\sim& -\frac{x^{-2\ep}}{2} + \frac{x^{-3\ep}}{2} N_1, \nn
\eea} \nn \\
\picturepage{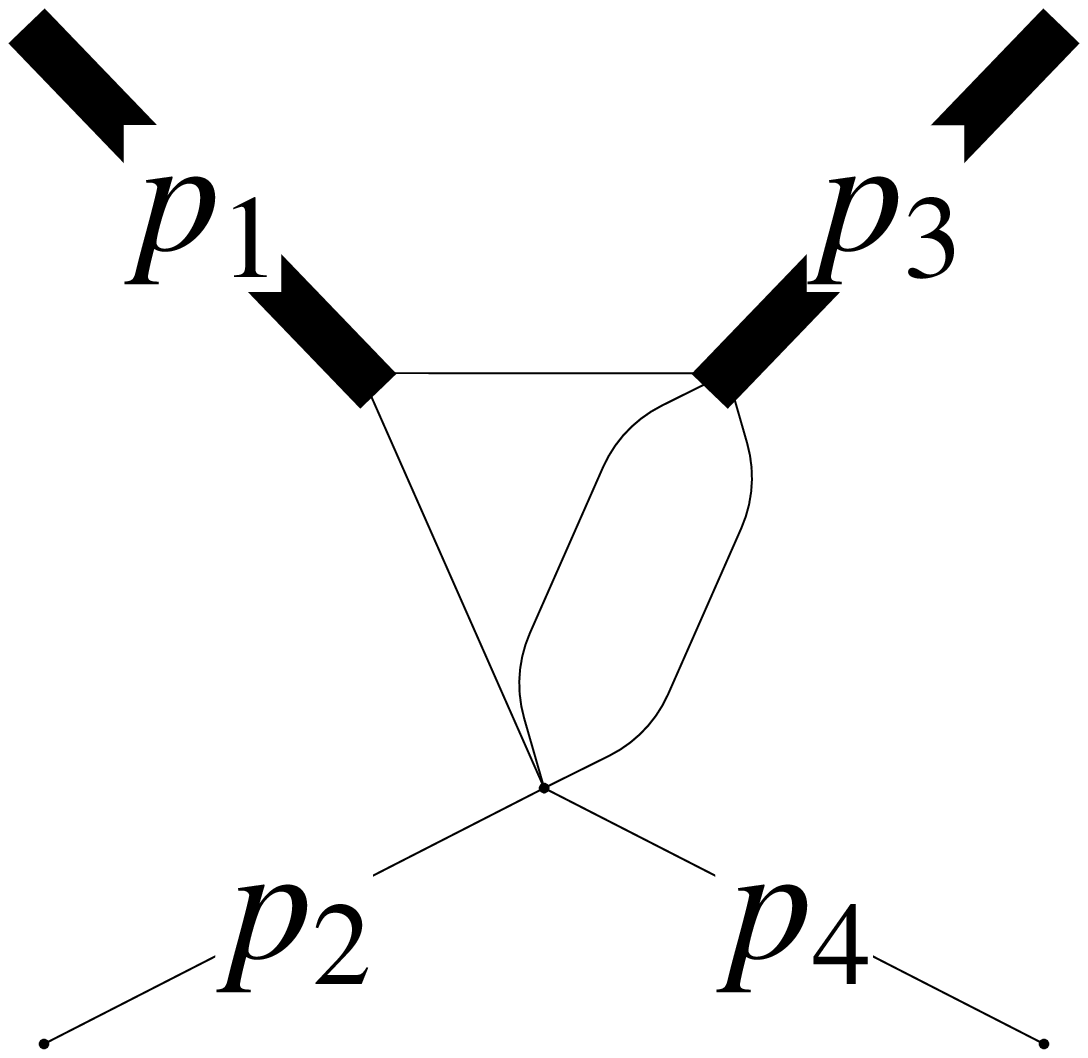}
{  \bea      g^{\rm P13}_{9} &=&   \eps^3 R_{13} G_{1, 0, 0, 0, 1, 2, 1, 0, 0}
\,,~~~~~~~~~~~~~~~~~~~~~~~~~~~~~~~~~~~~~~~~~~~~~
 \\
    f^{\rm P13}_{9} &\sim&   0,  \nn
\eea} \nn  \\
%\end{align} 
%\end{small}
%%%%%%%%%%%%%%%%%%%%%%%%%%%%%%%%%
%\begin{small}
%\begin{align}
\picturepage{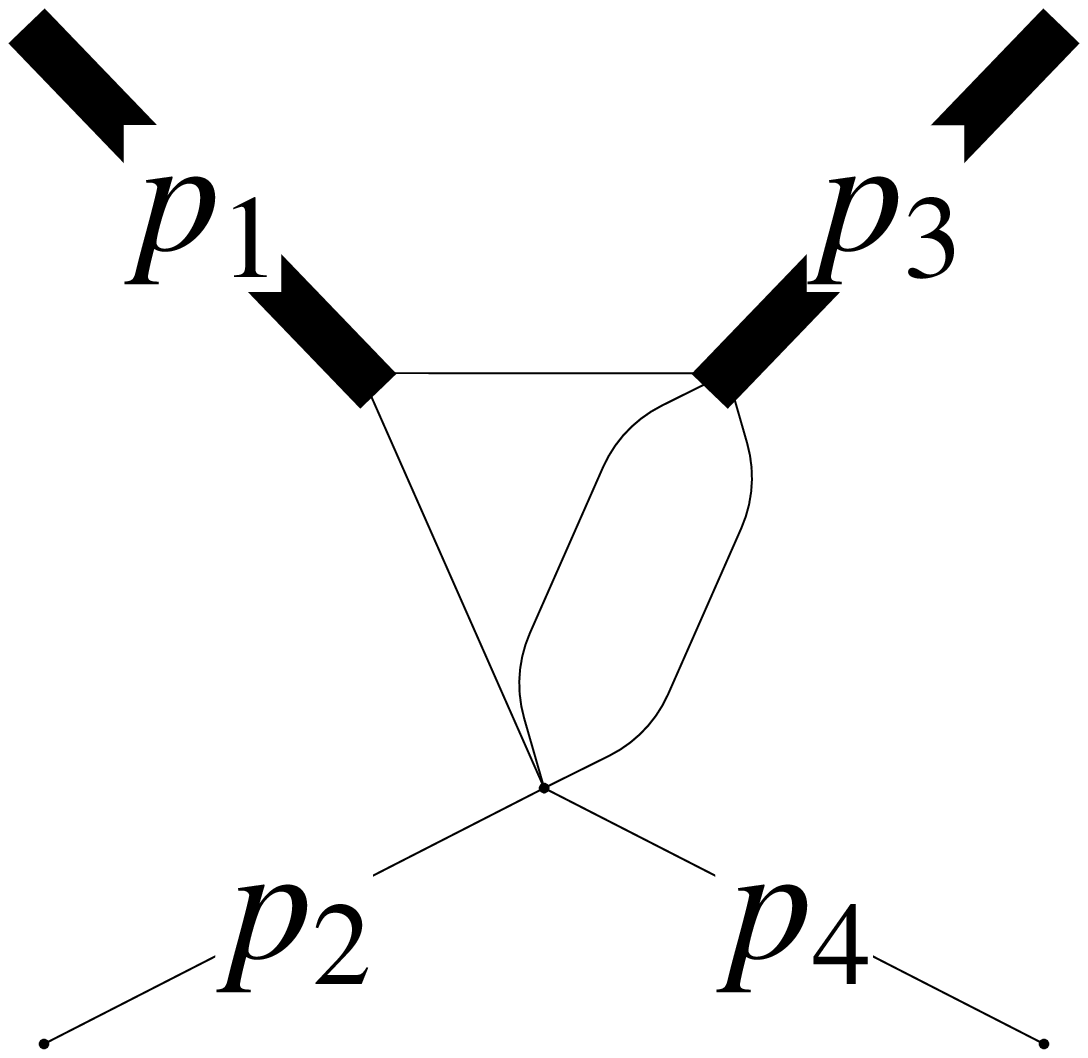}
{  \bea      g^{\rm P13}_{10} &=&  
  \frac{1}{2} \eps^2 \Big[ \eps (p_1^2  + p_3^2  - t) G_{1, 0, 0, 0, 1, 2, 1, 0, 0} + 
    2 p_1^2  G_{1, 0, 0, 0, 2, 2, 1, 0, -1}\Big]
\,,
 \\
    f^{\rm P13}_{10} &\sim& -e^{2i\pi \ep}, \nn
\eea} \nn 
\\
\picturepage{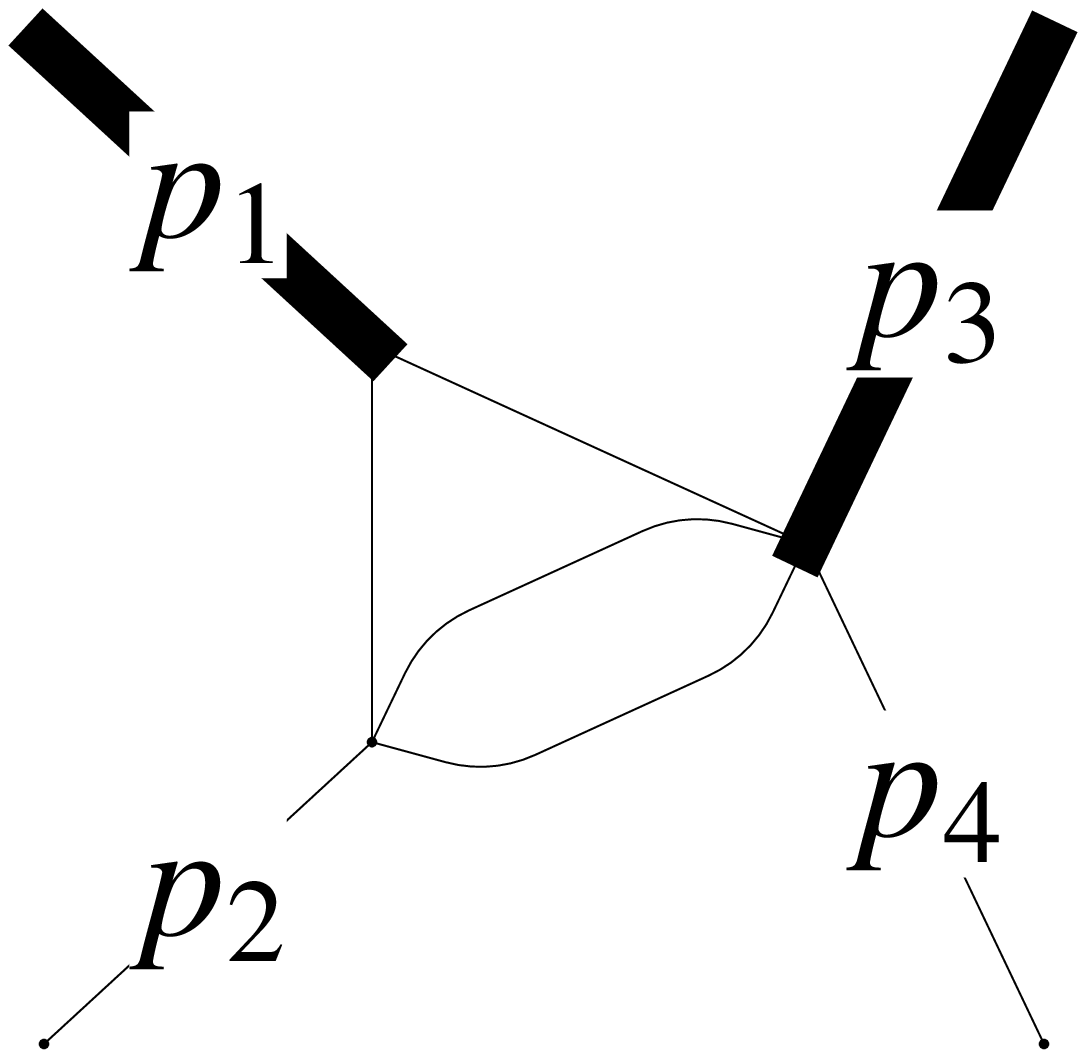}
{  \bea      g^{\rm P13}_{11} &=&  \eps^3 (p_1^2  - s) G_{1, 0, 0, 1, 1, 2, 0, 0, 0}
\;,~~~~~~~~~~~~~~~~~~~~~~~~~~~~~~~~~~~~~~~
 \\
    f^{\rm P13}_{11} &\sim& -\frac{x^{-2\ep}}{4} + 
   e^{2i\pi \ep} \left ( \frac{1}{4} + \frac{ \pi^2 \ep^2 }{12} 
      + \frac{ \zeta_3 \ep^3 }{2} 
+ \frac{ \pi^4 \ep^4 }{40}  
\right ),  \nn
\eea} \nn 
\\
\picturepage{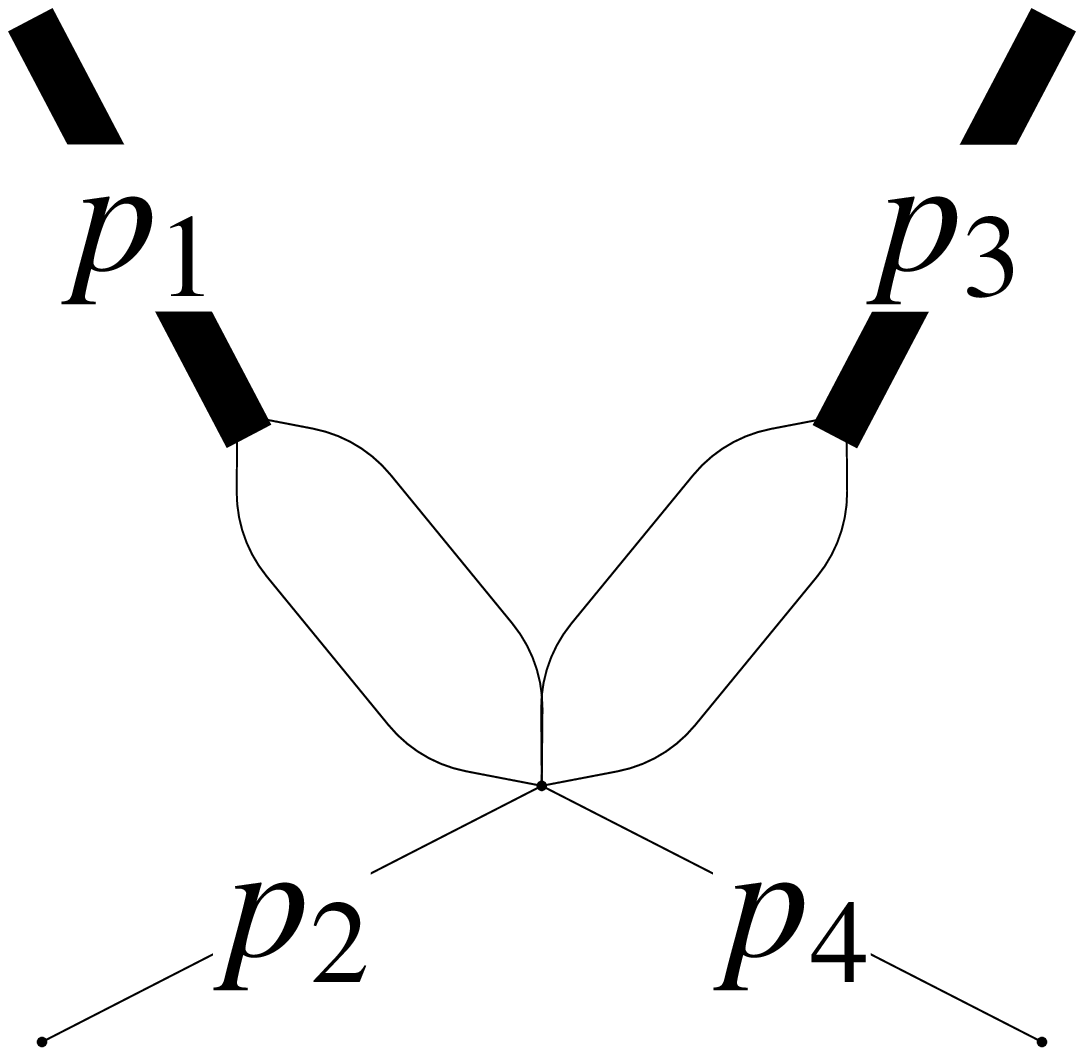}
{  \bea      g^{\rm P13}_{12} &=&  \eps^2 p_1^2  p_3^2  G_{2, 0, 2, 0, 1, 0, 1, 0, 0}
\,,~~~~~~~~~~~~~~~~~~~~~~~~~~~~~~~~~~~~~~~~~~~~~
 \\
    f^{\rm P13}_{12} &\sim& e^{2i\pi \ep} x^{-2\ep} \left (1 + 6 \zeta_3\ep^3 
+ \frac{\pi^4 \ep^4}{10}  \right ),  \nn
\eea} \nn 
\\
\picturepage{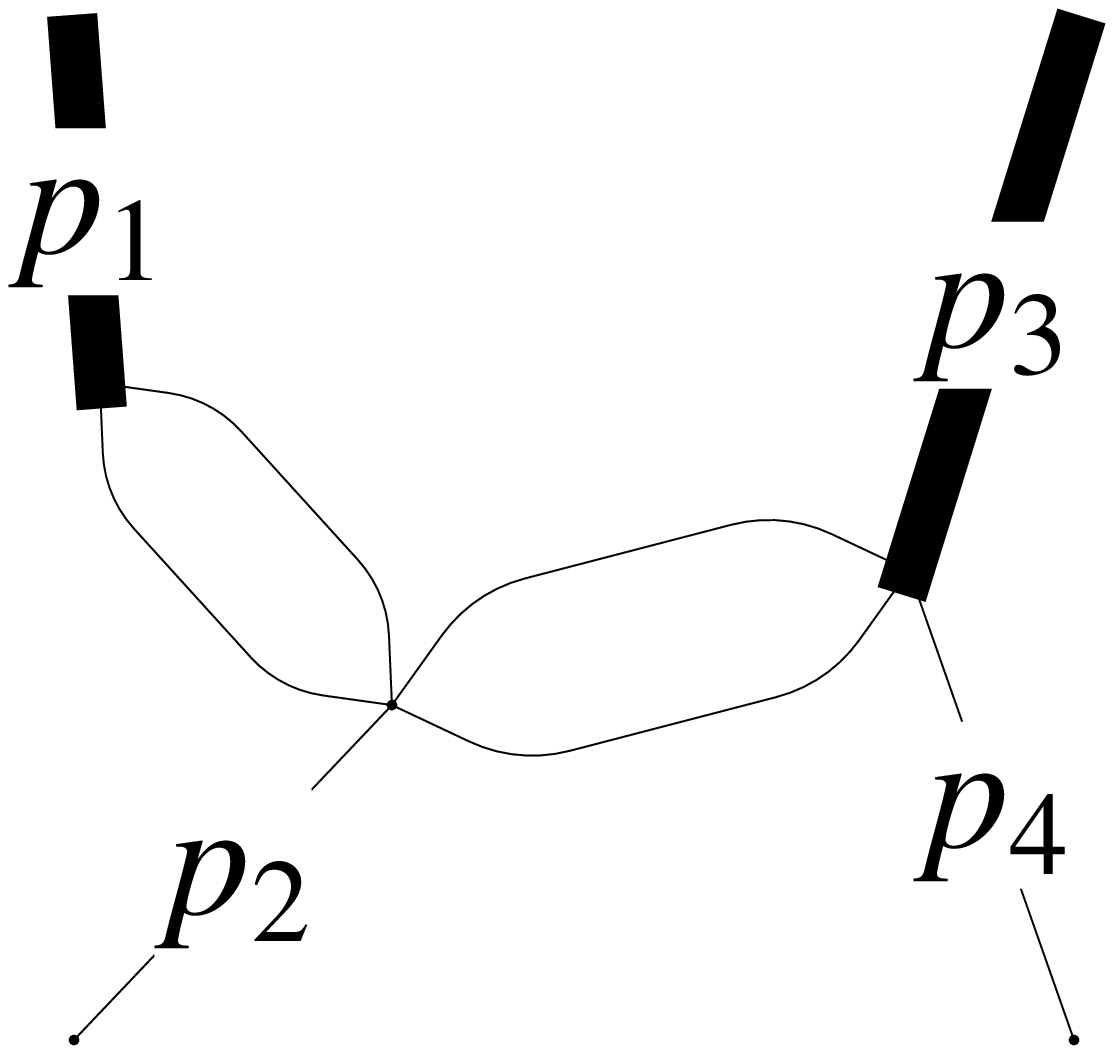}
{  \bea      g^{\rm P13}_{13} &=&  \eps^2 p_1^2  s G_{2, 0, 2, 1, 1, 0, 0, 0, 0})
\,,~~~~~~~~~~~~~~~~~~~~~~~~~~~~~~~~~~~~~~~~~~~~
 \\
    f^{\rm P13}_{13} &\sim& e^{i \pi \ep} x^{-\ep} \left (1 + 6 \zeta_3\ep^3 
+ \frac{\pi^4 \ep^4 }{10} \right ),  \nn
\eea} \nn 
\\
\picturepage{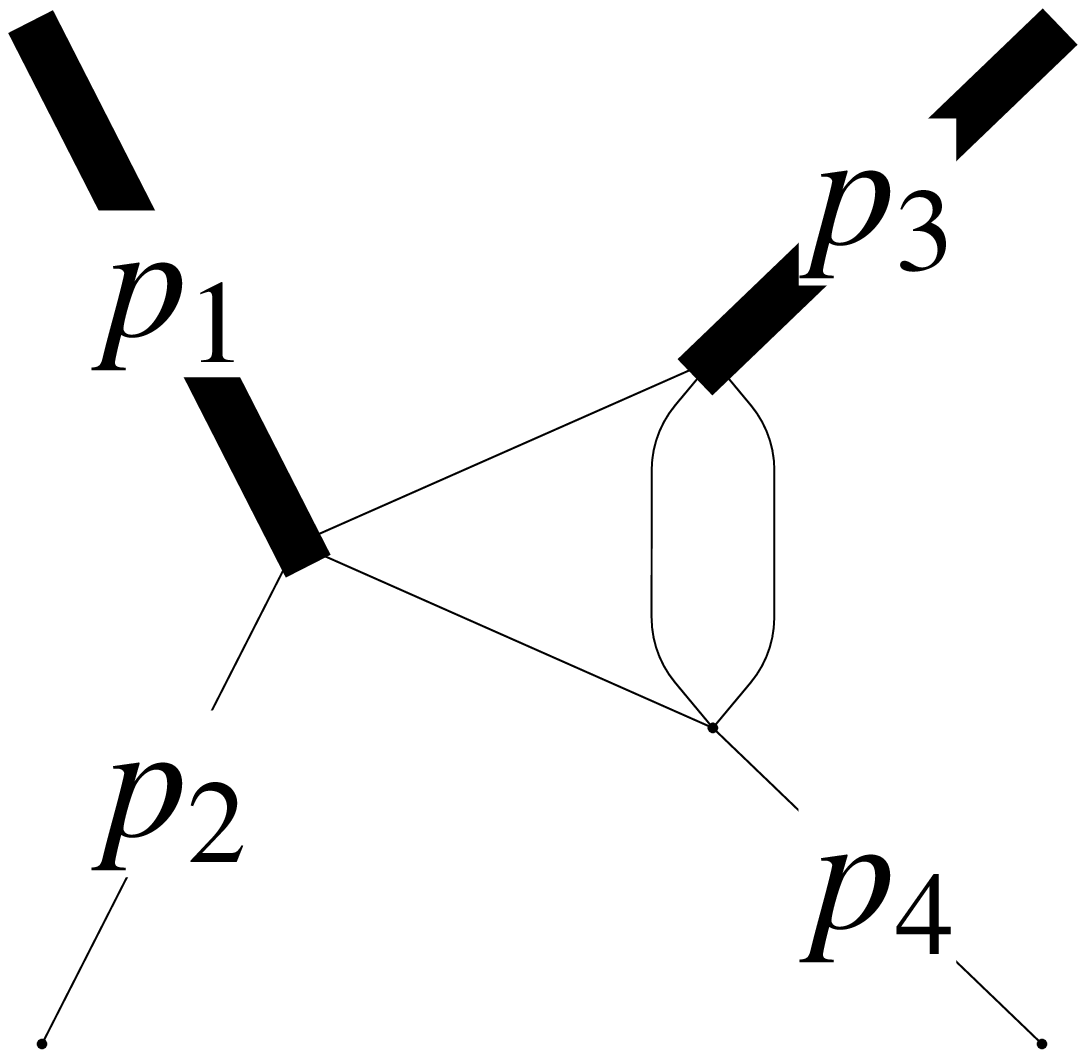}
{  \bea      g^{\rm P13}_{14} &=&  \eps^3 (p_3^2  - s) G_{1, 1, 0, 0, 0, 1, 2, 0, 0}
\,,~~~~~~~~~~~~~~~~~~~~~~~~~~~~~~~~~~~~~
\;,
 \\
    f^{\rm P13}_{14} &\sim&   \frac{1}{4} e^{2i\pi \ep} x^{-4\ep} 
 - x^{-2\ep} \left ( \frac{1}{4} + \frac{ \pi^2 \ep^2}{12} + \frac{\zeta_3 \ep^3}{2}
 + \frac{\pi^4 \ep^4  }{40}   \right ), \nn
\eea} \nn 
\\
\picturepage{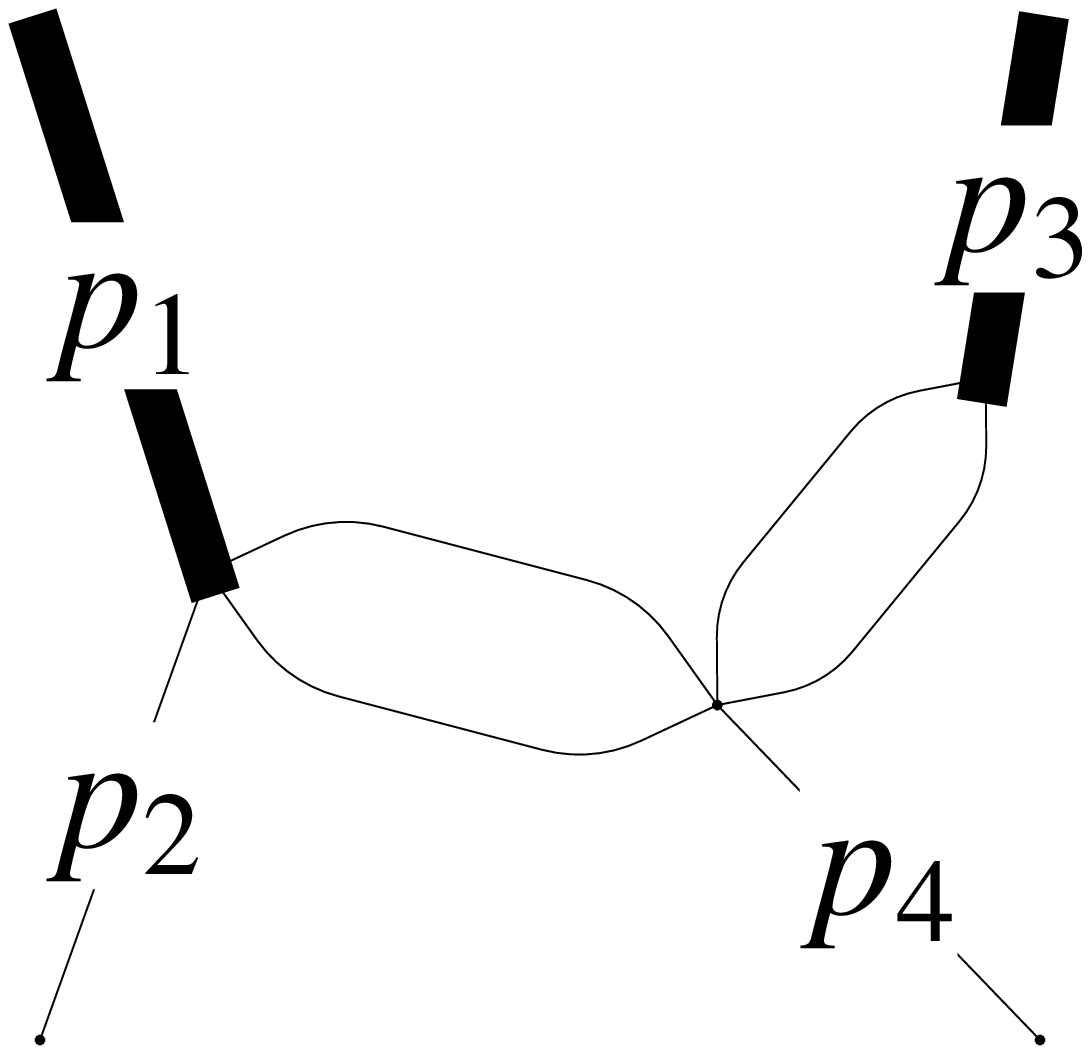}
{  \bea      g^{\rm P13}_{15} &=&  \eps^2 p_3^2  s G_{1, 2, 1, 0, 0, 0, 2, 0, 0}
\,,~~~~~~~~~~~~~~~~~~~~~~~~~~~~~~~~~~~~~~~~~~~
 \\
    f^{\rm P13}_{15} &\sim&  e^{i\ep \pi} x^{-3\ep} \left ( 1 + 6 \zeta_3\ep^3 + \frac{\pi^4 \ep^4}{10} \right ),   \nn
\eea} \nn \\
\picturepage{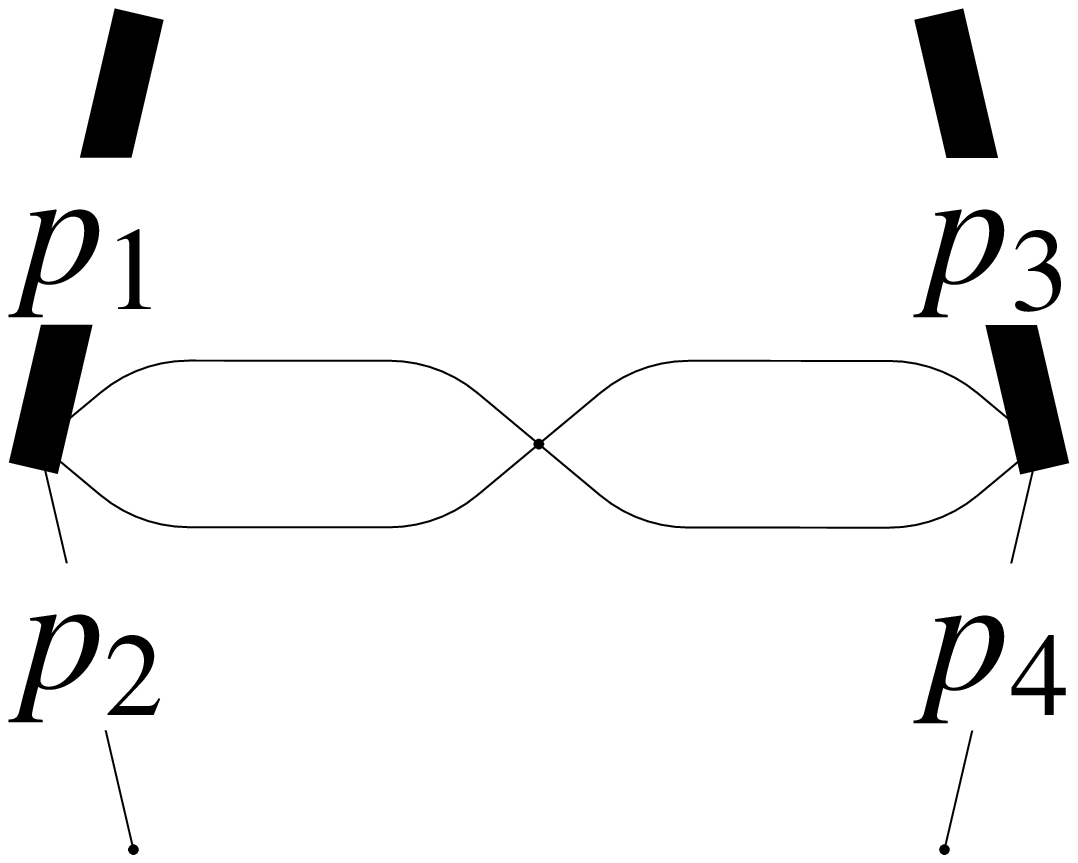}
{  \bea      g^{\rm P13}_{16} &=&  \eps^2 s^2 G_{1, 2, 1, 2, 0, 0, 0, 0, 0}
\;,~~~~~~~~~~~~~~~~~~~~~~~~~~~~~~~~~~~~~~~~~~~
 \\
    f^{\rm P13}_{16} &\sim&  x^{-2\ep} \left (1 + 6 \zeta_3\ep^3 + \frac{\pi^4 \ep^4}{10} \right ),\;\;\;    \nn
\eea} \nn 
\\
\picturepage{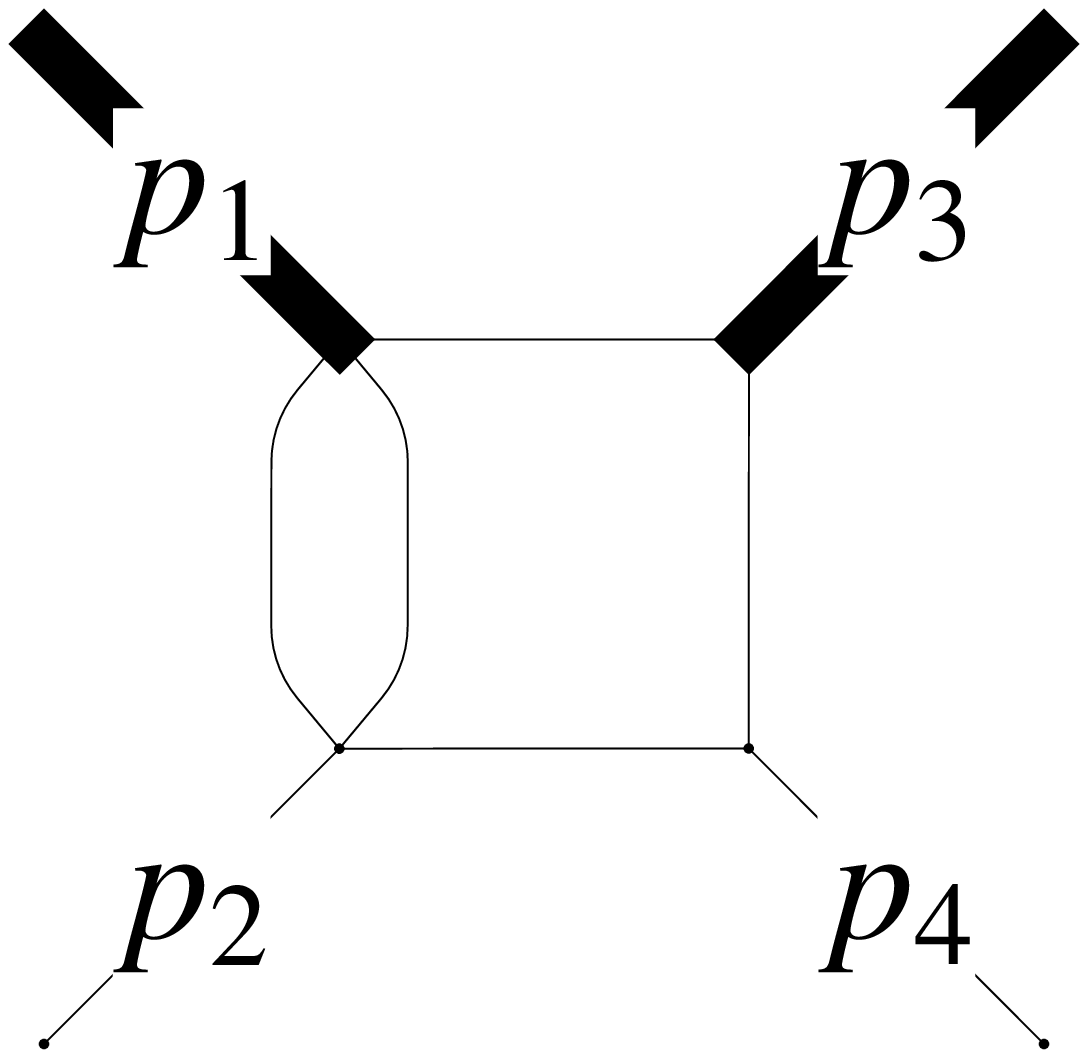}
{  \bea      g^{\rm P13}_{17} &=&  \eps^3 s t G_{0, 0, 1, 1, 1, 2, 1, 0, 0} 
\;,~~~~~~~~~~~~~~~~~~~~~~~~~~~~~~~~~~~~~~~~~~~
 \\
    f^{\rm P13}_{17} &\sim&    -\frac{3}{2}x^{-\ep} N_1
+ x^{-2\ep} e^{2i\pi \ep}  \left ( 1  + 6  \zeta_3 \ep^3 + \frac{\pi^4 \ep^4}{10}   \right ),    \nn
\eea} \nn 
\\
\picturepage{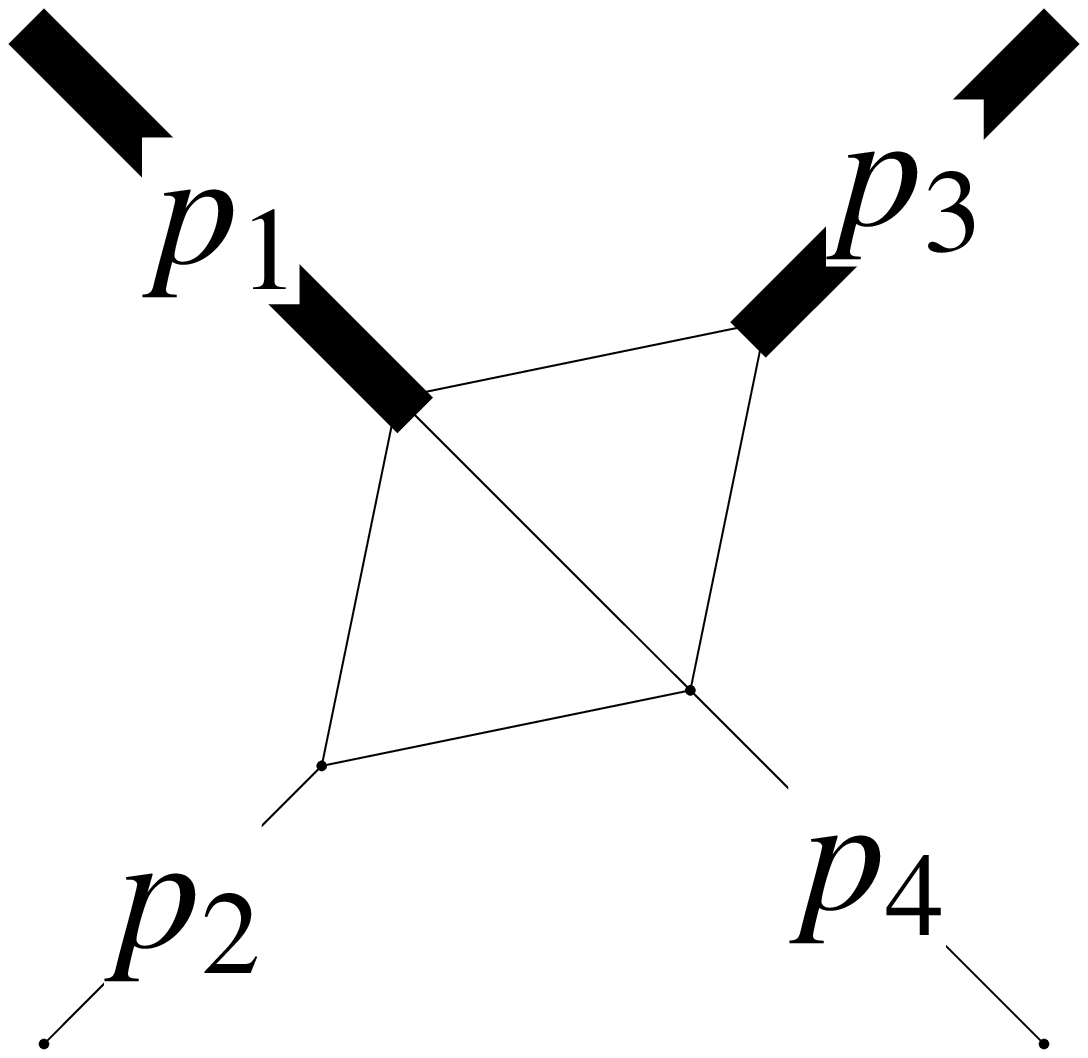}
{  \bea      g^{\rm P13}_{18} &=&  \eps^4 (p_1^2  - s - t) G_{0, 1, 1, 0, 1, 1, 1, 0, 0}
\;,~~~~~~~~~~~~~~~~~~~~~~~~~~~~
 \\
    f^{\rm P13}_{18} &\sim&   0, \nn
\eea} \nn \\
%\end{align}
%\end{small}
%
%\begin{small}
%\begin{align}
\picturepage{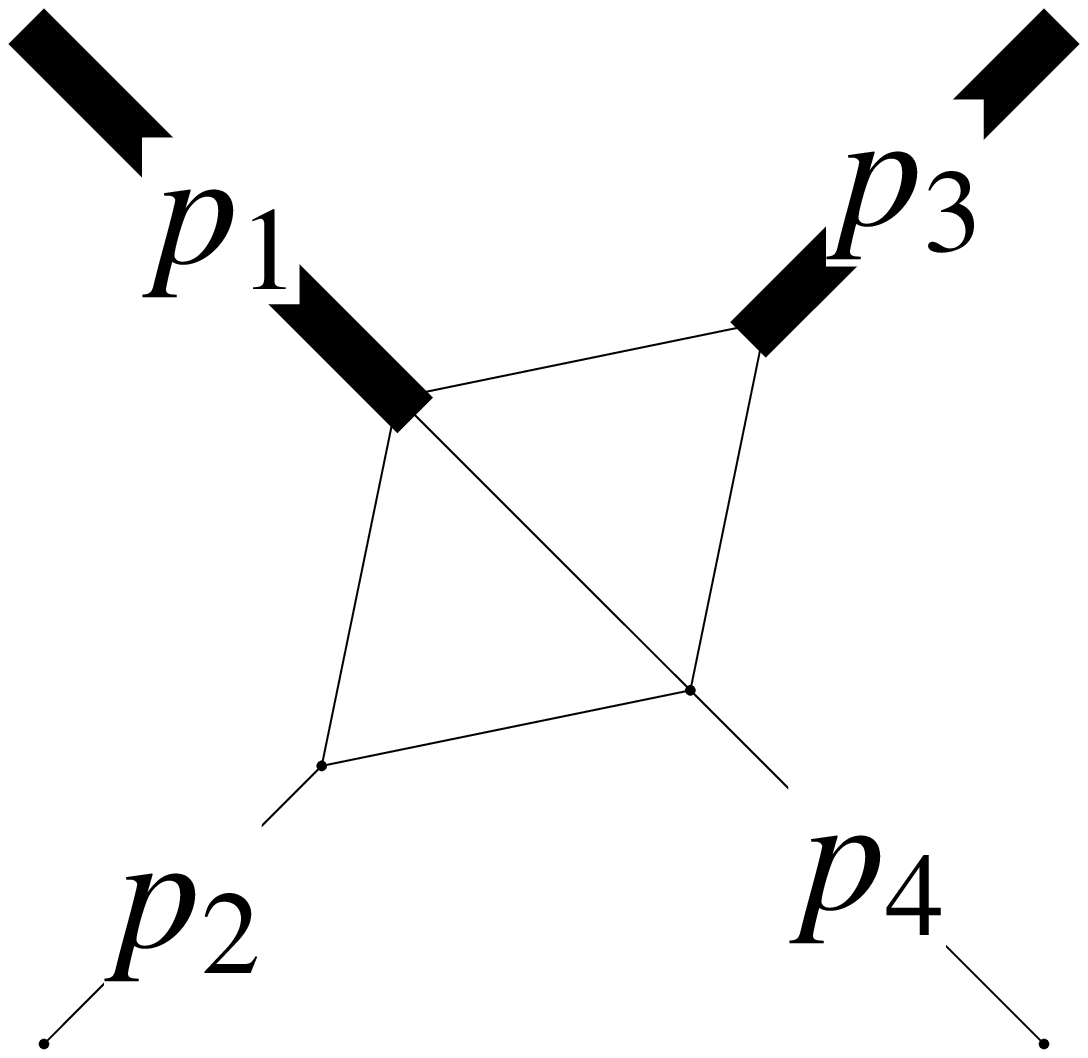}
{  \bea      g^{\rm P13}_{19} &=&  \eps^3 s t G_{0, 1, 1, 0, 1, 2, 1, 0, 0}
\;,~~~~~~~~~~~~~~~~~~~~~~~~~~~~~~~~~~~~~~~~~~~~~~~~~~~~~~~~~
 \\
    f^{\rm P13}_{19} &\sim&  -\frac{3}{2} x^{-2\ep} +
   x^{-3\ep} N_1,\;\;\;\;  
\nn
\eea} \nn 
\\
\picturepage{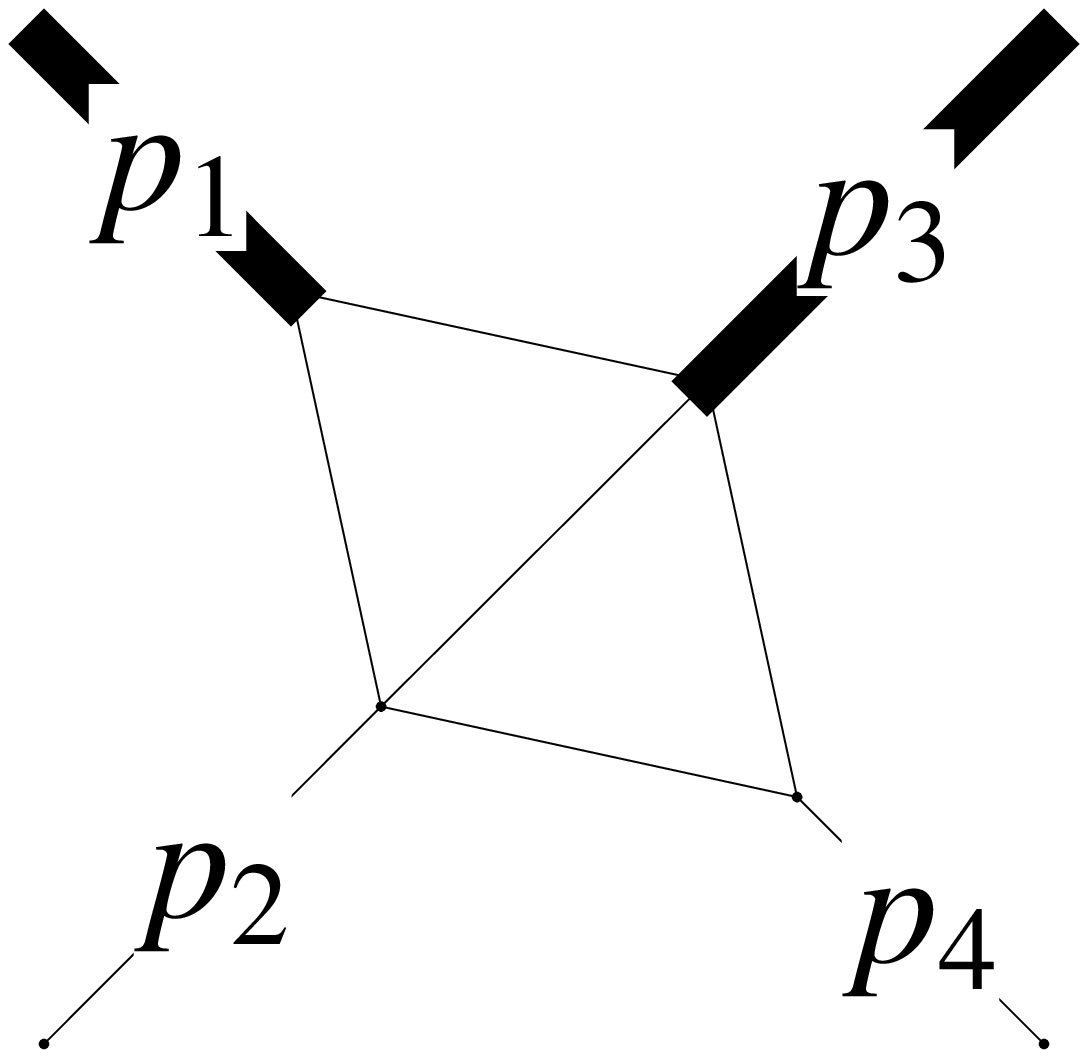}
{  \bea      g^{\rm P13}_{20} &=&  \eps^4 (p_3^2  - s - t) G_{1, 0, 0, 1, 1, 1, 1, 0, 0}
\;,
 \\
    f^{\rm P13}_{20} &\sim&    e^{2i\pi \ep}  \left ( \frac{\pi^2 \ep^2}{12} + \frac{\zeta_3 \ep^3  }{2}
         +  \frac{  \pi^4 \ep^4 }{40}  \right )
- \frac{x^{-2 \ep}}{4} 
 \left ( \frac{\pi^2 \ep^2 }{3} + 14\zeta_3\ep^3 + \frac{2 \pi^4 \ep^4}{3} \right ),  \nn
\eea} \nn 
\\
\picturepage{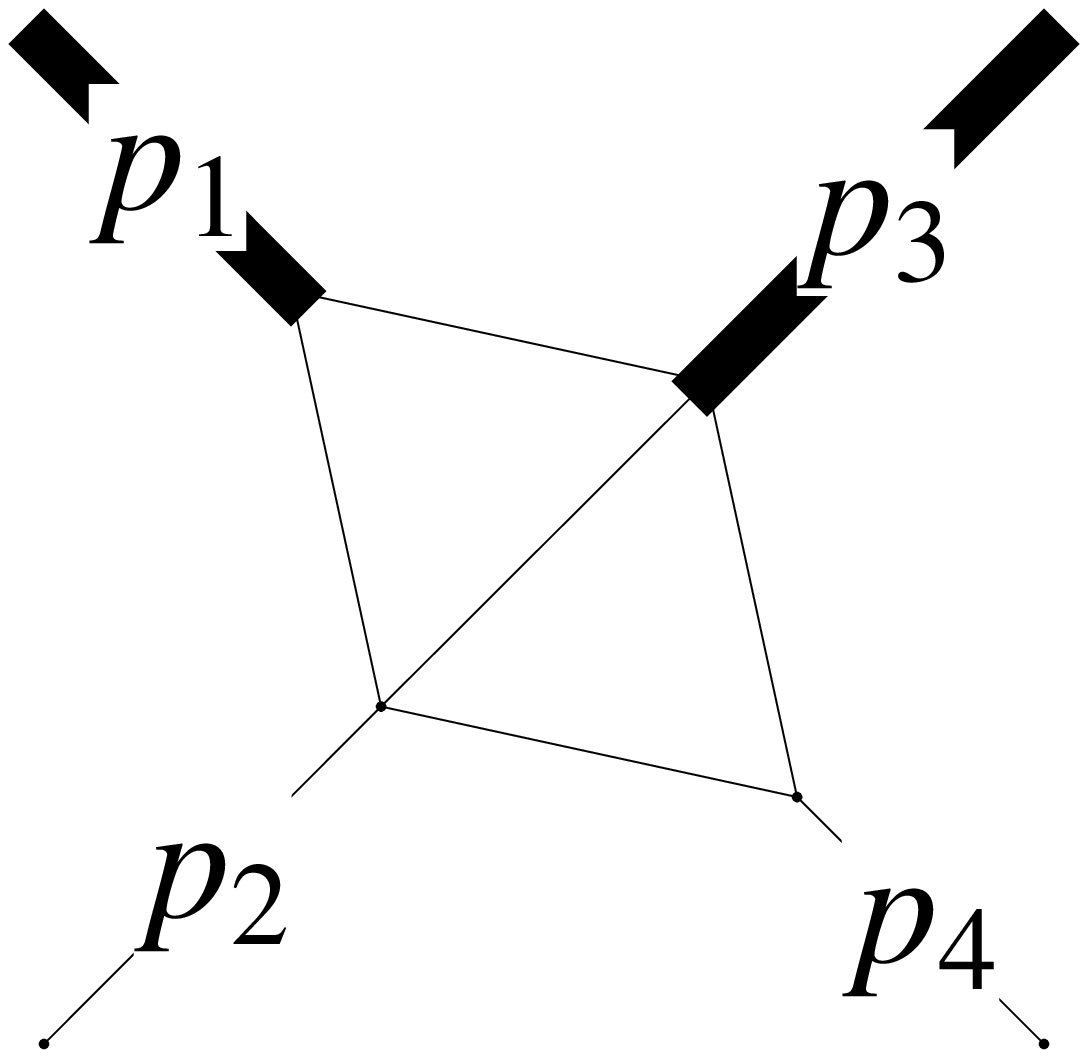}
{  \bea      g^{\rm P13}_{21} &=&  \eps^3 s t G_{1, 0, 0, 1, 1, 2, 1, 0, 0}
\;,~~~~~~~~~~~~~~~~~~~~~~~~~~~~~~~~~~~~~~~~~~~~~~~~~~~~~~~~~
 \\
    f^{\rm P13}_{21} &\sim&  
  \frac{e^{2i\pi \ep}}{2} x^{-4 \ep} 
 - \frac{3 x^{-2 \ep}}{2} + 
   \frac{x^{-2 \ep}}{2} \left ( 1 + \frac{\pi^2 \ep^2}{3} + 14 \zeta_3\ep^3 +
      \frac{2 \pi^4 \ep^4}{3} \right ),\nn
\eea} \nn 
\\
\picturepage{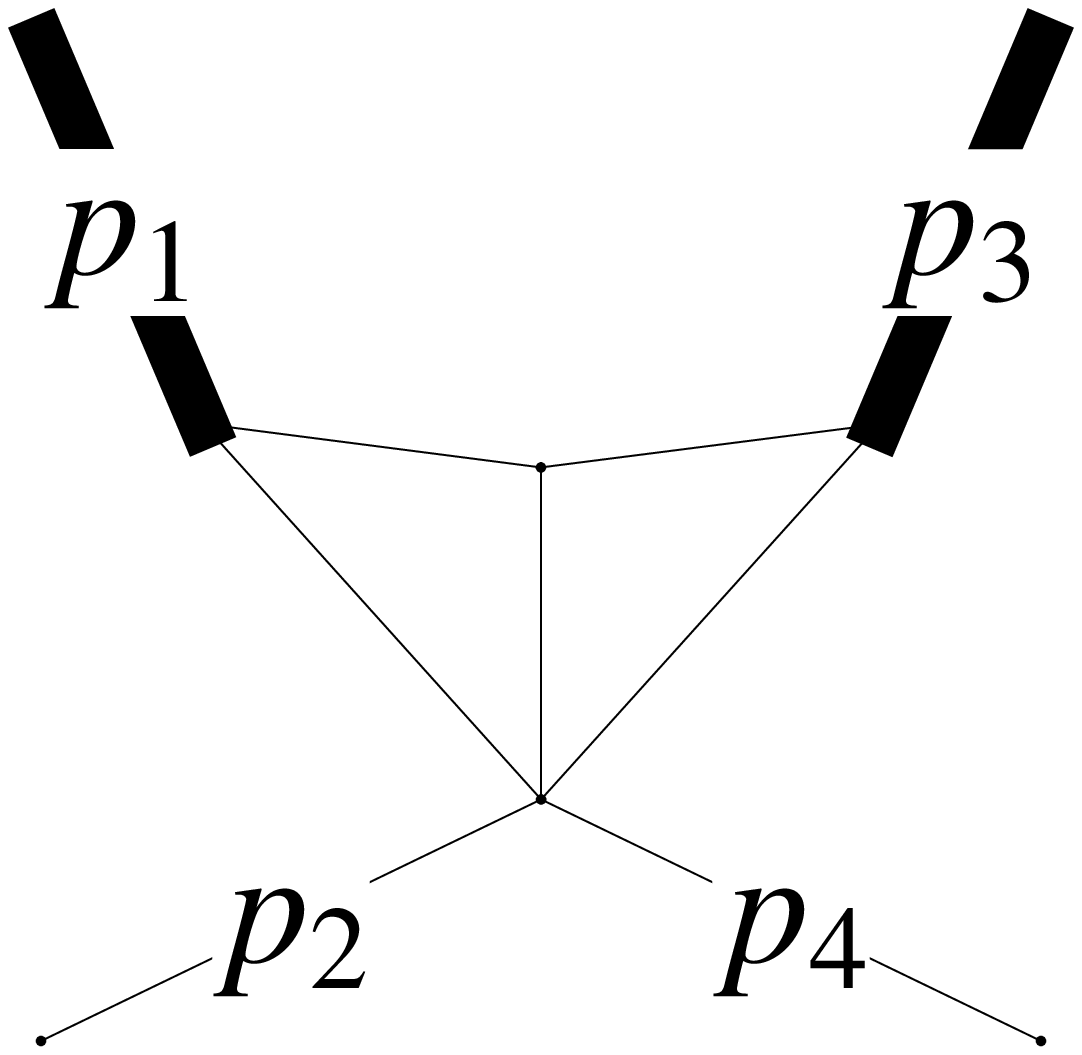}
{  \bea      g^{\rm P13}_{22} &=&  \eps^4 R_{13} G_{1, 0, 1, 0, 1, 1, 1, 0, 0}
\,,~~~~~~~~~~~~~~~~~~~~~~~~~~~~~~~~~~~~~~~~~~~~~~~~~~~~~~~
 \\
    f^{\rm P13}_{22} &\sim& 0,  \nn 
\eea} \nn \\
\picturepage{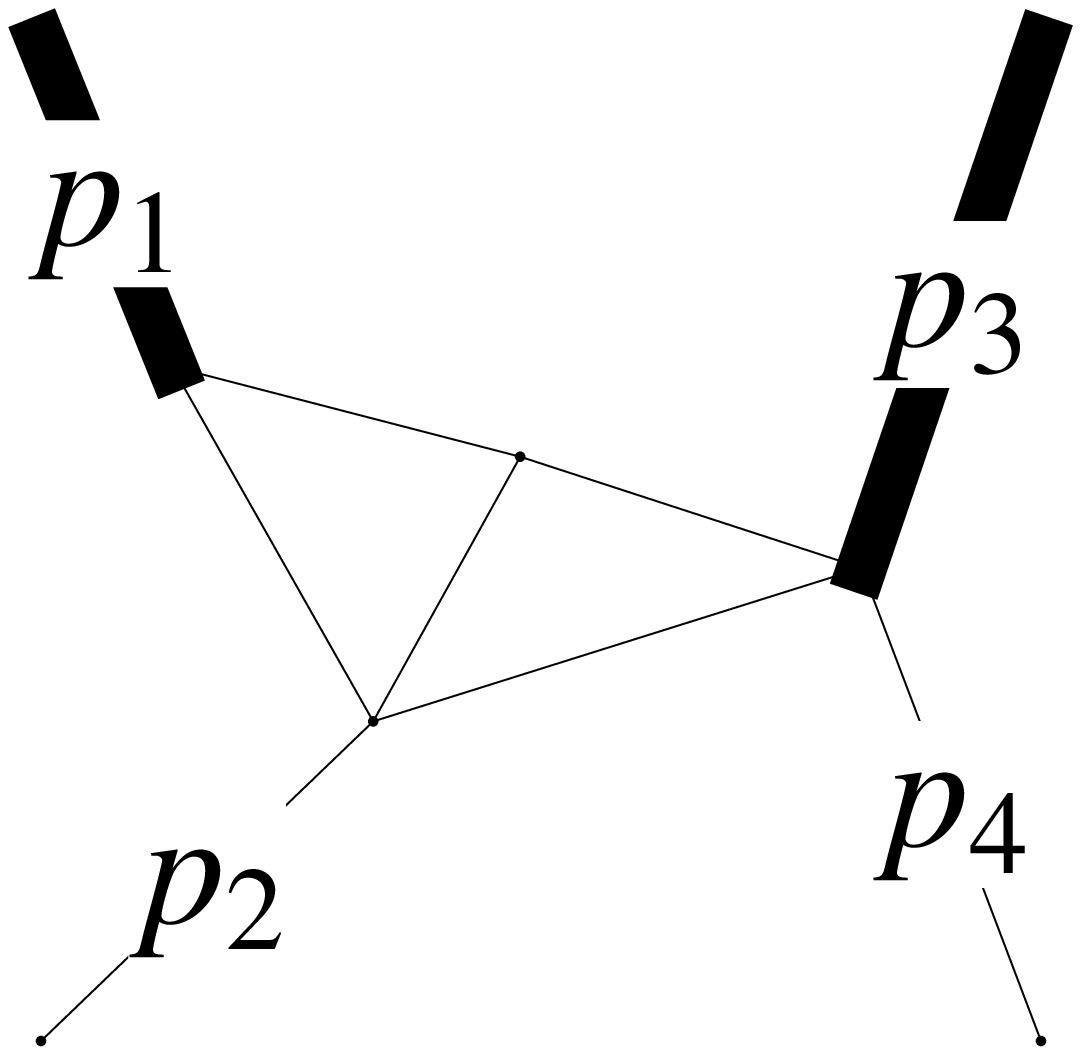}
{  \bea      g^{\rm P13}_{23} &=& \eps^4 (p_1^2  - s) G_{1, 0, 1, 1, 1, 1, 0, 0, 0}
\;,
 \\
    f^{\rm P13}_{23} &\sim& 
x^{-2 \ep} \left (-\frac{\pi^2 \ep^2}{12} - \frac{7 \zeta_3\ep^3 }{2} 
   - \frac{\pi^4 \ep^4}{6} \right ) + 
   \frac{e^{2i\pi \ep}}{2} \left ( -\frac{\pi^2 \ep^2}{6} -  \zeta_3 \ep^3 - \frac{\pi^4 \ep^4}{20} 
\right ) \nn
\\
&-& 
   x^{-\ep} \left ( -\frac{\pi^2 \ep^2}{6} + \left ( - \frac{i \pi^3}{6} - 4 \zeta_3 \right ) 
  - \left ( \frac{\pi^4}{24} + 4i \pi \zeta_3 \right ) \ep^4 
     \right ),
\nn
\eea} \nn 
\\
\picturepage{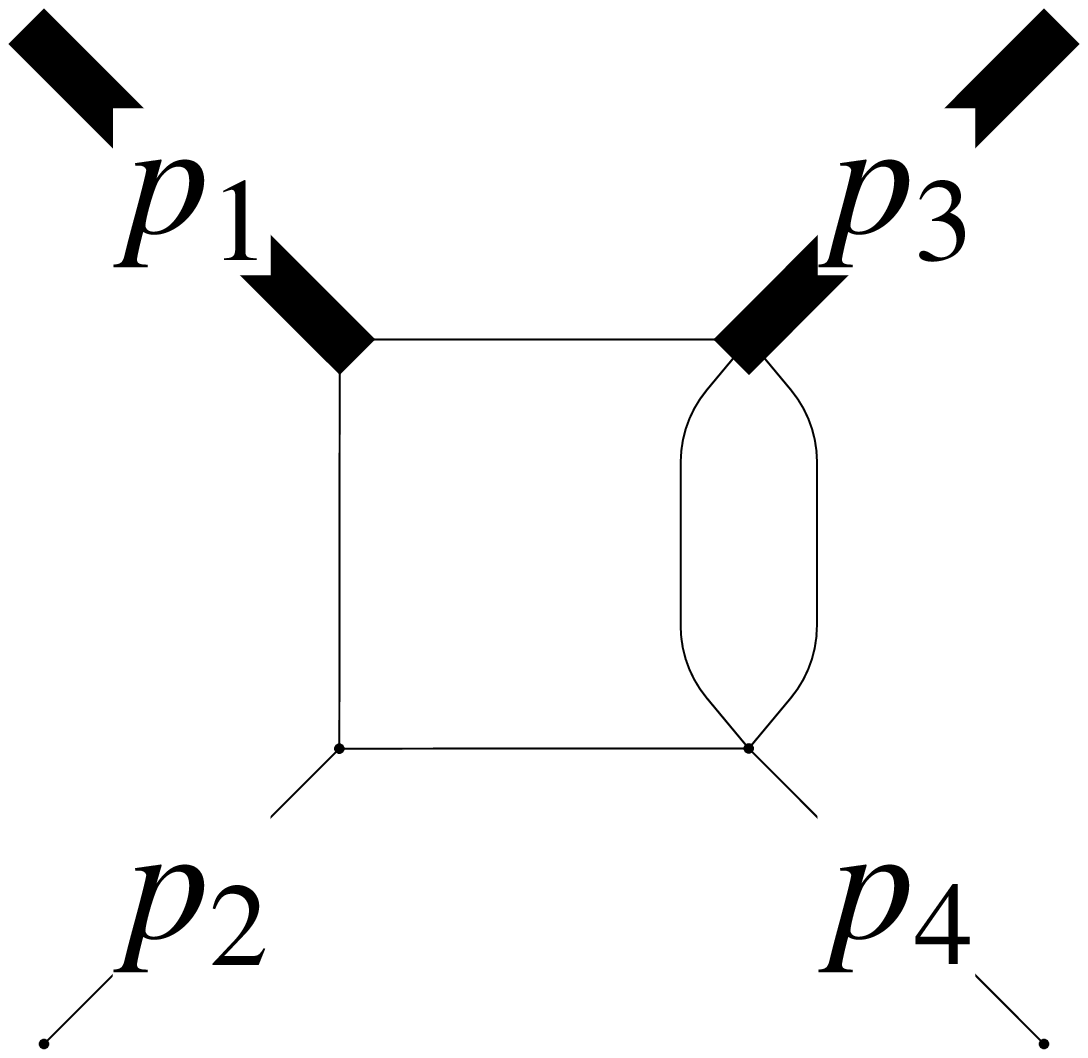}
{  \bea      g^{\rm P13}_{24} &=&  \ep^3 s t G_{1, 1, 0, 0, 1, 1, 2, 0, 0}
\;,~~~~~~~~~~~~~~~~~~~~~~~~~~~~~~~~~~~~~~~~~~~~~~~~~~~~~~~~~~~
 \\
    f^{\rm P13}_{24} &\sim&   \frac{e^{2i\pi \ep} x^{-4\ep}}{4} 
 -  \frac{3x^{-2\ep}}{4} \left ( 1 + \frac{ \pi^2 \ep^2 }{3} +2 \zeta_3 \ep^3 
    + \frac{ \pi^4 \ep^4 }{10}  \right ),   \nn
\eea} \nn 
\\
\picturepage{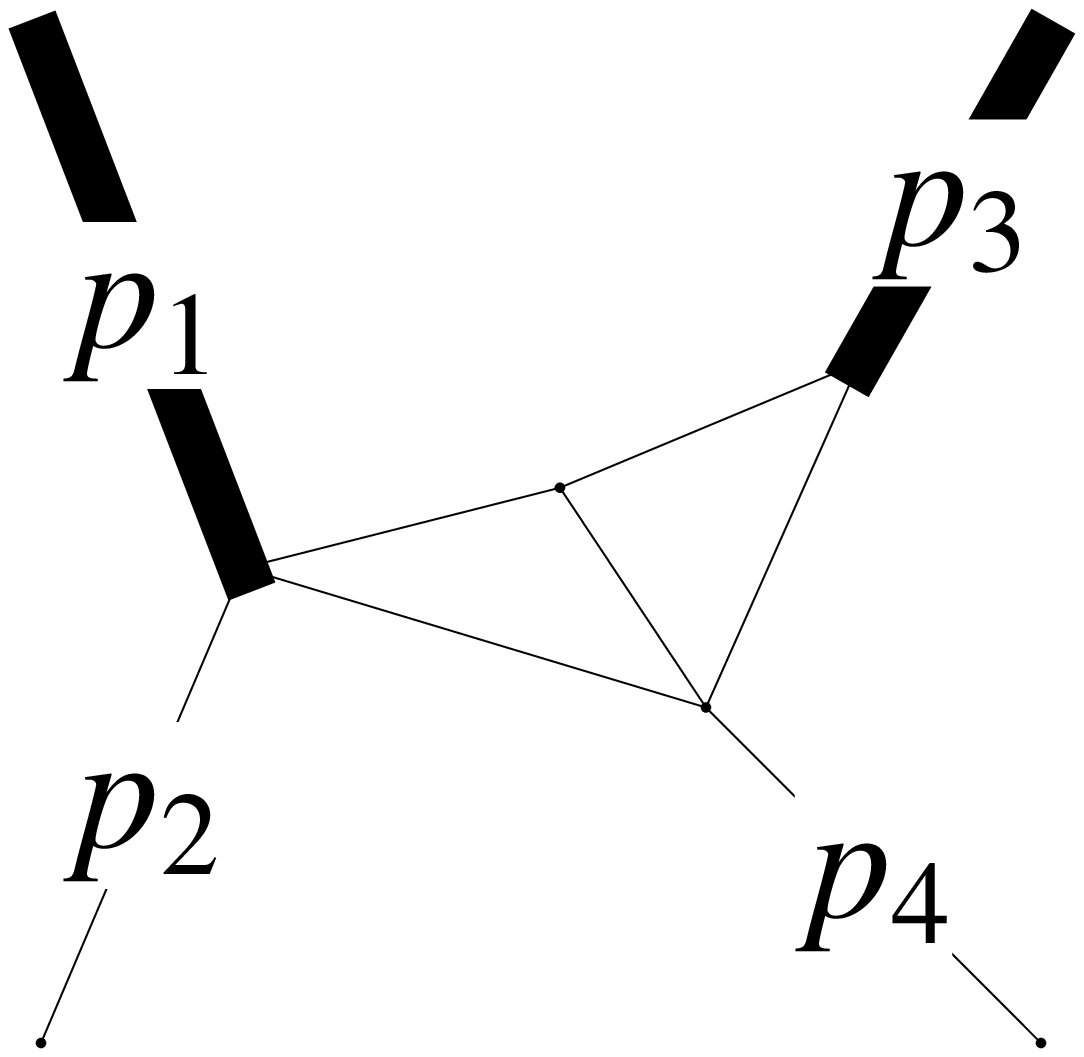}
{  \bea      g^{\rm P13}_{25} &=&  \eps^4 (p_3^2  - s) G_{1, 1, 1, 0, 0, 1, 1, 0, 0} \;,\;\;\;
~~~~~~~~~~~~~~~~~~~~~~~~~~~~~~~~~~~~~~~~~~~~~~~
 \\
    f^{\rm P13}_{25} &\sim& 
   \frac{x^{-2\ep}}{2}
    \left (  \frac{\pi^2 \ep^2}{6}  + 
       \zeta_3 \ep^3 + \frac{ \pi^4 \ep^4}{20}
 \right ) \nn
\\
& + & x^{-4\ep} \left ( \frac{\pi^2 \ep^2 }{12} 
 + \left (\frac{i\pi^3}{6} + \frac{7 \zeta_3  }{2}\right ) \ep^3 + 
      7 i \pi \zeta_3\ep^4 \right ) 
 \nn
\\
& -& x^{-3\ep} \Big (  
  \frac{ \pi^2 \ep^2 }{6} +  \left ( \frac{i\pi^3}{6}  + 4 \zeta_3 \right ) \ep^3
 + \left ( \frac{\pi^4}{24} + 4 i \pi  \zeta_3 \right )   \ep^4 \Big ), 
 \nn
\eea} \nn \\
\picturepage{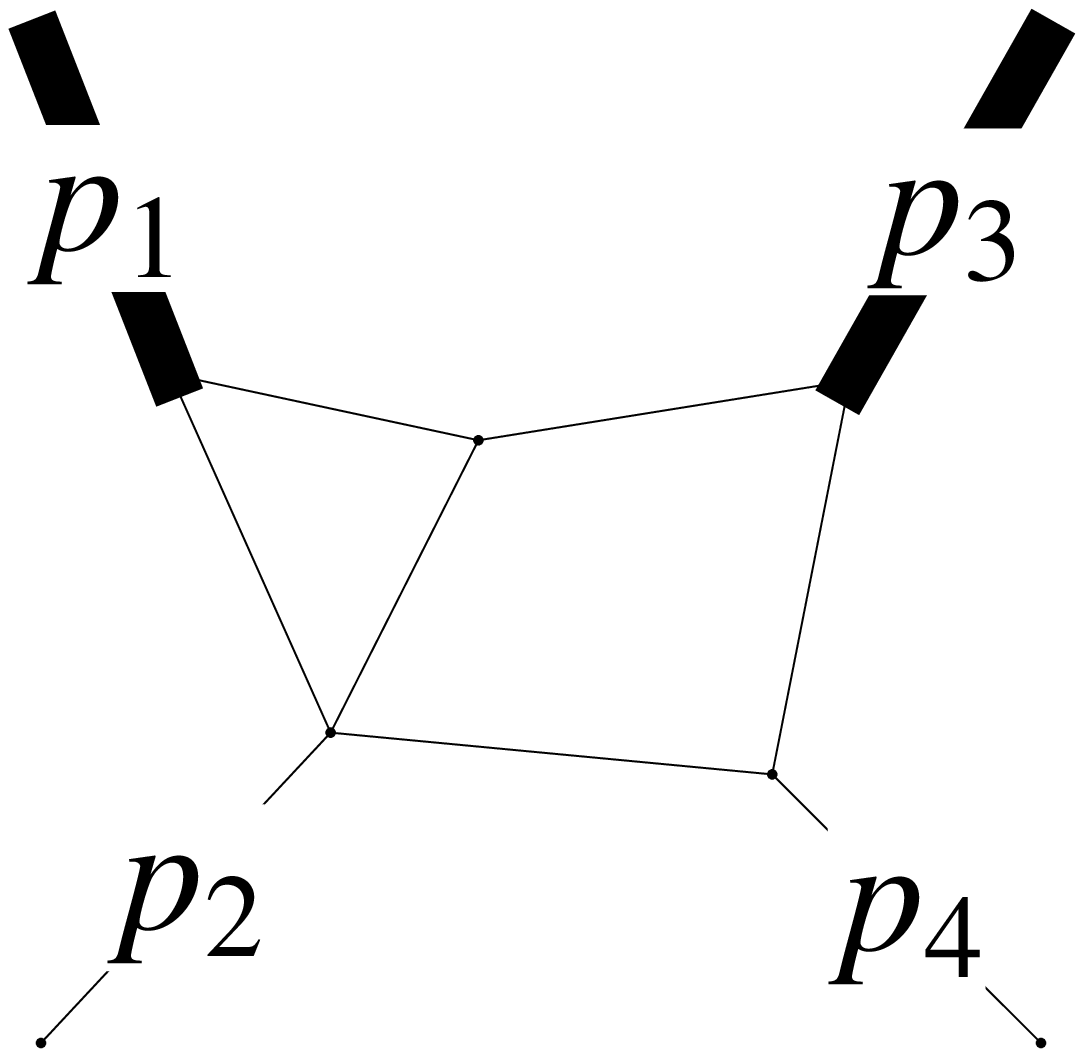}
{  \bea      g^{\rm P13}_{26} &=&  \eps^4 (p_1^2  (p_3^2  - s) + s t) G_{1, 0, 1, 1, 1, 1, 1, 0, 0}
\;,~~~~~~~~~~~~~~~~~~~~~~~~~~~~~~~~~~~~~~~~
 \\
    f^{\rm P13}_{26} &\sim& 0\;,  \nn
\eea} \nn \\
%\end{align}
%\end{small}
%
%\begin{small}
%\begin{align}
\picturepage{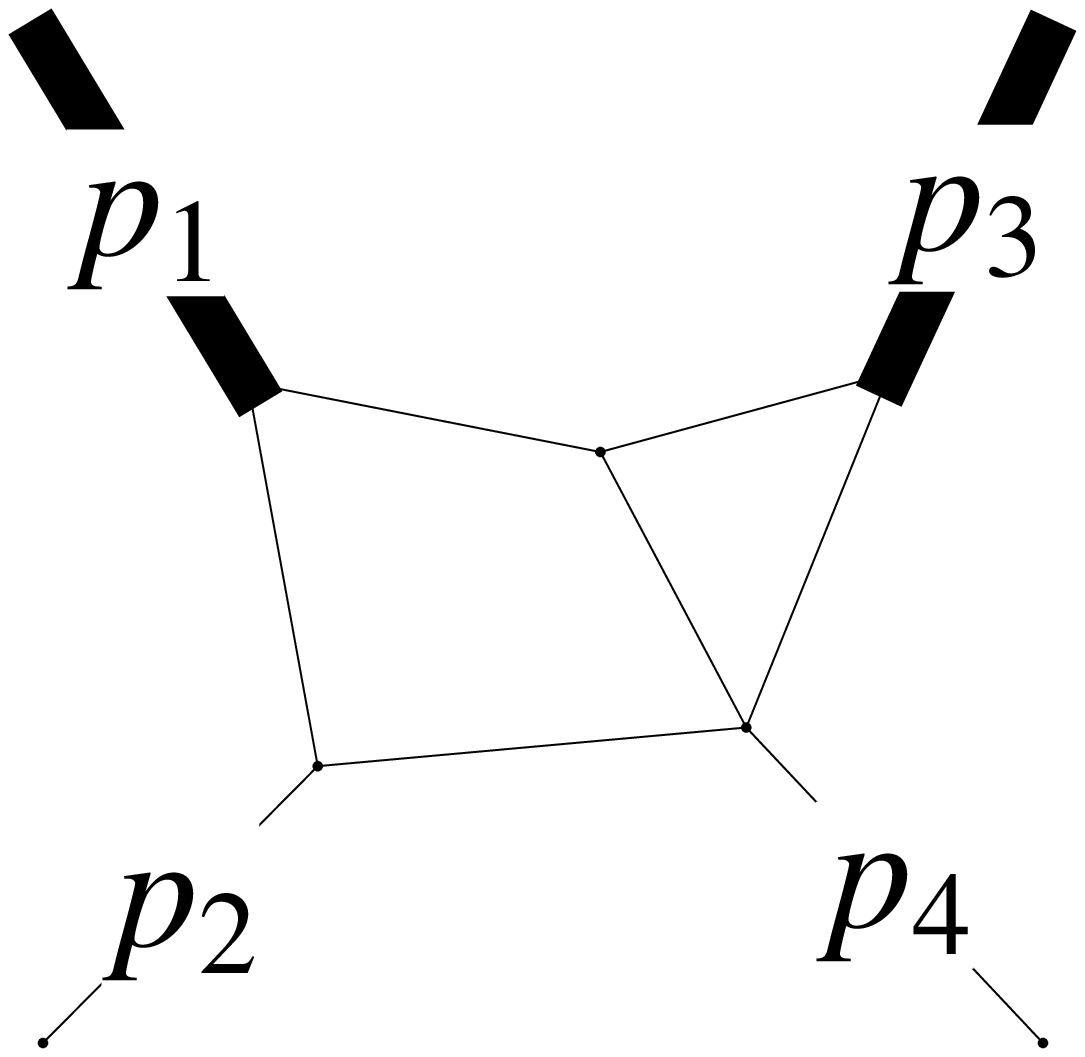}
{  \bea      g^{\rm P13}_{27} &=&  \eps^4 [p_1^2  p_3^2  + s ( t-p_3^2 )] G_{1, 1, 1, 0, 1, 1, 1, 0, 0}
\;,~~~~~~~~~~~~~~~~~~~~~~~~~~~~~~~~~~~~~~~~~~
 \\
    f^{\rm P13}_{27} &\sim& 
      - x^{-3\ep} \left  ( \frac{\pi^2 \ep^2}{3} +  \left ( 8 \zeta_3 + \frac{i \pi^3}{3} \right ) \ep^3
   +  \left ( \frac{\pi^4}{12} + 8 i \pi \zeta_3 \right ) \ep^4  \right ) \nn
\\
&+& x^{-2\ep} \left ( \frac{\pi^2 \ep^2}{4} + \frac{3 \zeta_3 \ep^3 }{2} 
   + \frac{3\pi^4 \ep^4}{40} \right ) \nn
\\
&+& \frac{x^{-4\ep}}{2} \left (  \frac{\pi^2 \ep^2 }{6} 
 +  \left ( \frac{i \pi^3}{3} + 7 \zeta_3 \right ) \ep^3
       + 14 i\pi \zeta_3 \ep^4  \right ),
 \nn
\eea} \nn 
\\
\picturepage{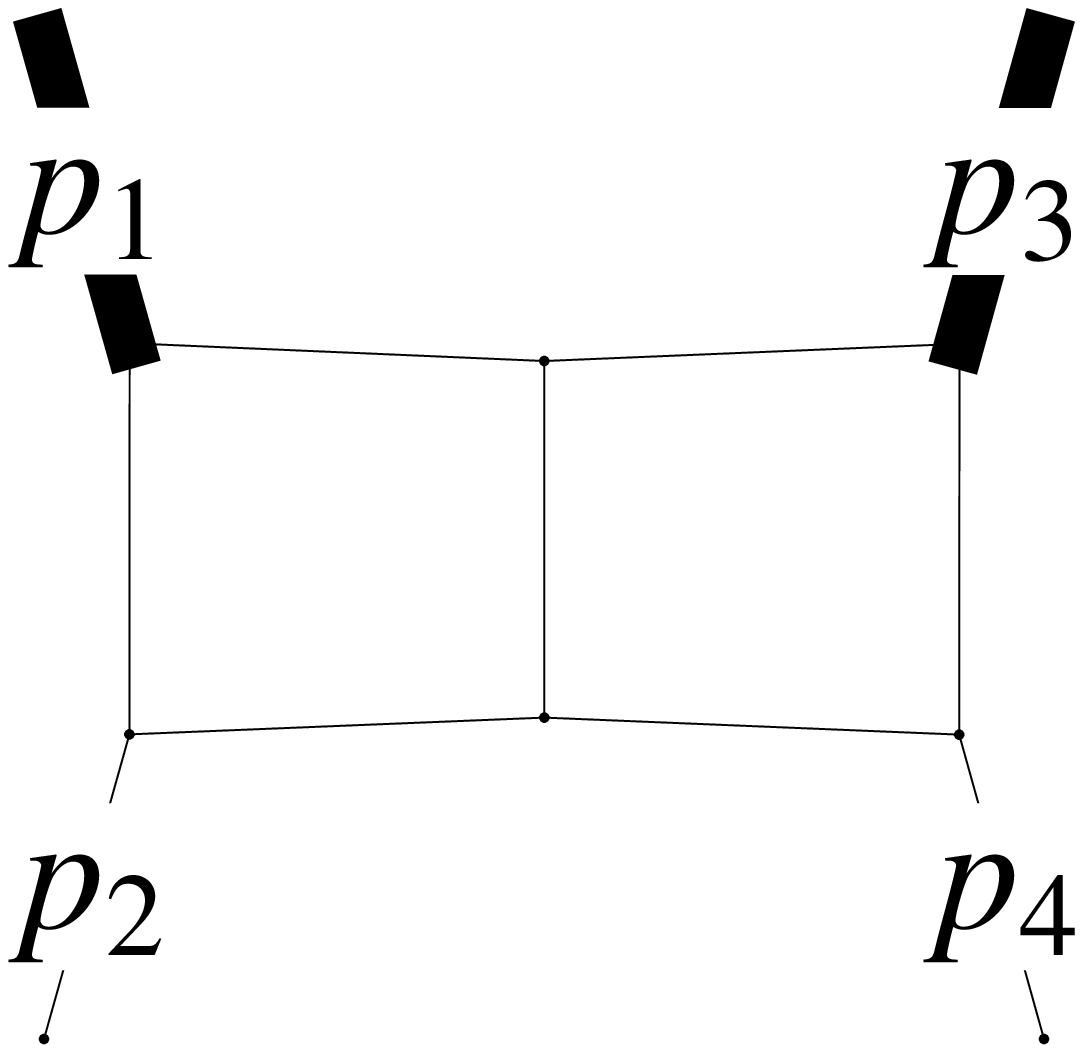}
{  \bea      g^{\rm P13}_{28} &=&  \eps^4 s^2 t G_{1, 1, 1, 1, 1, 1, 1, 0, 0}
\;,~~~~~~~~~~~~~~~~~~~~~~~~~~~~~~~~~~~~~~~~~~~~~~~~~~~~~~~~~~~~~~
 \\
    f^{\rm P13}_{28} &\sim& 
-x^{-2\ep} \left ( 1 + \frac{5 \pi^2 \ep^2}{12} 
+ \frac{29  \zeta_3  \ep^3 }{2} + \frac{71 \pi^4 \ep^4 }{360}  
     \right ) \nn
\\
&+&  x^{-3\ep} \left ( 1 + i \pi \ep  +  \left ( \frac{i \pi^3}{3} + 18 \zeta_3 \right ) \ep^3 + 
     \left ( \frac{4\pi^4}{15}  + 18 i \pi \zeta_3 \right ) \ep^4 \right )
\nn
\\
&-&x^{-4\ep} \left ( \frac{1}{4} + \frac{i\pi \ep}{2} - \frac{ 5  \pi^2 \ep^2 }{12} - 
     \left ( \frac{i \pi^3}{6} - \frac{7 \zeta_3}{2} \right ) \ep^3 + 
     \left (\frac{\pi^4}{6} + 7 i \pi \zeta_3 \right ) \ep^4 
\right ),
 \nn
\eea} \nn \\ 
\picturepage{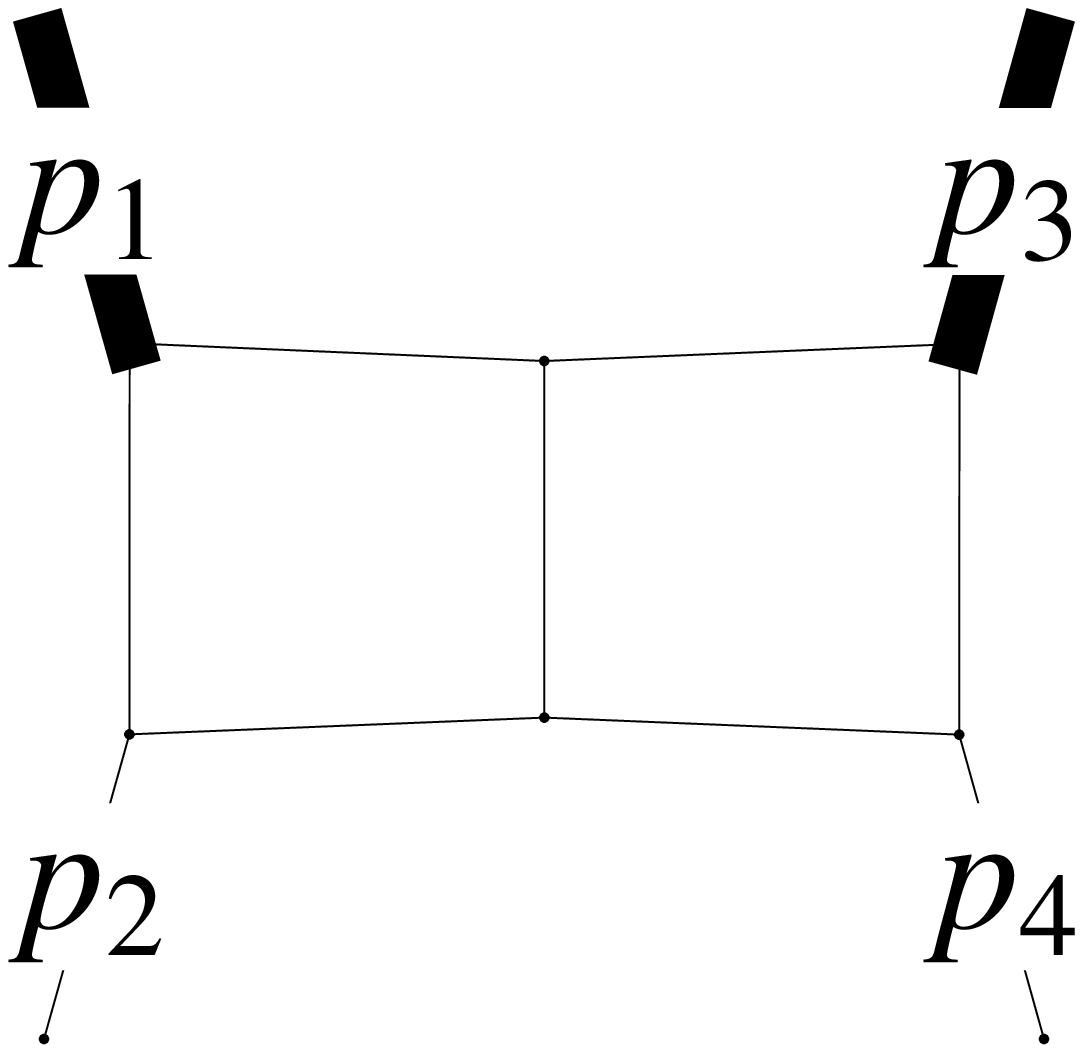}
{  \bea      g^{\rm P13}_{29} &=&  \eps^4 s \Big[ (p_1^2  - p_3^2  + t) G_{1, 1, 1, 0, 1, 1, 1, 0, 0} 
+ (p_3^2  - s) G_{1, 1, 1, 1, 1, 1, 1, -1, 0}\Big]
\;, 
 \\
    f^{\rm P13}_{29} &\sim& x^{-2\ep}  \left (- \frac{3}{4} +  \frac{\pi^2 \ep^2}{3} 
 - 4 \zeta_3 \ep^3 + \frac{23 \pi^4 \ep^4}{180} \right ) \nn
\\
&+& x^{-3\ep} \left (1 + i  \pi \ep  - \frac{2 \pi^2 \ep^2}{3} + 
     \left ( -\frac{i \pi^3}{3} + 2 \zeta_3 \right ) \ep^3 + 
     \left ( \frac{\pi^4}{10} + 2 i \pi \zeta_3 \right ) \ep^4  \right )  
\nn 
\\
&+& 
 x^{-4 \ep} \left (-\frac{1}{4} - \frac{i \pi \ep}{2} + \frac{7 \pi^2 \ep^2 }{12} + 
     \left ( \frac{i \pi^3}{2} + \frac{7 \zeta_3}{2} \right ) \ep^3 + 
     \left ( -\frac{\pi^4}{6} + 7 i \pi \zeta_3 \right )  \right ) \ep^4 . \nn
\eea} \nn 
\end{align}
\end{small}

Finally, for the  family $P_{23}$  a convenient set of master integrals and the corresponding 
boundary conditions are 

\begin{small}
 \begin{align}
\picturepage{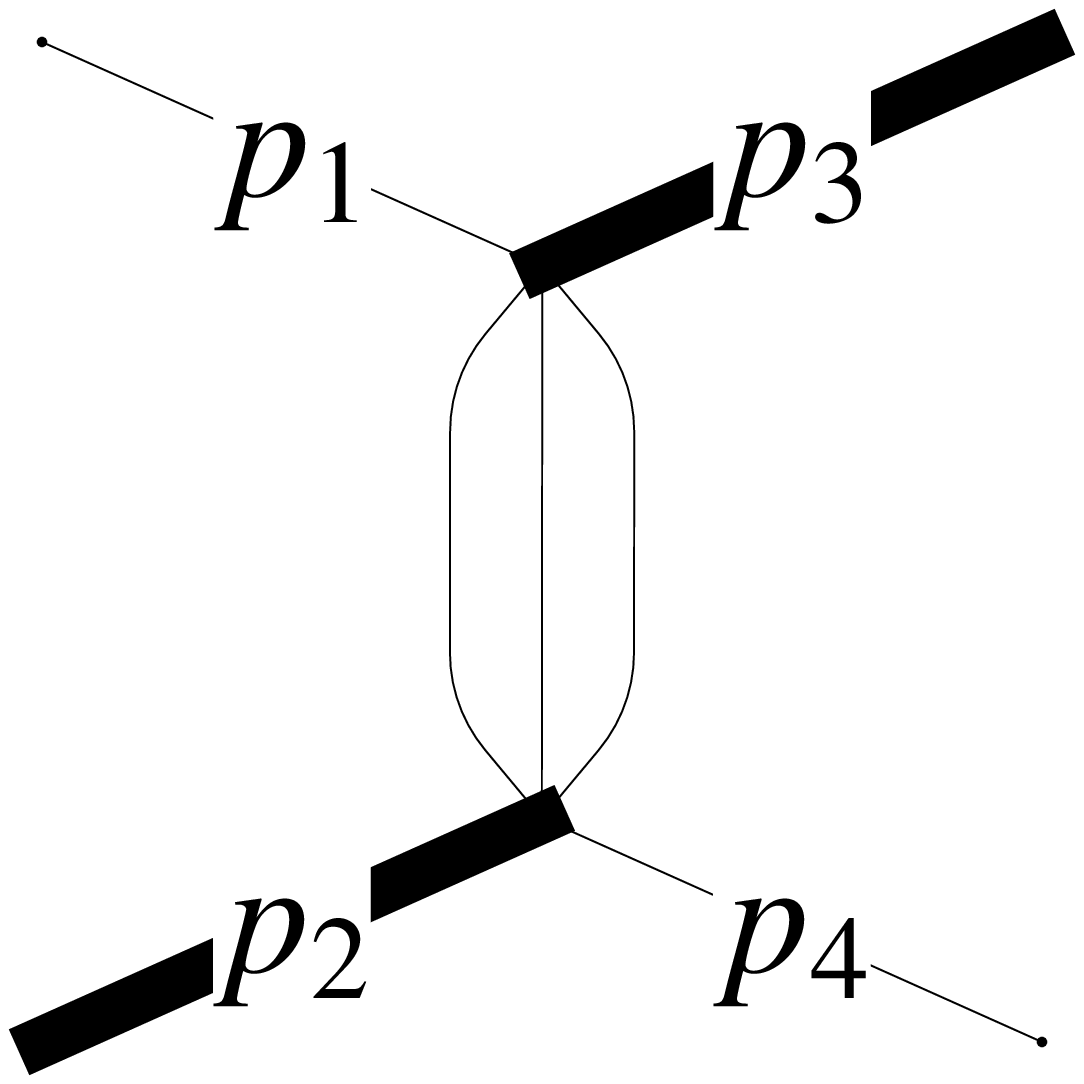}
{  \bea     
  g^{\rm P23}_1 &=& \eps^2 t G_{0, 0, 0, 0, 1, 2, 2, 0, 0} 
\,,~~~~~~~~~~~~~~~~~~~~~~~~~~~~~~~~~~~~~~~~~~~~~~~~~~~~~~~~
  \\
     f^{\rm P23}_{1} &\sim & -x^{-2\ep},  \nn 
\eea} \nn \\
\picturepage{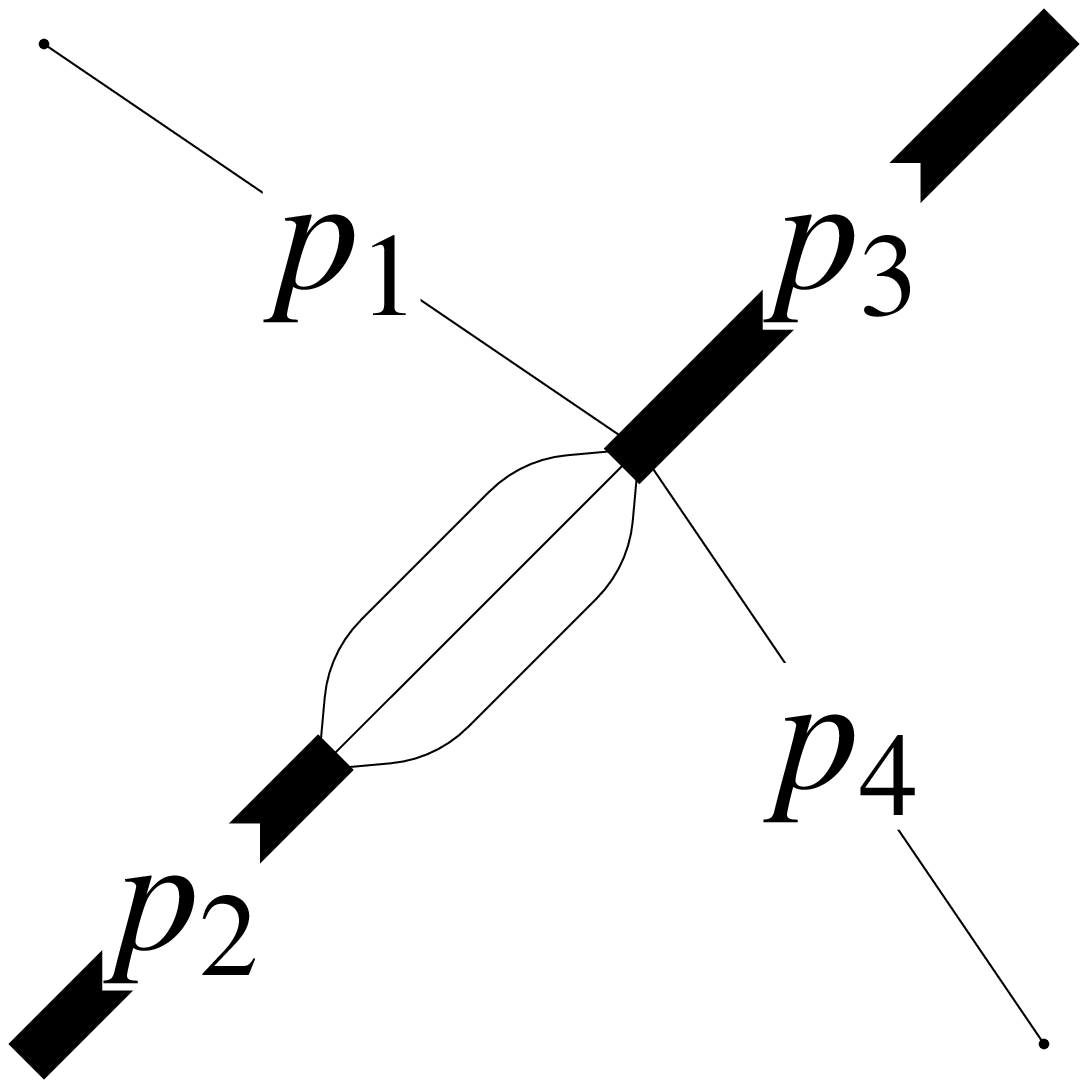}
{  \bea     
  g^{\rm P23}_2 &=& \eps^2 p_2^2  G_{0, 0, 0, 1, 2, 2, 0, 0, 0}
\,,~~~~~~~~~~~~~~~~~~~~~~~~~~~~~~~~~~~~~~~~~~~~~~~~~~~~~~~
  \\
     f^{\rm P23}_{2} &\sim &  -x^{-4\ep}e^{2i\pi \ep},  \nn 
\eea} \nn \\
\picturepage{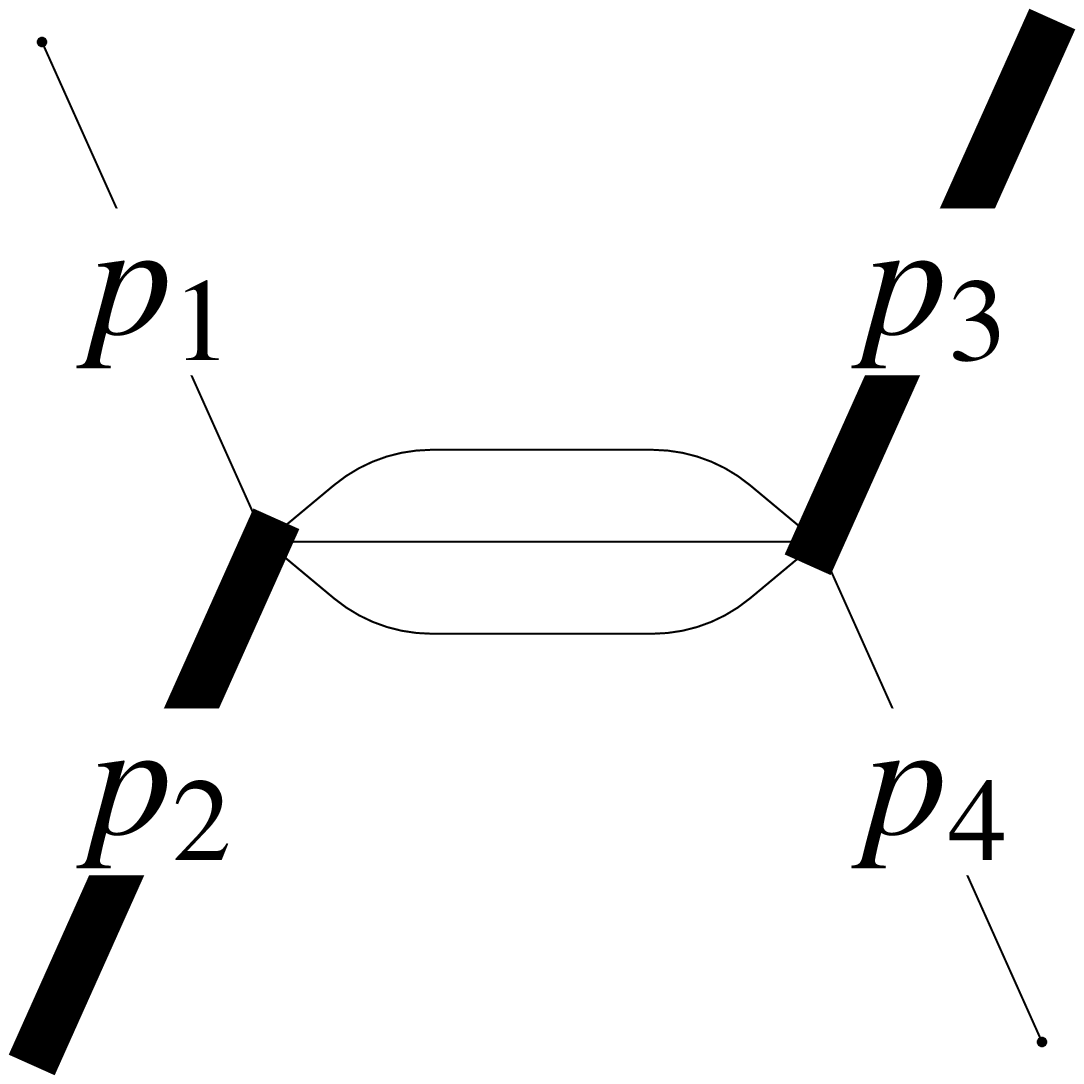}
{  \bea     
  g^{\rm P23}_3 &=& \eps^2 s G_{0, 2, 2, 0, 0, 1, 0, 0, 0} 
\,,~~~~~~~~~~~~~~~~~~~~~~~~~~~~~~~~~~~~~~~~~~~~~~~~~~~~~~~~
  \\
     f^{\rm P23}_{3} &\sim &  -x^{-2\ep}, \nn 
\eea} \nn \\ 
%\end{align}
%\end{small}
%\begin{small}
% \begin{align}
\picturepage{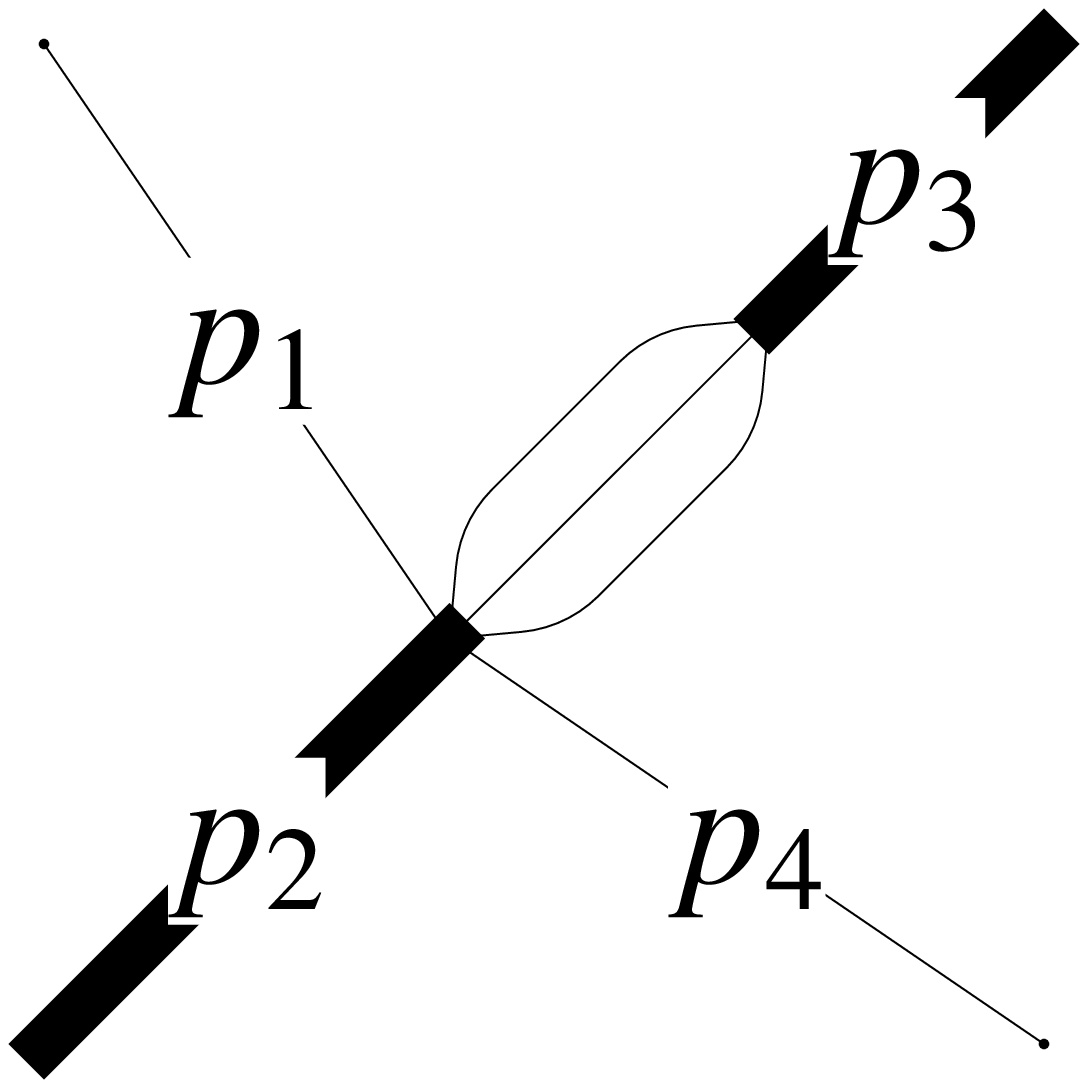}
{  \bea     
  g^{\rm P23}_4 &=& \eps^2 p_3^2  G_{1, 0, 0, 0, 0, 2, 2, 0, 0},
\,,~~~~~~~~~~~~~~~~~~~~~~~~~~~~~~~~~~~~~~~~~~~~~~~~~~~~~
  \\
     f^{\rm P23}_{4} &\sim &  -e^{2i\pi \ep},  \nn 
\eea} \nn \\
\picturepage{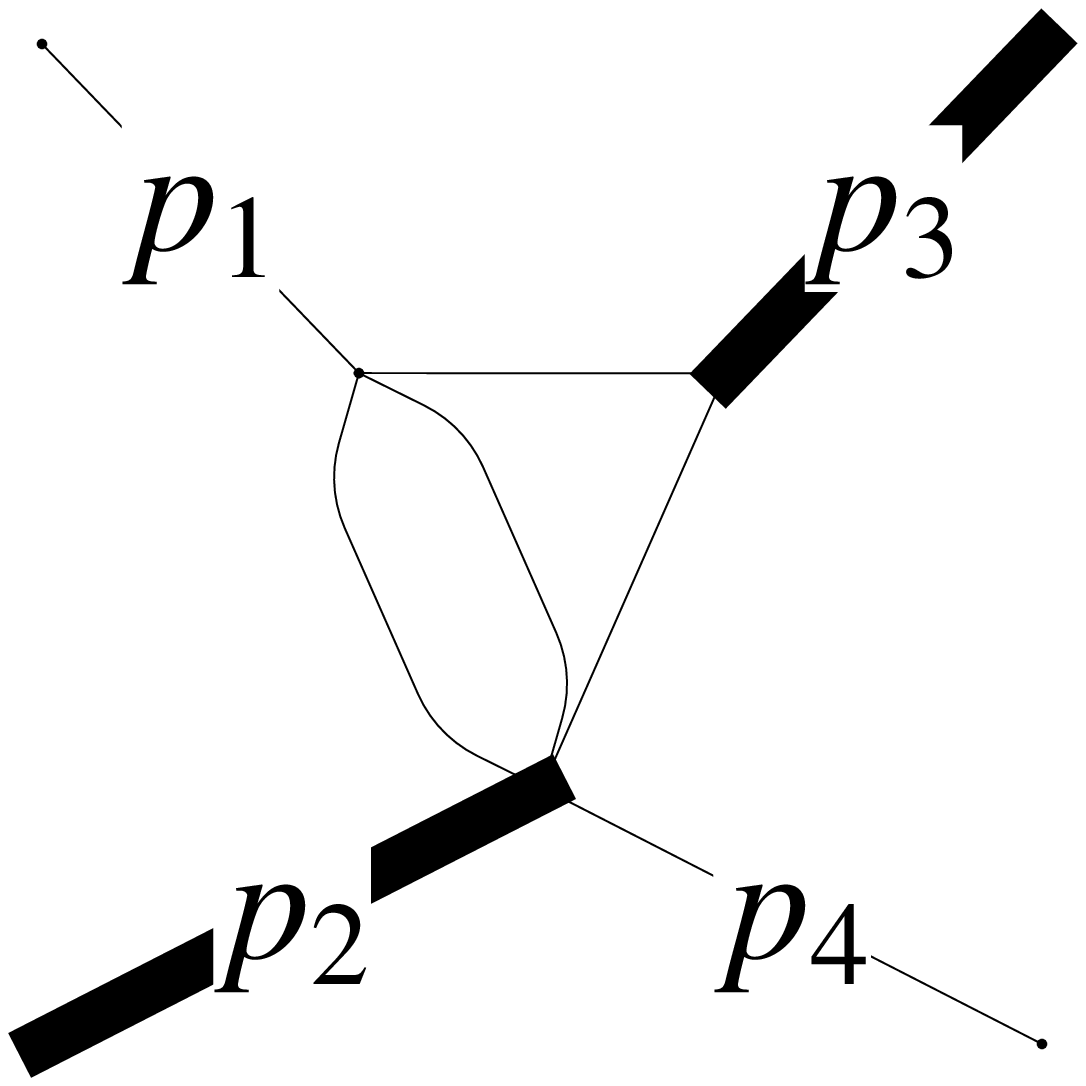}
{  \bea     
  g^{\rm P23}_5 &=& -2 \eps^3 (p_3^2  - t) G_{0, 0, 1, 0, 2, 1, 1, 0, 0}
\,,~~~~~~~~~~~~~~~~~~~~~~~~~~~~~~~~~~~~~~~~~~~~
  \\
     f^{\rm P23}_{5} &\sim &  \frac{x^{-2\ep}}{2}  
  - \frac{e^{2i\pi \ep}}{2} \left ( 1 + \frac{\pi^2 \ep^2  }{3} + 2 \zeta_3 \ep^3  
+ \frac{\pi^4 \ep^4}{10} \right ),  \nn 
\eea} \nn \\
\picturepage{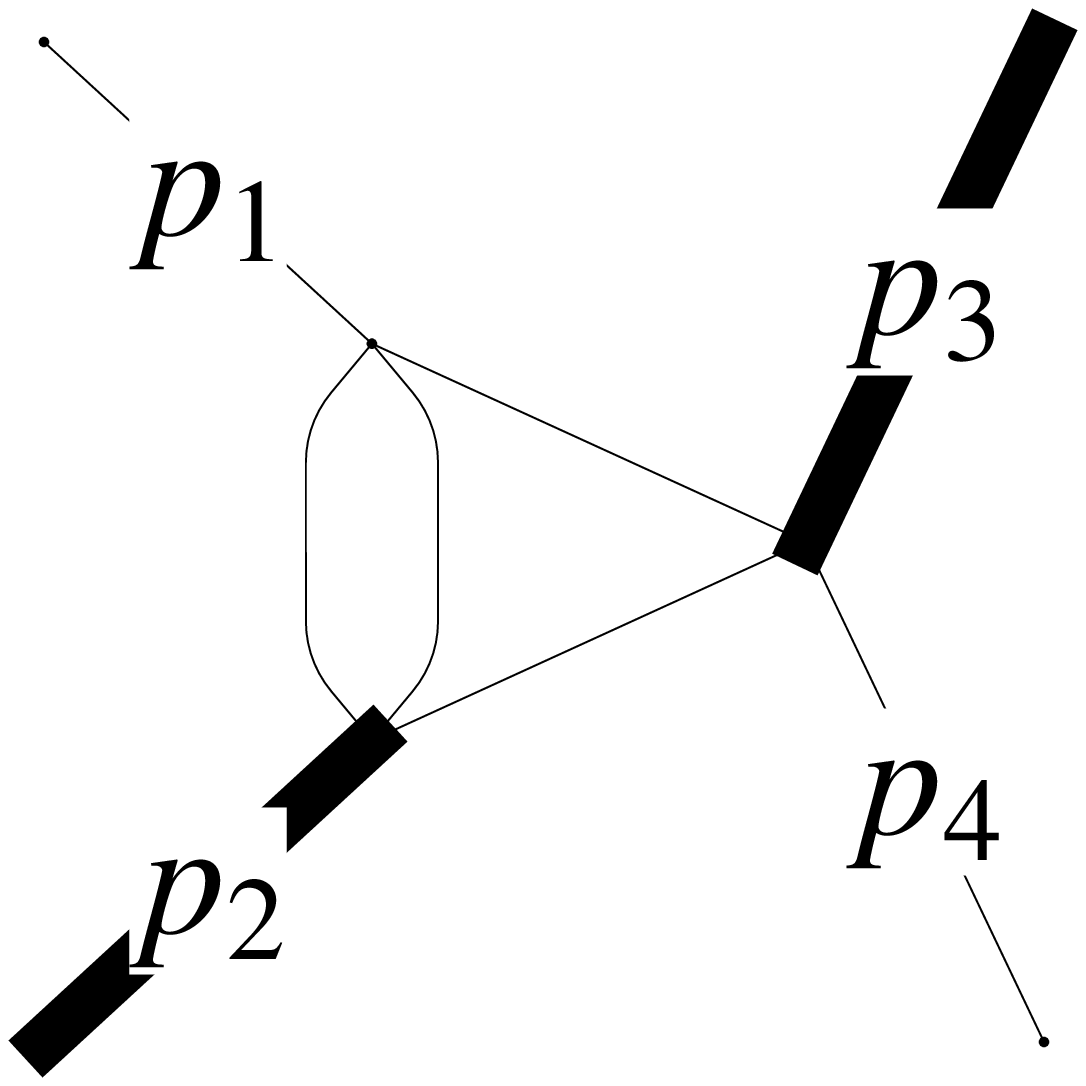}
{  \bea     
  g^{\rm P23}_6 &=& -2 \eps^3 (p_2^2  - s) G_{0, 0, 1, 1, 1, 2, 0, 0, 0}
\,,~~~~~~~~~~~~~~~~~~~~~~~~~~~~~~~~~~~~~~~~~~~~
  \\
     f^{\rm P23}_{6} &\sim &  -\frac{1}{2}e^{2i\pi \ep} x^{-4\ep} + 
   x^{-2\ep} \left ( \frac{1}{2} + \frac{\pi^2 \ep^2 }{6} 
+ \zeta_3 \ep^3  + \frac{ \pi^4 \ep^4 }{20} \right ),  \nn 
\eea} \nn \\
\picturepage{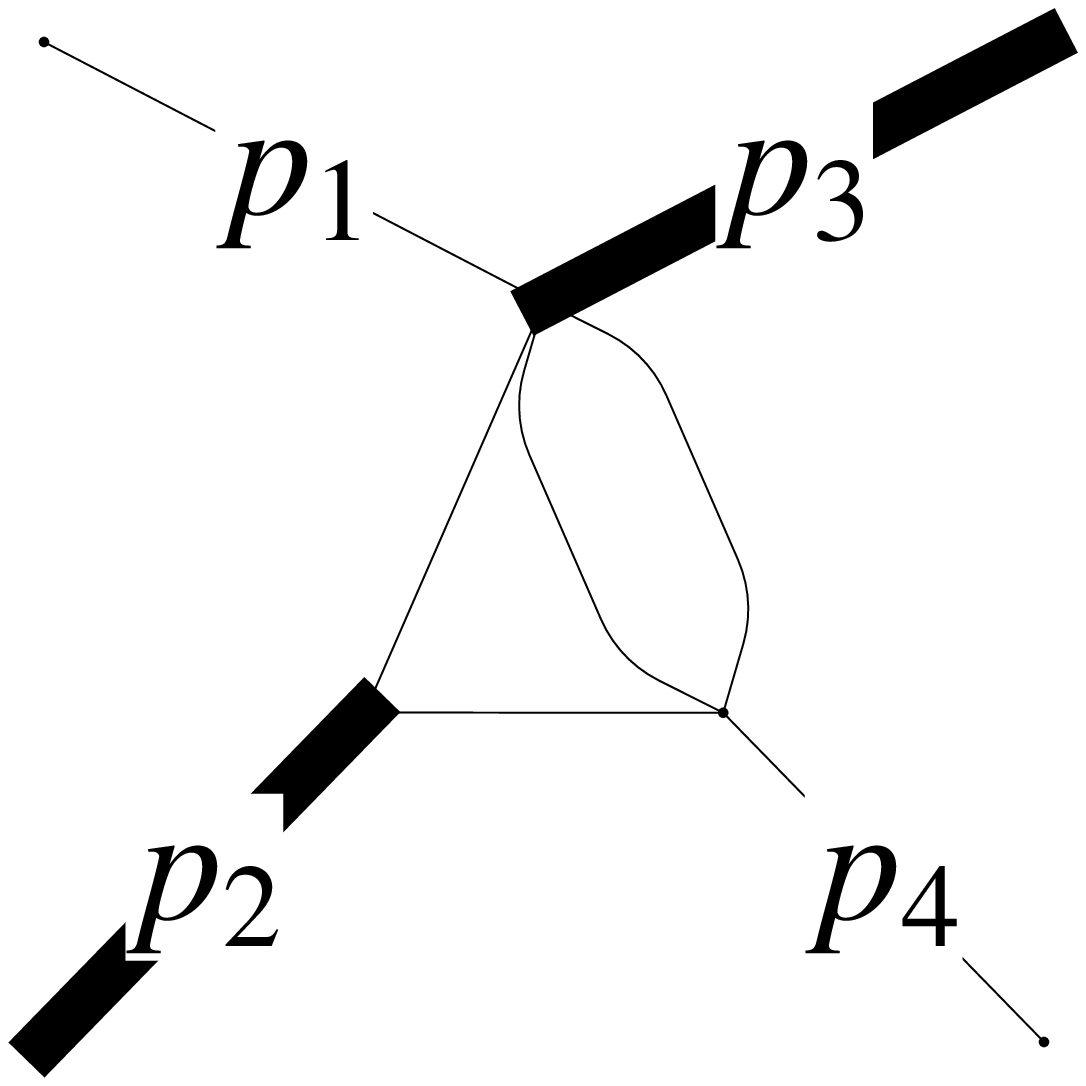}
{  \bea     
  g^{\rm P23}_7 &=& -2 \eps^3 (p_2^2  - t) G_{0, 1, 0, 0, 1, 1, 2, 0, 0}
\,,~~~~~~~~~~~~~~~~~~~~~~~~~~~~~~~~~~~~~~~~~~~~
  \\
     f^{\rm P23}_{7} &\sim &  x^{-2\ep}  - x^{-3\ep} N_1, \nn 
\eea} \nn \\
\picturepage{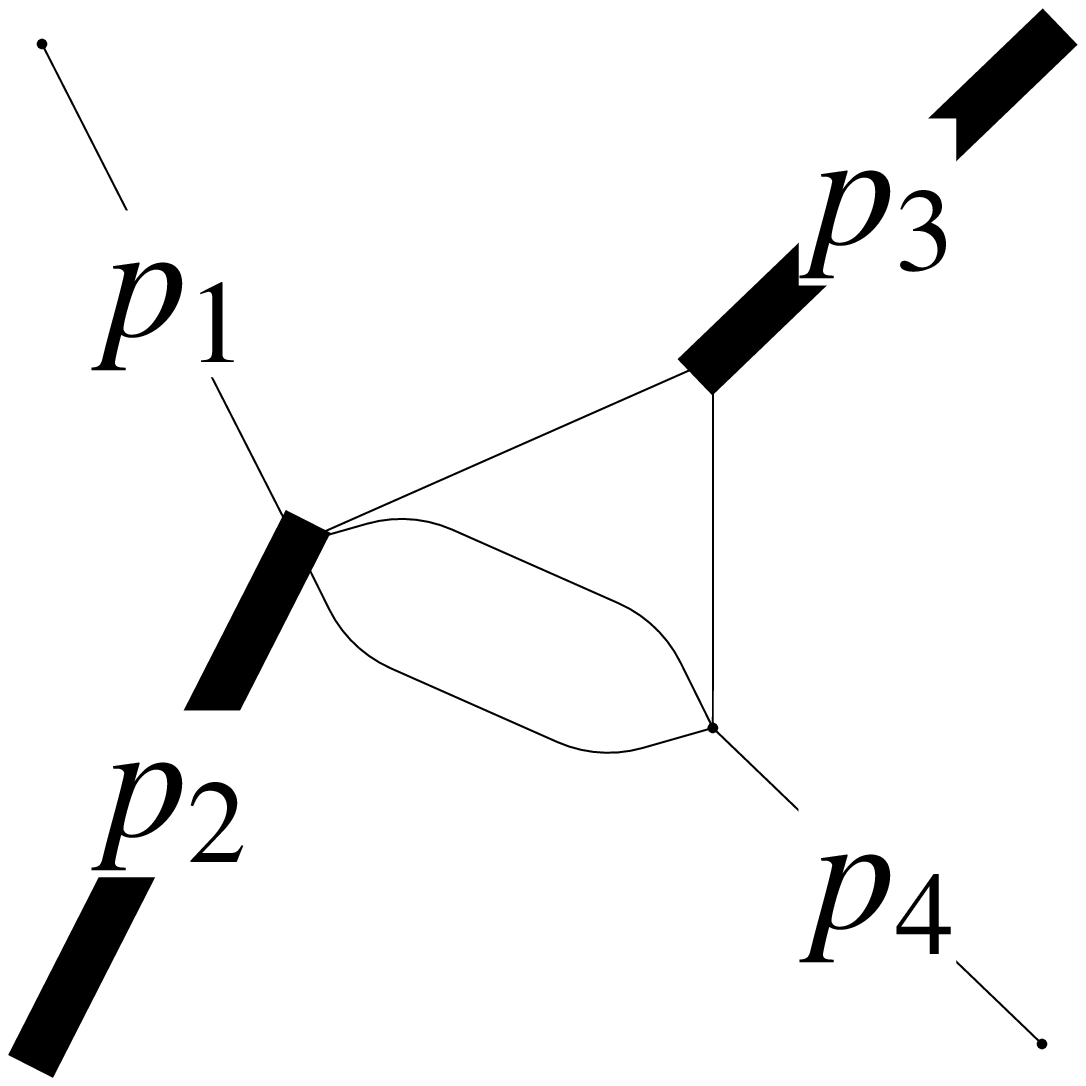}
{  \bea     
  g^{\rm P23}_8 &=& -2 \eps^3 (p_3^2  - s) G_{0, 2, 1, 0, 0, 1, 1, 0, 0}
\,,~~~~~~~~~~~~~~~~~~~~~~~~~~~~~~~~~~~~~~~~~~~~
  \\
     f^{\rm P23}_{8} &\sim &  \frac{x^{-2\ep}}{2} + 
  - \frac{e^{2i\pi \ep}}{2} \left ( 1 + \frac{ \pi^2 \ep^2 }{3} 
+ 2 \zeta_3 \ep^3  + \frac{\pi^4 \ep^4 }{10} \right ),
 \nn 
\eea} \nn \\
\picturepage{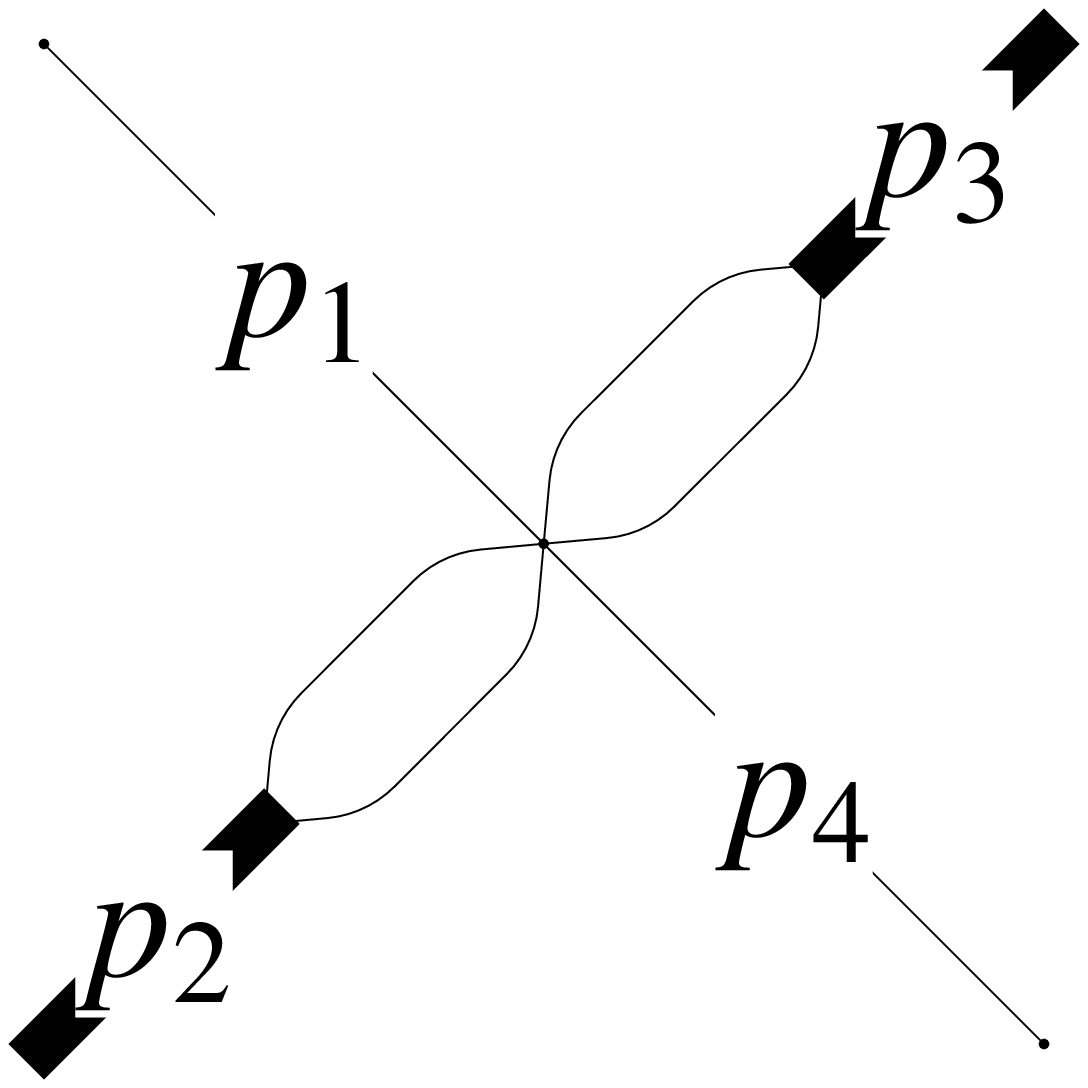}
{  \bea     
  g^{\rm P23}_9 &=& \eps^2 p_2^2  p_3^2  G_{0, 2, 2, 0, 1, 0, 1, 0, 0}
\,,~~~~~~~~~~~~~~~~~~~~~~~~~~~~~~~~~~~~~~~~~~~~~~~~~~~~
  \\
     f^{\rm P23}_{9} &\sim &  x^{-2\ep}e^{2i\pi \ep} N_2,  \nn 
\eea} \nn \\
\picturepage{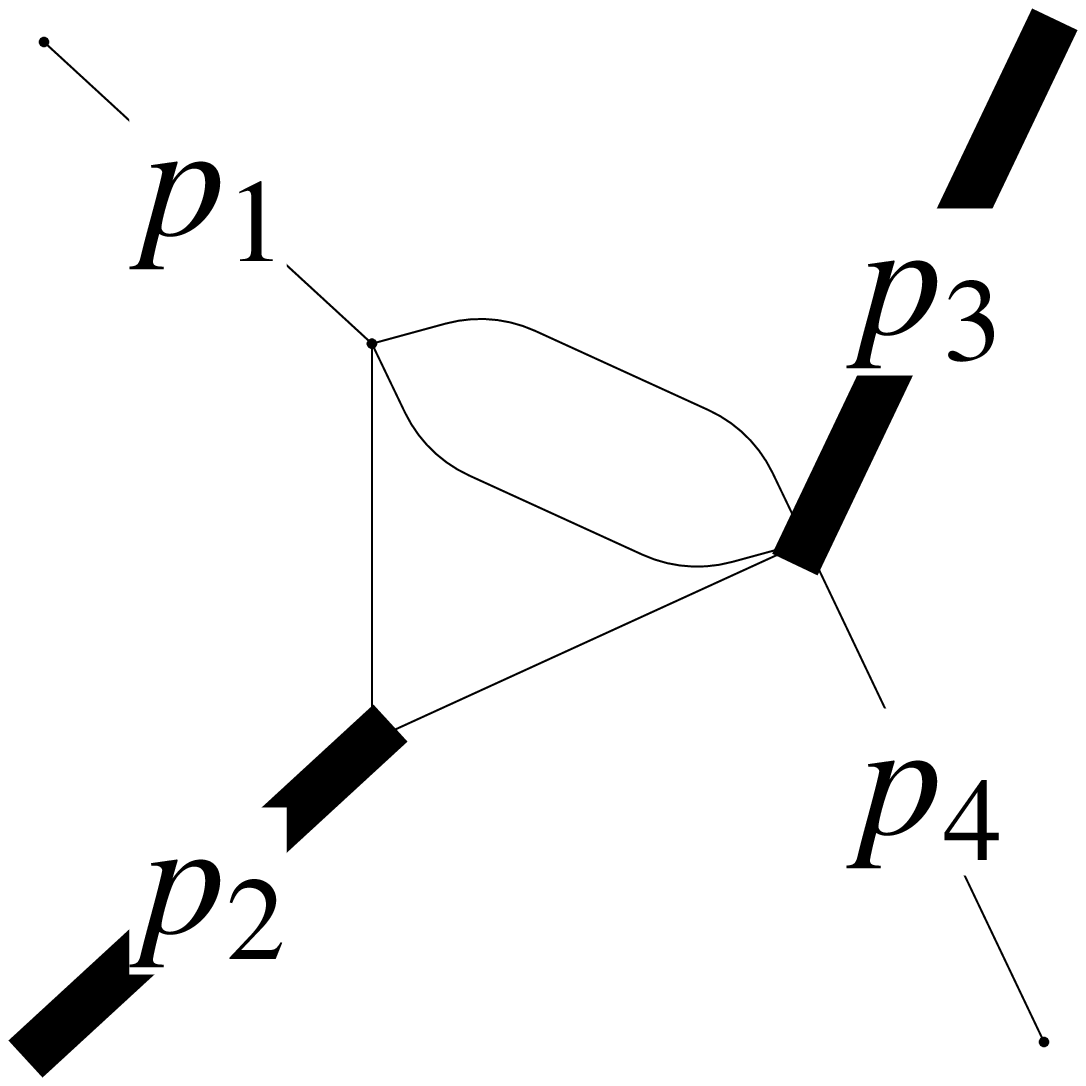}
{  \bea     
  g^{\rm P23}_{10} &=& -2 \eps^3 (p_2^2  - s) G_{0, 1, 2, 0, 1, 1, 0, 0, 0}
\,,~~~~~~~~~~~~~~~~~~~~~~~~~~~~~~~~~~~~~~~~~~~~
  \\
     f^{\rm P23}_{10} &\sim &   x^{-2\ep}  - x^{-3\ep} N_1, \nn 
\eea} \nn \\
\picturepage{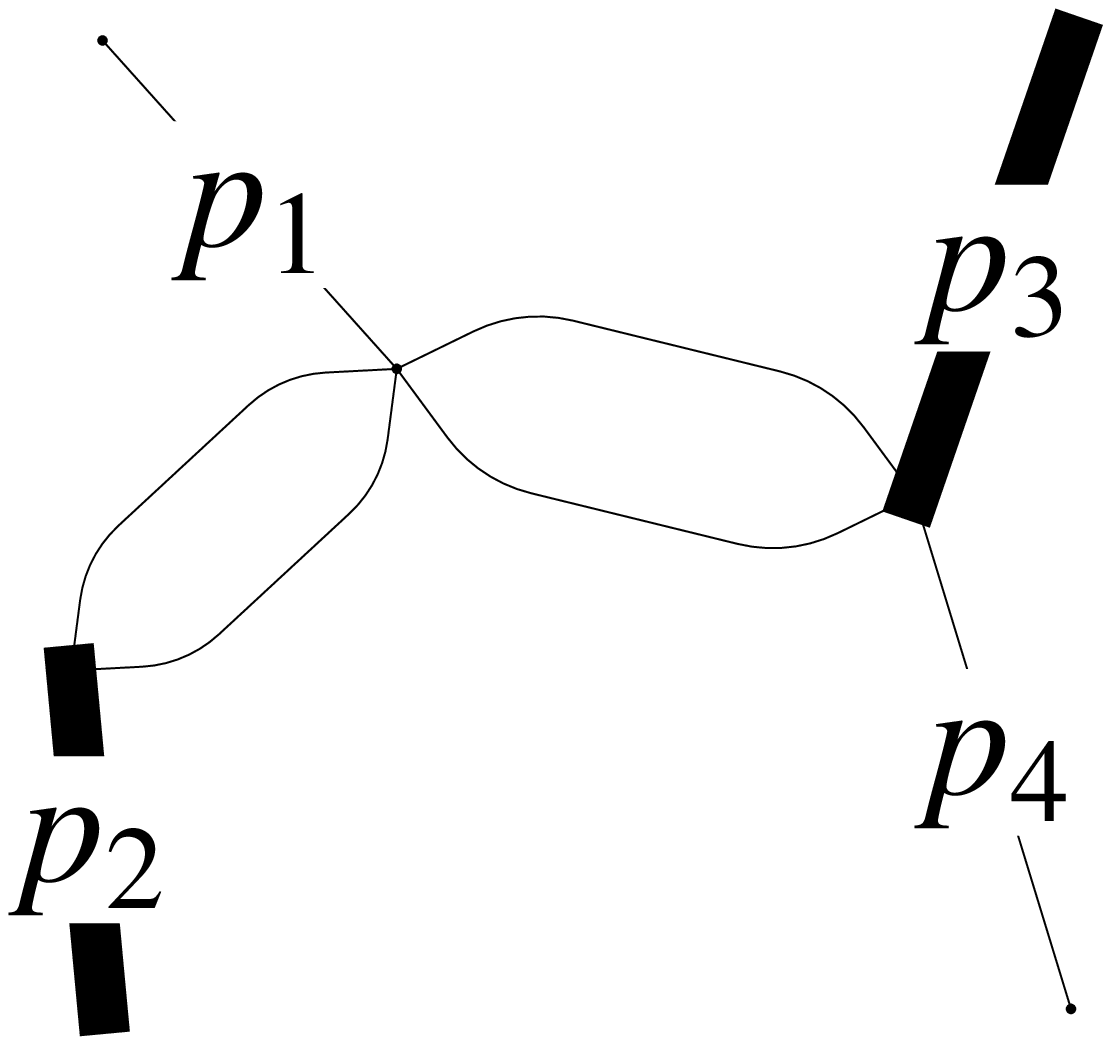}
{  \bea     
  g^{\rm P23}_{11} &=& \eps^2 p_2^2  s G_{0, 2, 2, 1, 1, 0, 0, 0, 0}
\,,~~~~~~~~~~~~~~~~~~~~~~~~~~~~~~~~~~~~~~~~~~~~~~~~~~~~~
  \\
     f^{\rm P23}_{11} &\sim &  x^{-3\ep}e^{i\pi \ep} N_2,  \nn 
\eea} \nn \\
\picturepage{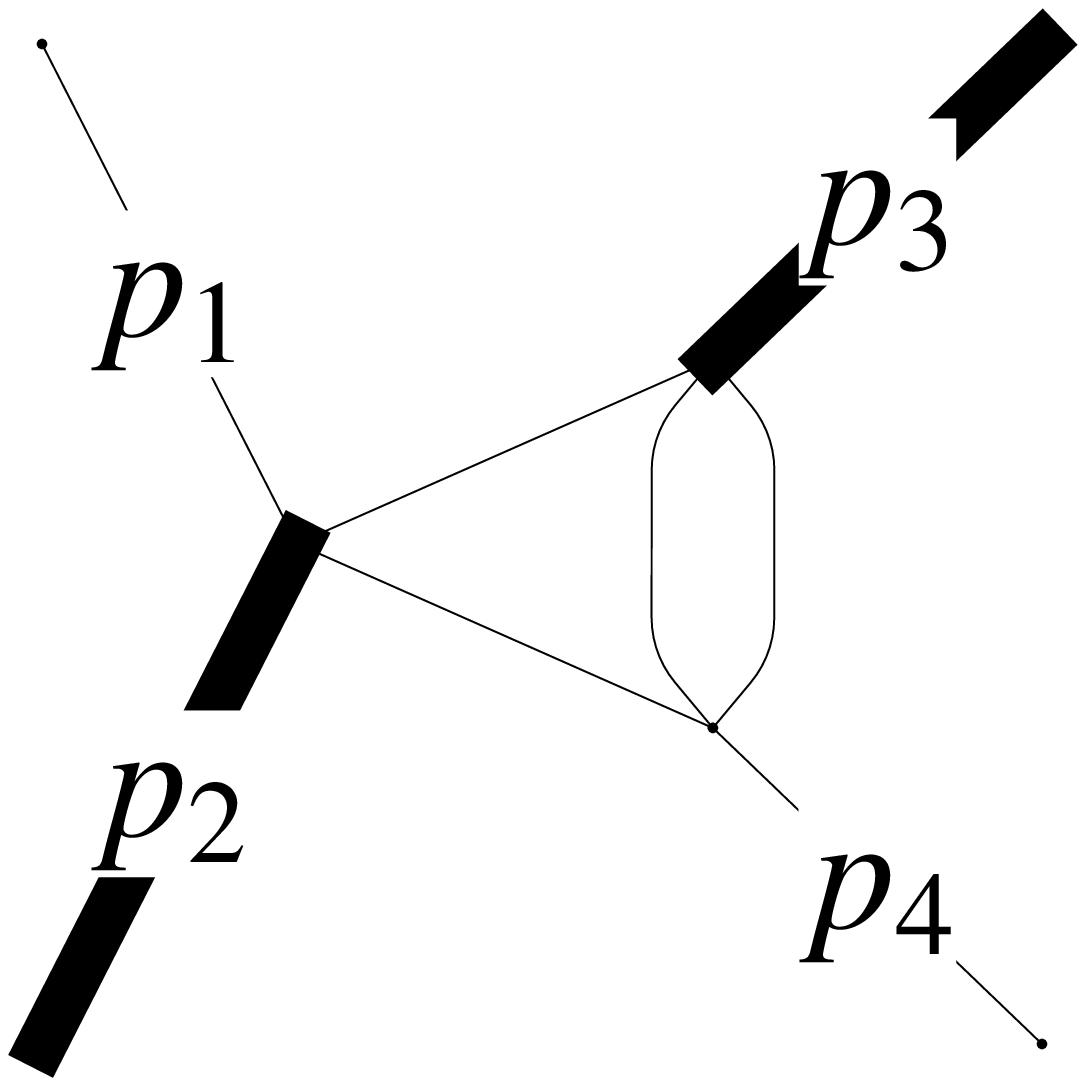}
{  \bea     
  g^{\rm P23}_{12} &=& -2 \eps^3 (p_3^2  - s) G_{1, 1, 0, 0, 0, 2, 1, 0, 0}
\,,~~~~~~~~~~~~~~~~~~~~~~~~~~~~~~~~~~~~~~~~~~~~
  \\
     f^{\rm P23}_{12} &\sim &   -e^{2i\pi \ep} + 
  e^{i\pi \ep} x^{-\ep} \left (1 - \frac{\pi^2 \ep^2 }{6} + 2  \zeta_3 \ep^3 
- \frac{ \pi^4 \ep^4 }{40}  \right ),  \nn 
\eea} \nn \\
\picturepage{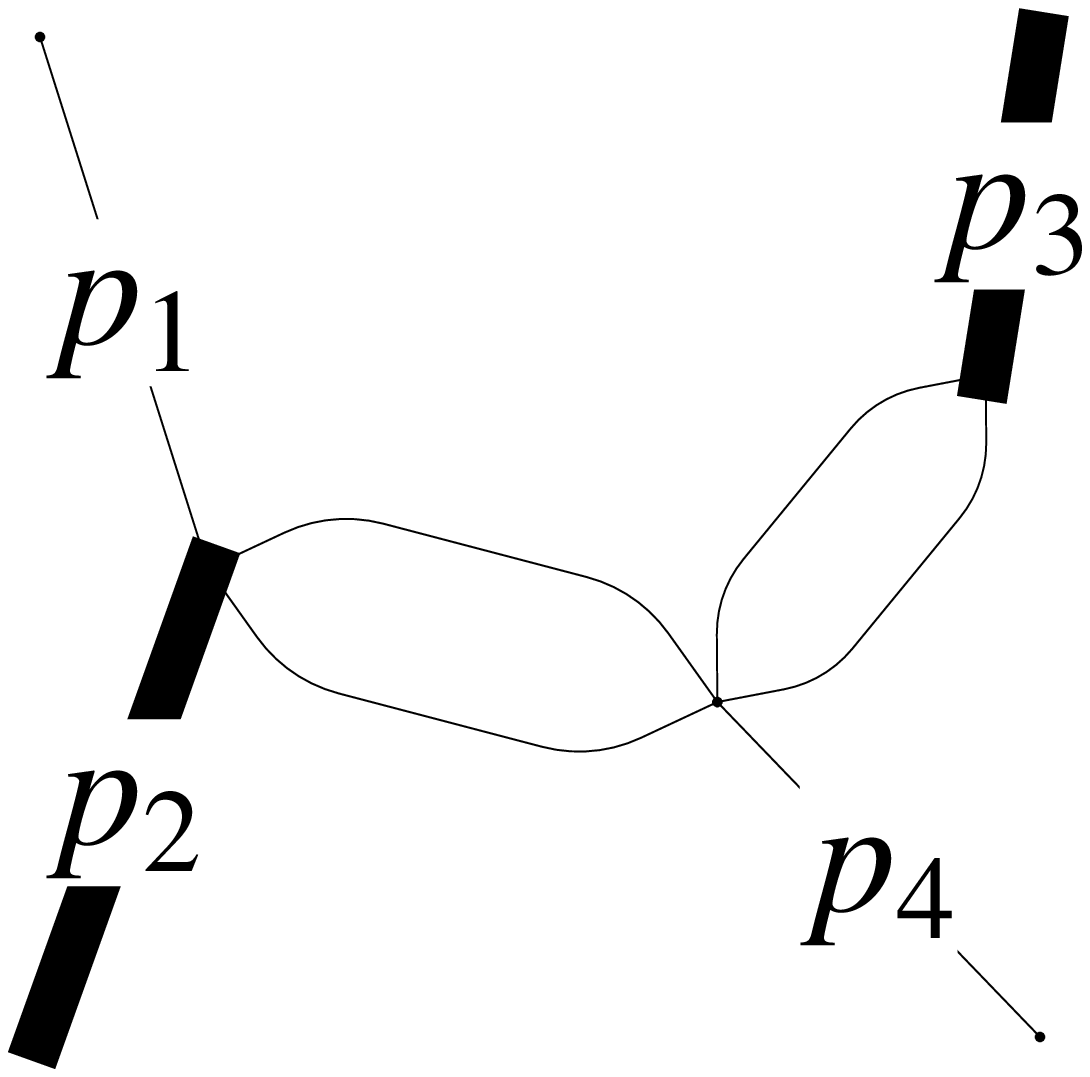}
{  \bea     
  g^{\rm P23}_{13} &=& \eps^2 p_3^2  s G_{1, 2, 2, 0, 0, 0, 1, 0, 0}
\,,~~~~~~~~~~~~~~~~~~~~~~~~~~~~~~~~~~~~~~~~~~~~~~~~~~~~~~
  \\
     f^{\rm P23}_{13} &\sim &  x^{-\ep} e^{i\pi \ep} N_2,  \nn 
\eea} \nn \\
%
%\end{align} 
%\end{small}
%
%\begin{small}
%\begin{align}
\picturepage{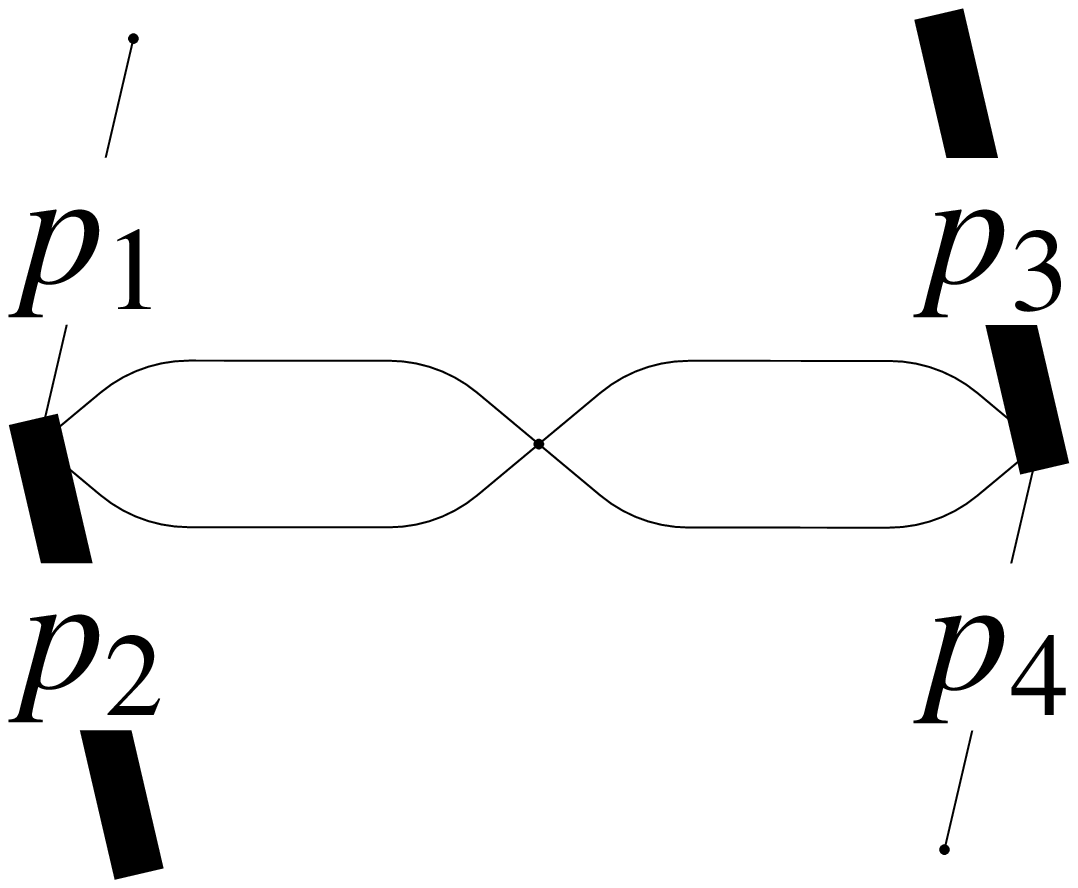}
{  \bea     
  g^{\rm P23}_{14} &=& \eps^2 s^2 G_{1, 2, 2, 1, 0, 0, 0, 0, 0}    
\,,~~~~~~~~~~~~~~~~~~~~~~~~~~~~~~~~~~~~~~~~~~~~~~~~~~~~~
  \\
     f^{\rm P23}_{14} &\sim &  x^{-2\ep} N_2,   \nn 
\eea} \nn \\
\picturepage{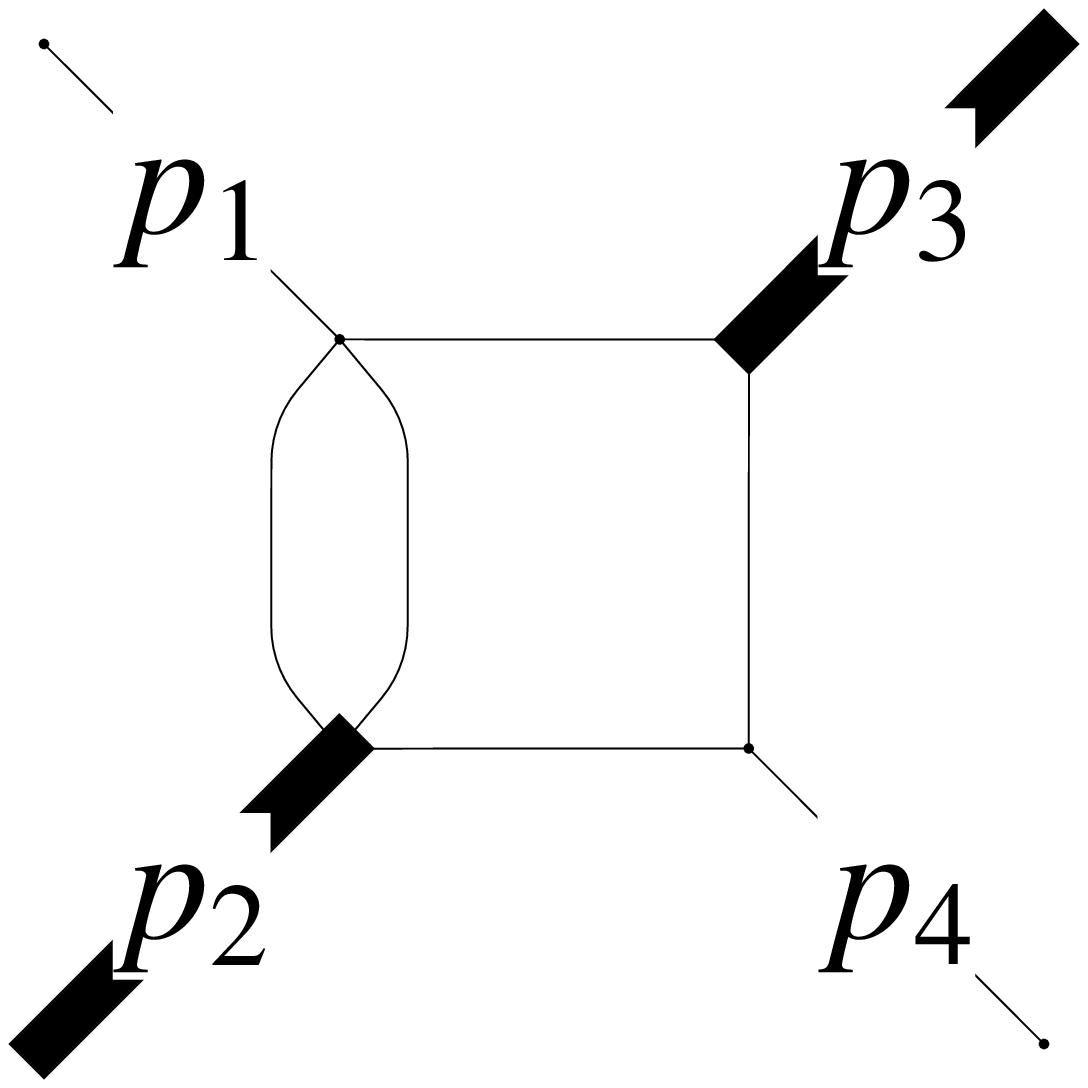}
{  \bea     
  g^{\rm P23}_{15} &=& -2 \eps^3 (p_2^2  p_3^2  - s t) G_{0, 0, 1, 1, 2, 1, 1, 0, 0}
\,,~~~~~~~~~~~~~~~~~~~~~~~~~~~~~~~~~~~~~~~~~
  \\
     f^{\rm P23}_{15} &\sim &  6i \pi \ep x^{-4\ep} [(z-y) (1-z) ]^{-2\ep},  \nn 
\eea} \nn \\
\picturepage{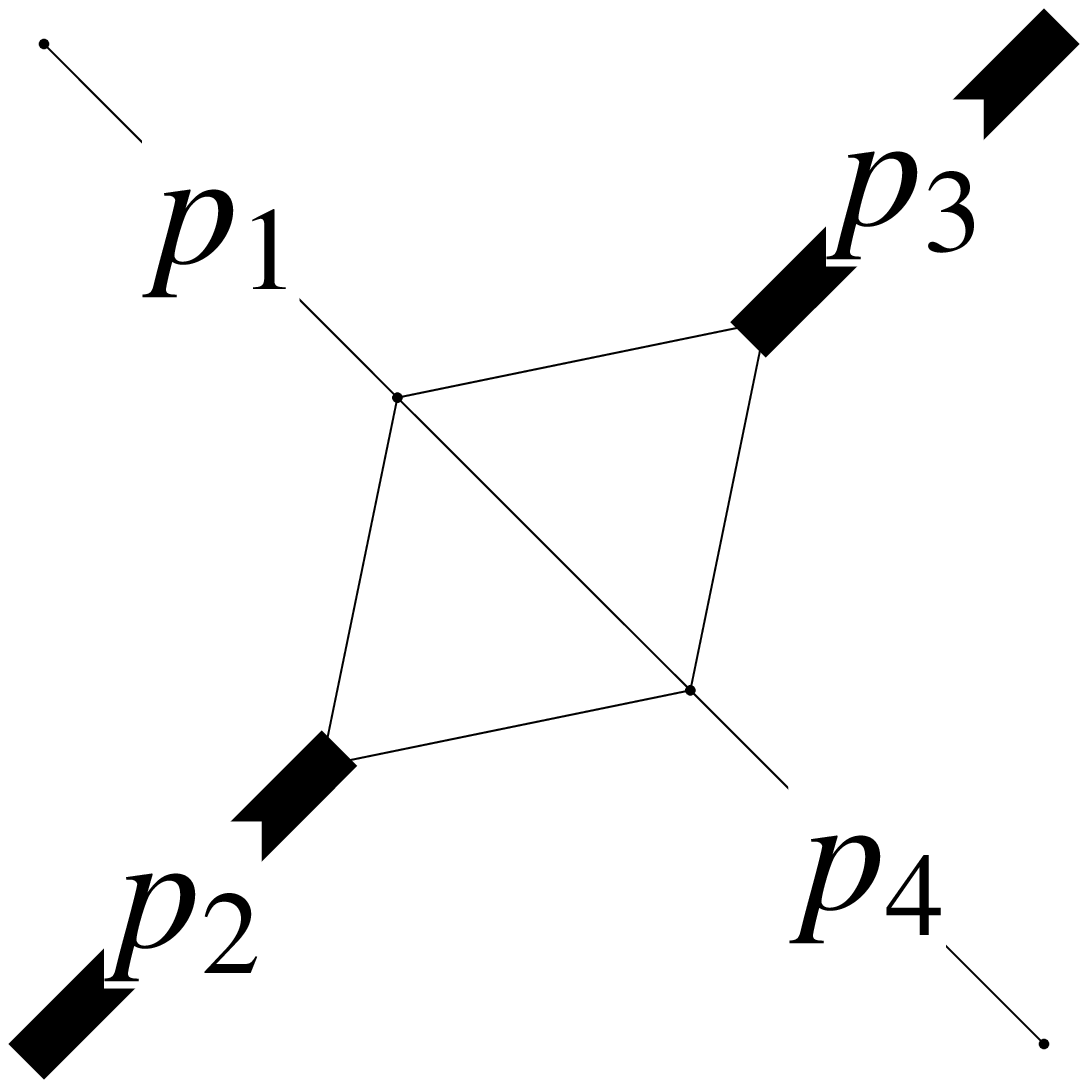}
{  \bea     
  g^{\rm P23}_{16} &=& 4 \eps^4 R_{23} G_{0, 1, 1, 0, 1, 1, 1, 0, 0}
\,,~~~~~~~~~~~~~~~~~~~~~~~~~~~~~~~~~~~~~~~~~~~~~~~~~~~~~~~
  \\
     f^{\rm P23}_{16} &\sim &  0, \nn 
\eea} \nn \\
\picturepage{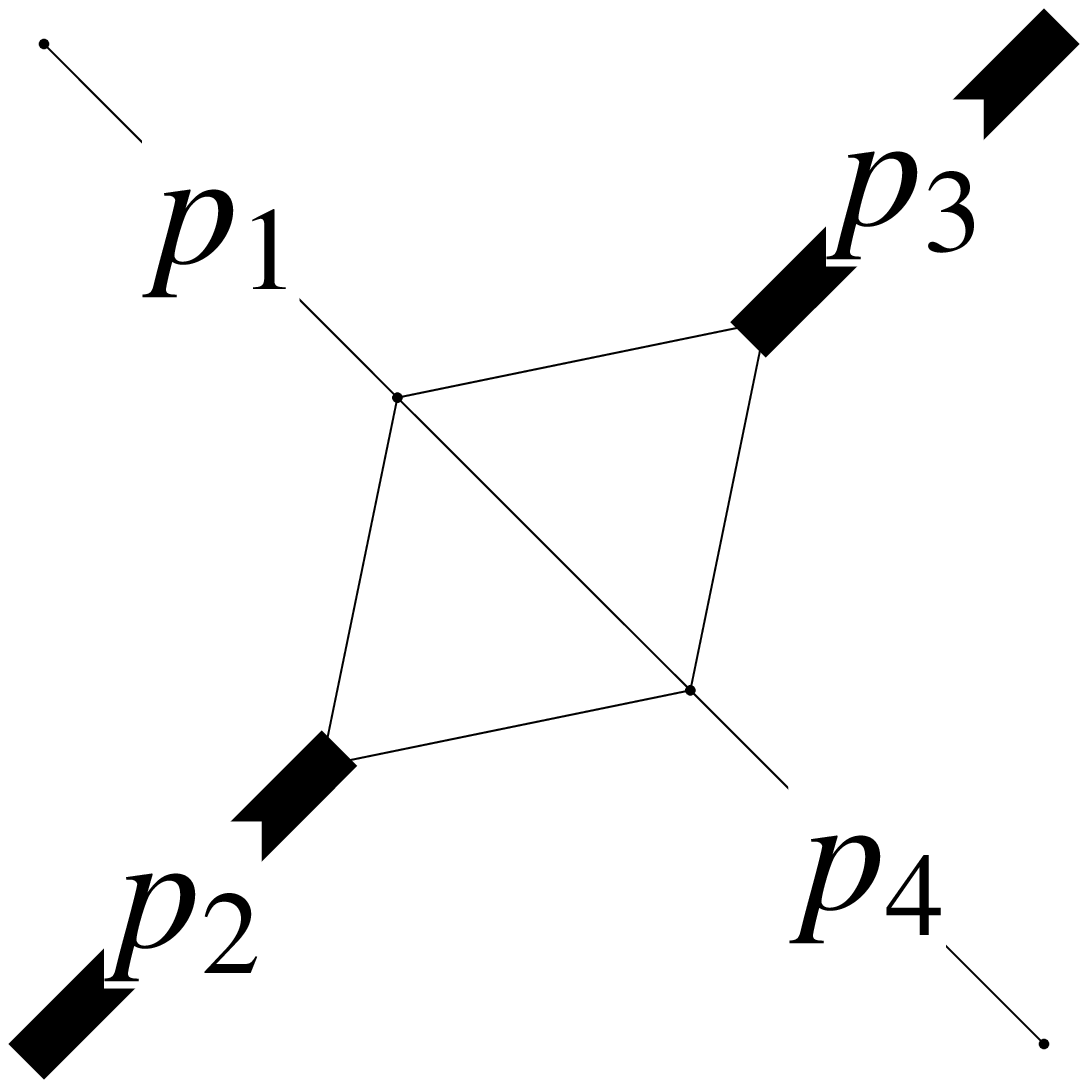}
{  \bea     
  g^{\rm P23}_{17} &=& -2 \eps^3 (p_2^2  p_3^2  - s t) G_{0, 1, 1, 0, 1, 2, 1, 0, 0}
\,,~~~~~~~~~~~~~~~~~~~~~~~~~~~~~~~~~~~~~~~~~~~~
  \\
     f^{\rm P23}_{17} &\sim &  4i \pi \ep N_3 x^{-4\ep} \left [ ( z-y) (1-z) \right ]^{-3\ep},  \nn 
\eea} \nn \\
\picturepage{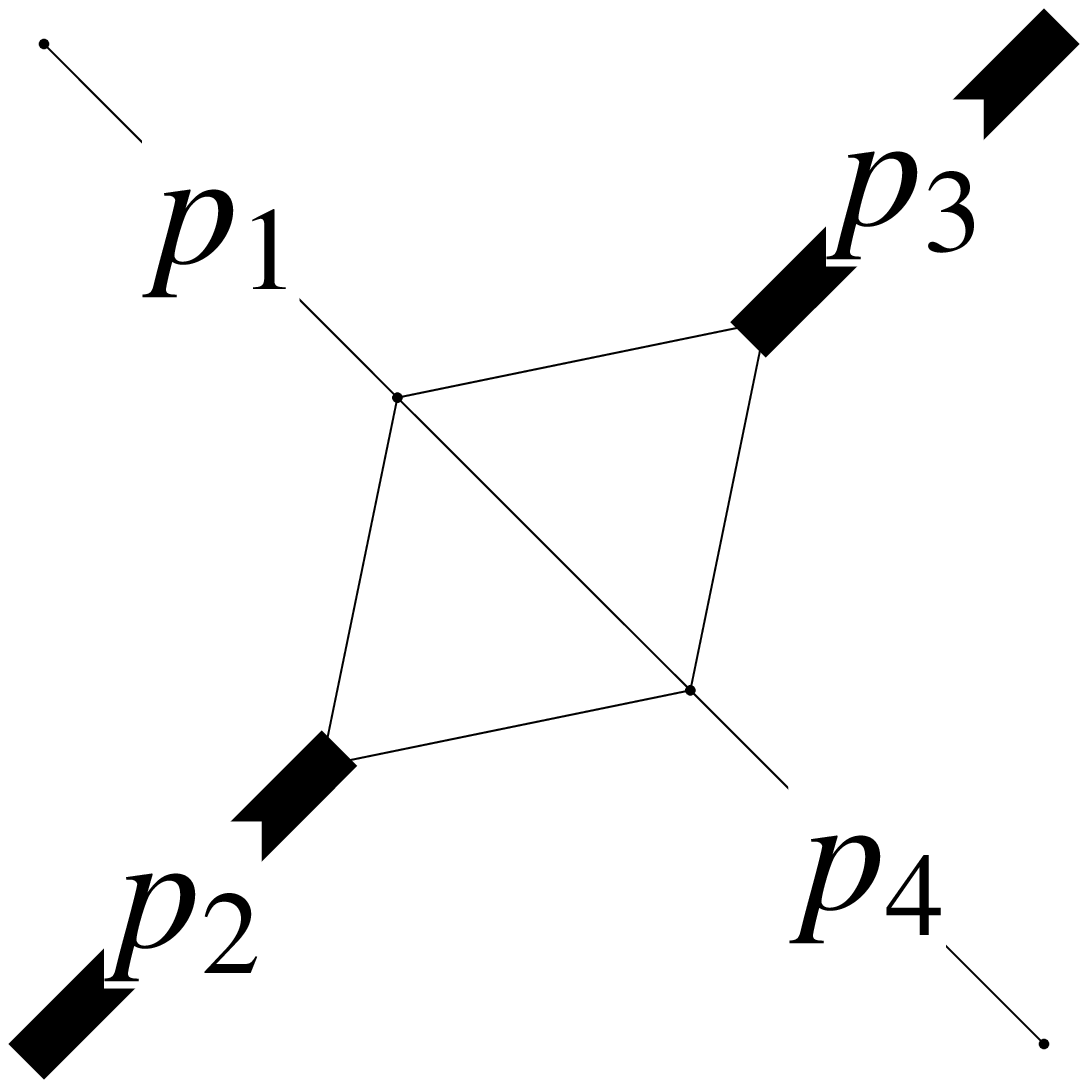}
{  \bea     
  g^{\rm P23}_{18} &=& -2 \eps^3 p_2^2  (p_3^2  - s) G_{0, 2, 1, 0, 1, 1, 1, 0, 0}
\,,
  \\
    ~~~~~~f^{\rm P23}_{18} &\sim &   
x^{-4\ep} \left ( 1 + \frac{2 i  \pi \ep}{3} + \frac{ \pi^2 \ep^2 }{3} + 
     \left ( \frac{2 i \pi^3}{9} - 2 \zeta(3) \right ) \ep^3 + 
     \left ( - \frac{7 \pi^4}{90} - \frac{4 i \pi \zeta_3}{3} \right ) \ep^4 \right ) 
  \nn
\\
& +&    x^{-2\ep}  \left ( 2 + 4i \pi \ep   - 4  \pi^2 \ep^2 + 
     \left ( - \frac{8i \pi^3}{3} + 12 \zeta_3 \right  ) \ep^3 + 
     \left (  \frac{ 23 \pi^4}{15} + 24 i \pi \zeta_3 \right ) \ep^4
\right ) \nn
\\
&-& 3 x^{-3\ep}  N_1
-\frac{2 i \pi \ep}{3} N_3 x^{-4\ep}[( z-y) (1-z)]^{-3\ep},
\nn 
\eea} \nn \\
\picturepage{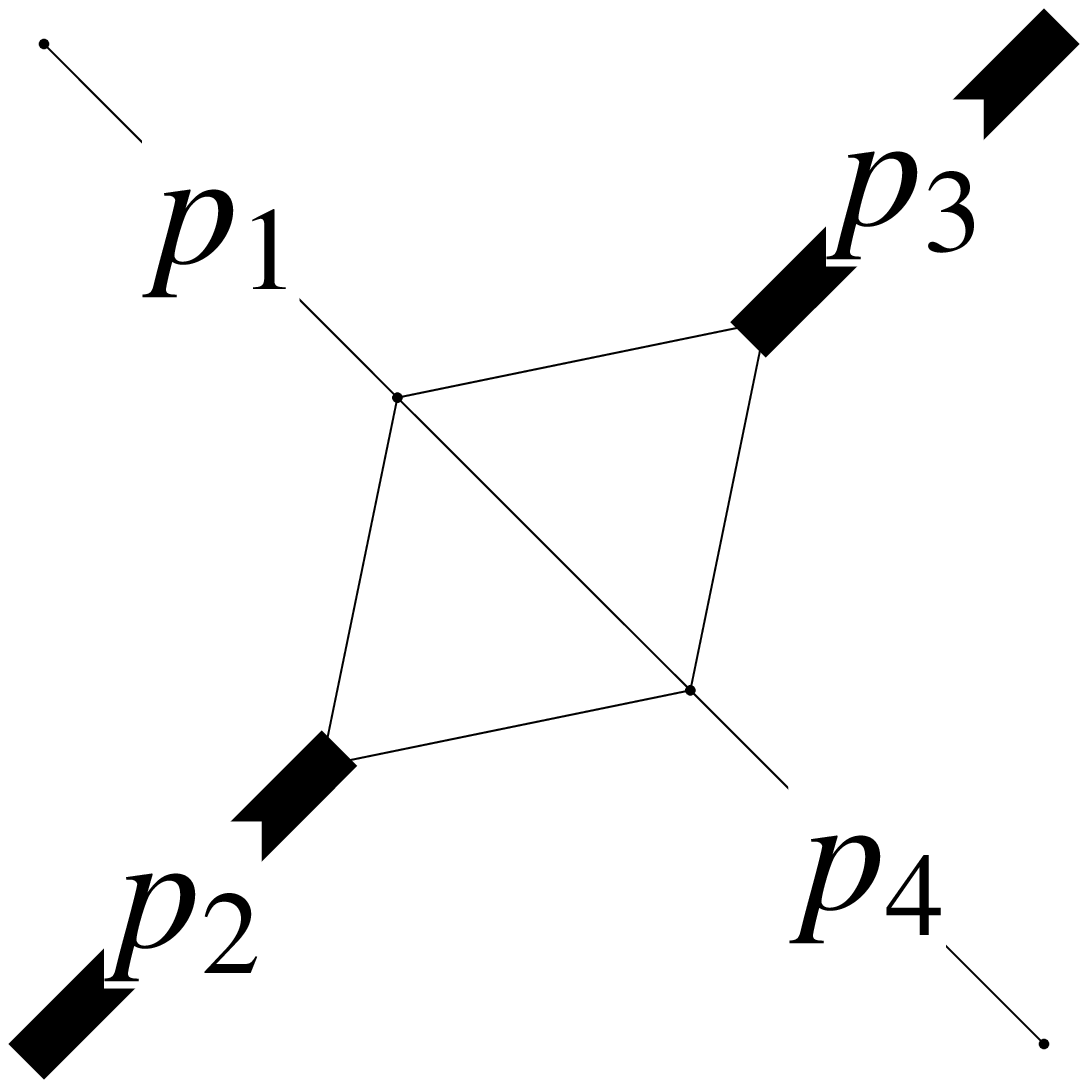}
{  \bea     
  g^{\rm P23}_{19} &=& -2 \eps^3 p_3^2  (p_2^2  - s) G_{0, 1, 2, 0, 1, 1, 1, 0, 0}
\,,
  \\
     f^{\rm P23}_{19} &\sim &  
 x^{-4\ep}  \left (  -\frac{1}{2}  - \frac{i\pi \ep}{3} - \frac{ \pi^2 \ep^2}{6} + 
     \left (- \frac{i \pi^3}{9} + \zeta_3 \right ) \ep^3 + 
     \left (  \frac{ 7 \pi^4}{180} + \frac{2 i \pi \zeta_3}{3} \right ) \ep^4 
  \right ) \nn
\\
& + &
x^{-3\ep} \left ( 2 + 2 i \pi \ep  - \frac{4  \pi^2 \ep^2 }{3} + 
     \left ( - \frac{2 i \pi^3}{3} + 4 \zeta_3 \right ) \ep^3 + 
     \left ( \frac{\pi^4}{5} + 4 i \pi \zeta_3 \right ) \ep^4 \right ) \nn
\\
  &-& \frac{3 x^{-2\ep} }{2} 
-\frac{2 i \pi \ep}{3} N_3 x^{-4\ep}[( z-y) (1-z)]^{-3\ep},
 \nn 
\eea} \nn \\
\picturepage{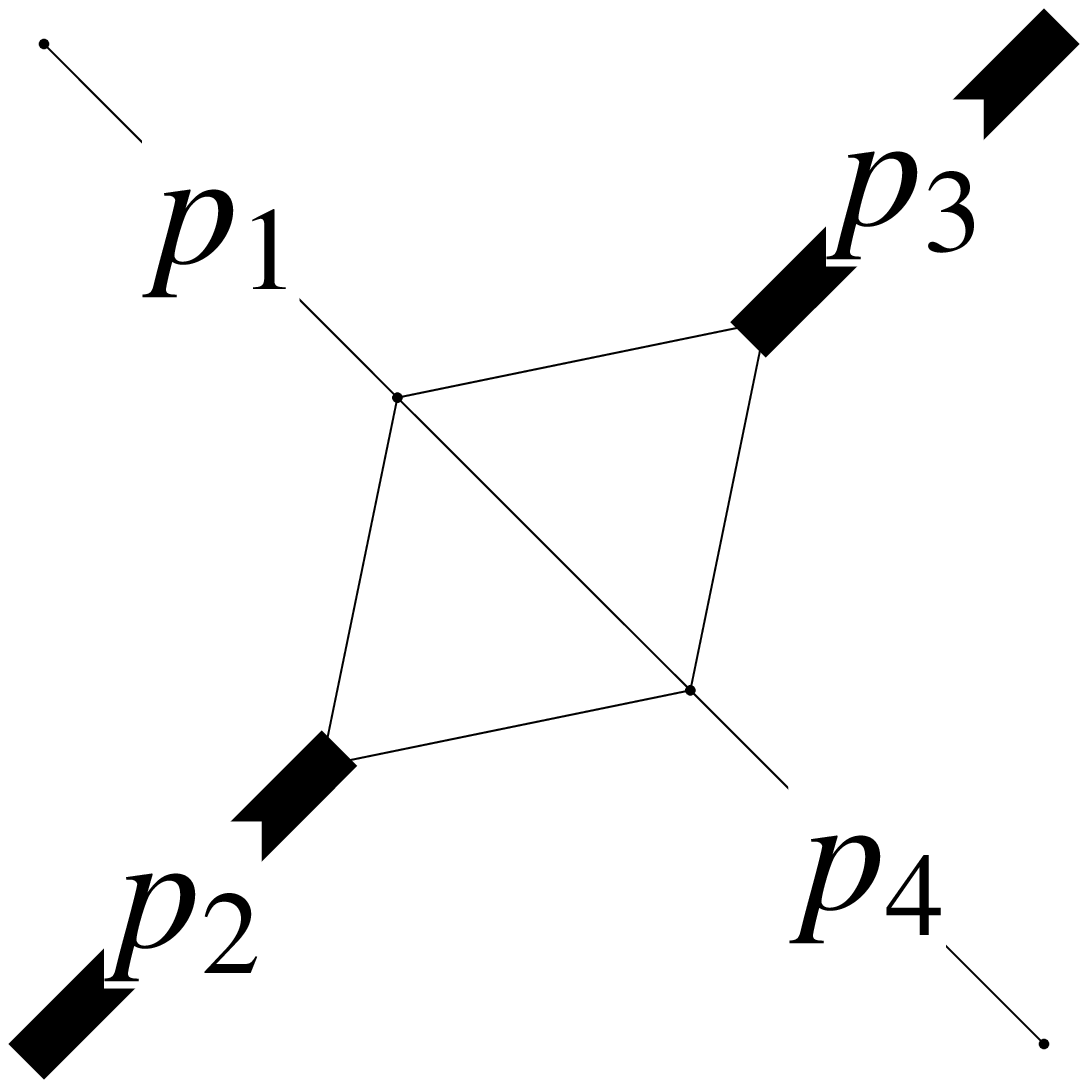}
{  \bea     
  g^{\rm P23}_{20} &=& -2 \eps^3 p_3^2  (p_2^2  - t) G_{0, 1, 1, 0, 1, 1, 2, 0, 0}
\,,~~~~~~~~~~~~~~~~~~~~~~~~~~~~~~~~~~~~~~~~~~~~
  \\
     f^{\rm P23}_{20} &\sim &  f^{\rm P23}_{19} ,
\nn 
\eea} \nn \\
%\end{align} 
%\end{small}
%
%\begin{small}
%\begin{align}
\picturepage{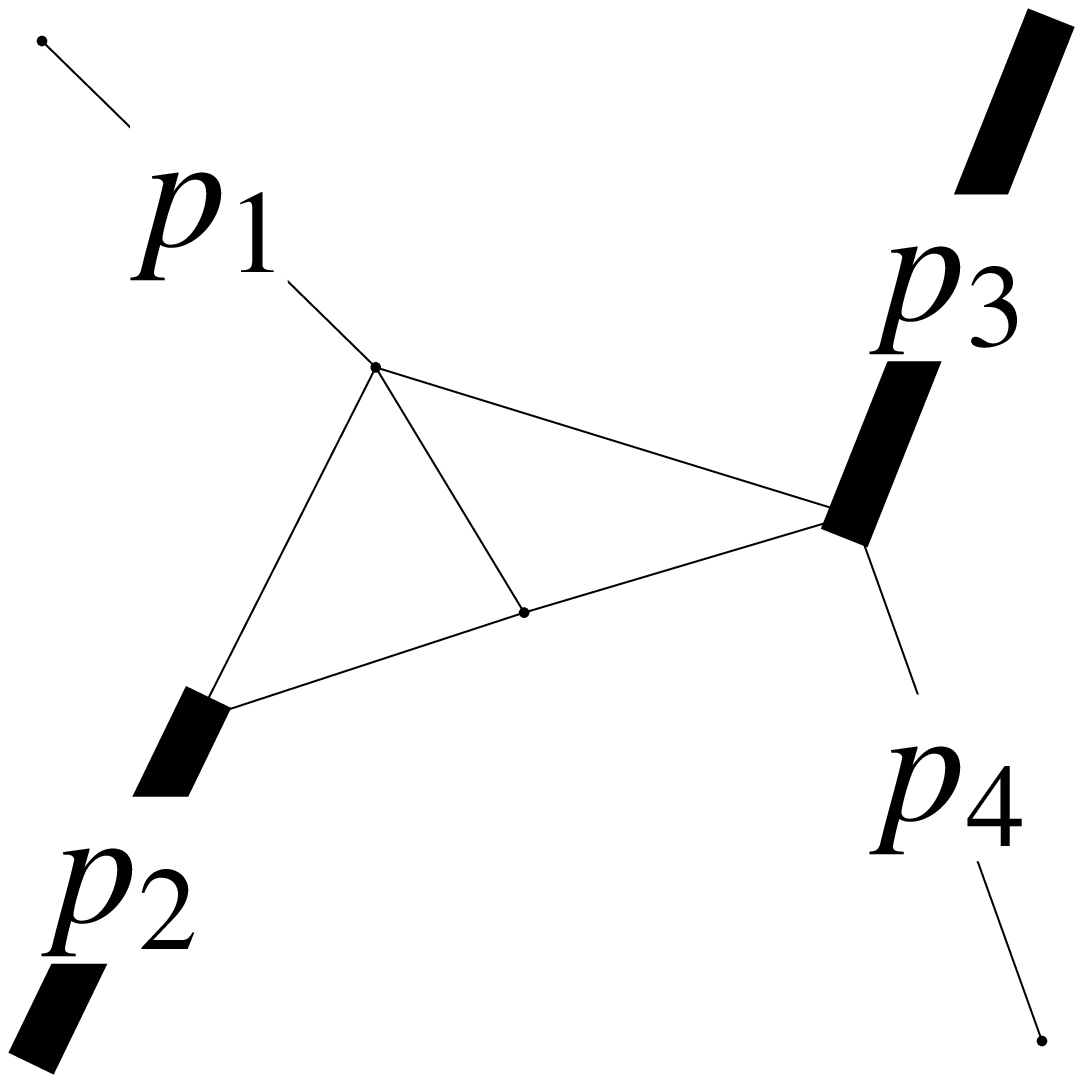}
{  \bea     
  g^{\rm P23}_{21} &=& 4 \eps^4 (p_2^2  - s) G_{0, 1, 1, 1, 1, 1, 0, 0, 0}
\,,~~~~~~~~~~~~~~~~~~~~~~~~~~~~~~~~~~~~~~~~~~~~~~~~~~~~
  \\
     f^{\rm P23}_{21} &\sim & 
x^{-2\ep}  \left (  \frac{ \pi^2 \ep^2 }{3} 
 + 2  \zeta_3 \ep^3   + \frac{ \pi^4 \ep^4 }{10} \right ) \nn
\\
&+&  x^{-4\ep}  \left ( \frac{ \pi^2 \ep^2 }{3} + 
     \left ( \frac{2 i \pi^3}{3} + 14 \zeta_3  \right ) \ep^3
+ 28 i  \pi \zeta_3 \ep^4  
\right ) \nn
\\
&+&  x^{-3\ep} \left ( -\frac{2  \pi^2 \ep^2 }{3} 
-  \left ( \frac{2i\pi^3}{3} + 16 \zeta_3 \right ) \ep^3
  -\left ( \frac{\pi^4}{6} + 16 i \pi \zeta_3 \right ) \ep^4 \right ),
  \nn 
\eea} \nn \\ 
\picturepage{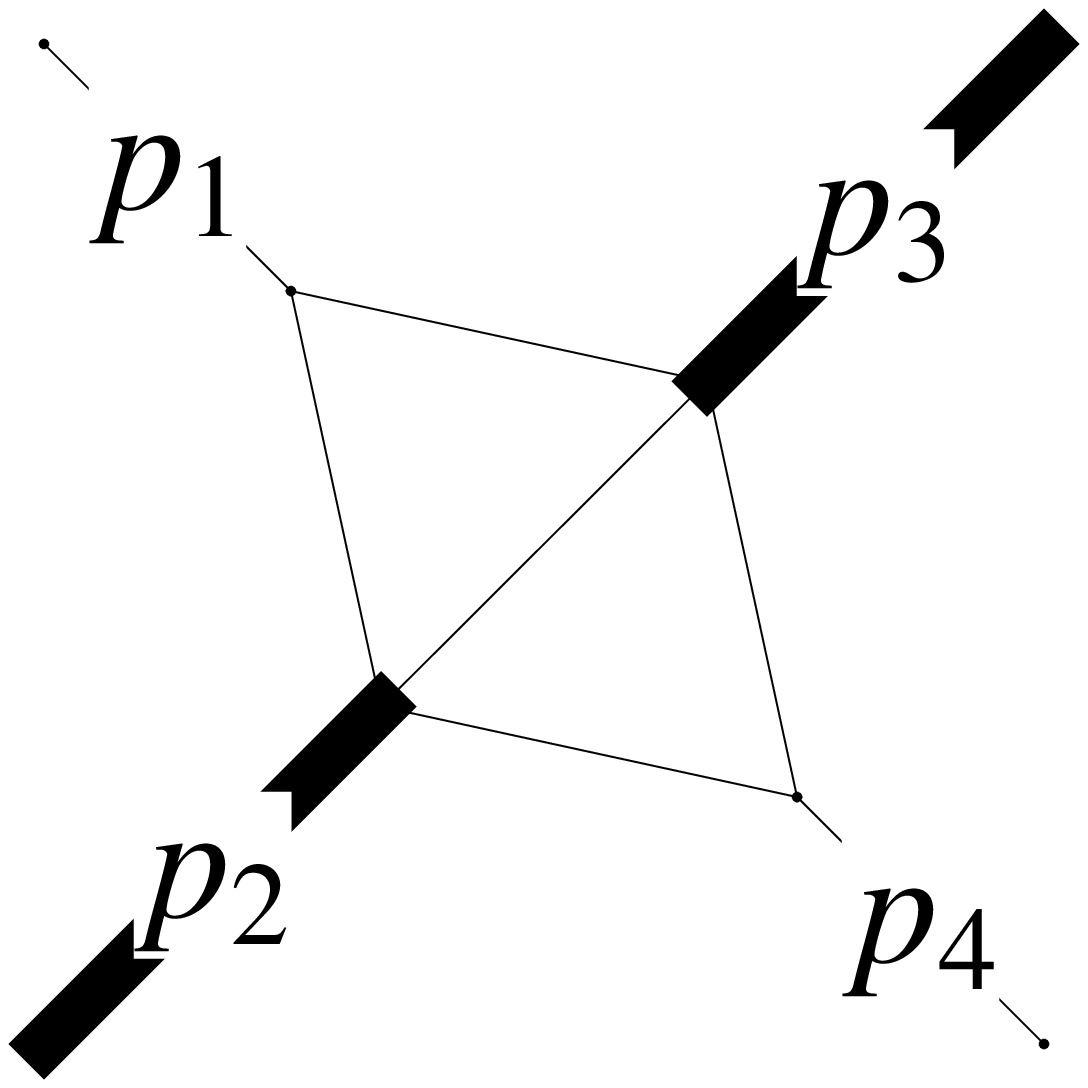}
{  \bea     
  g^{\rm P23}_{22} &=& -2 \eps^3 (p_2^2  p_3^2  - s t) G_{1, 0, 0, 1, 1, 2, 1, 0, 0}
\,,~~~~~~~~~~~~~~~~~~~~~~~~~~~~~~~~~~~~~~~~~~~~~~~~
  \\
     f^{\rm P23}_{22} &\sim & 12i\pi\ep x^{-4\ep} [ (z-y)^(1-z)]^{-2\ep},  \nn 
\eea} \nn \\
\picturepage{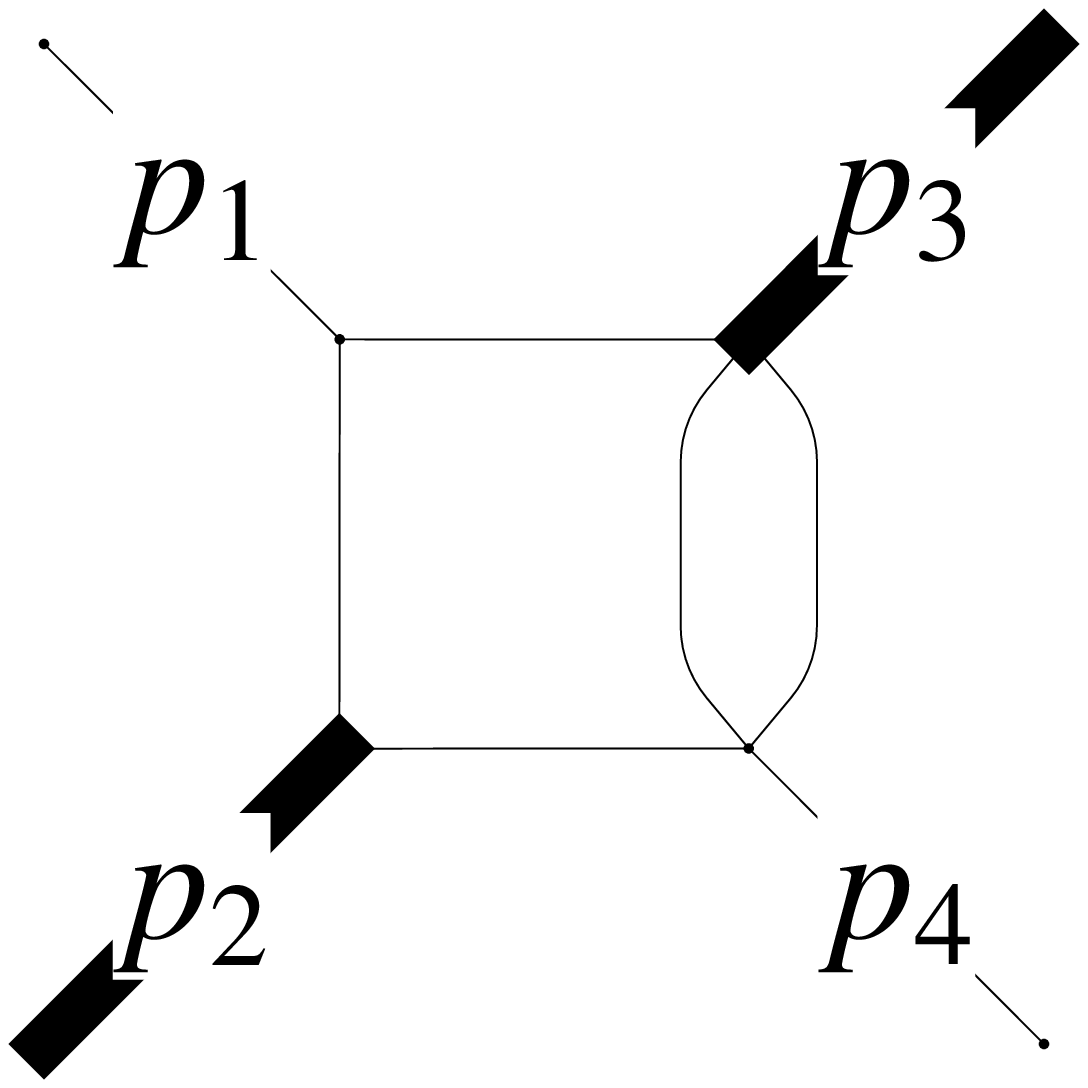}
{  \bea     
  g^{\rm P23}_{23} &=& -2 \eps^3 (p_2^2  p_3^2  - s t) G_{1, 1, 0, 0, 1, 1, 2, 0, 0}
\,,~~~~~~~~~~~~~~~~~~~~~~~~~~~~~~~~~~~~~~~~~~~~~~~
\\
     f^{\rm P23}_{23}  & \sim &  6i\pi\ep x^{-3\ep} [ (z-y) (1-z)]^{-2\ep}, \nn 
\eea} \nn \\
\picturepage{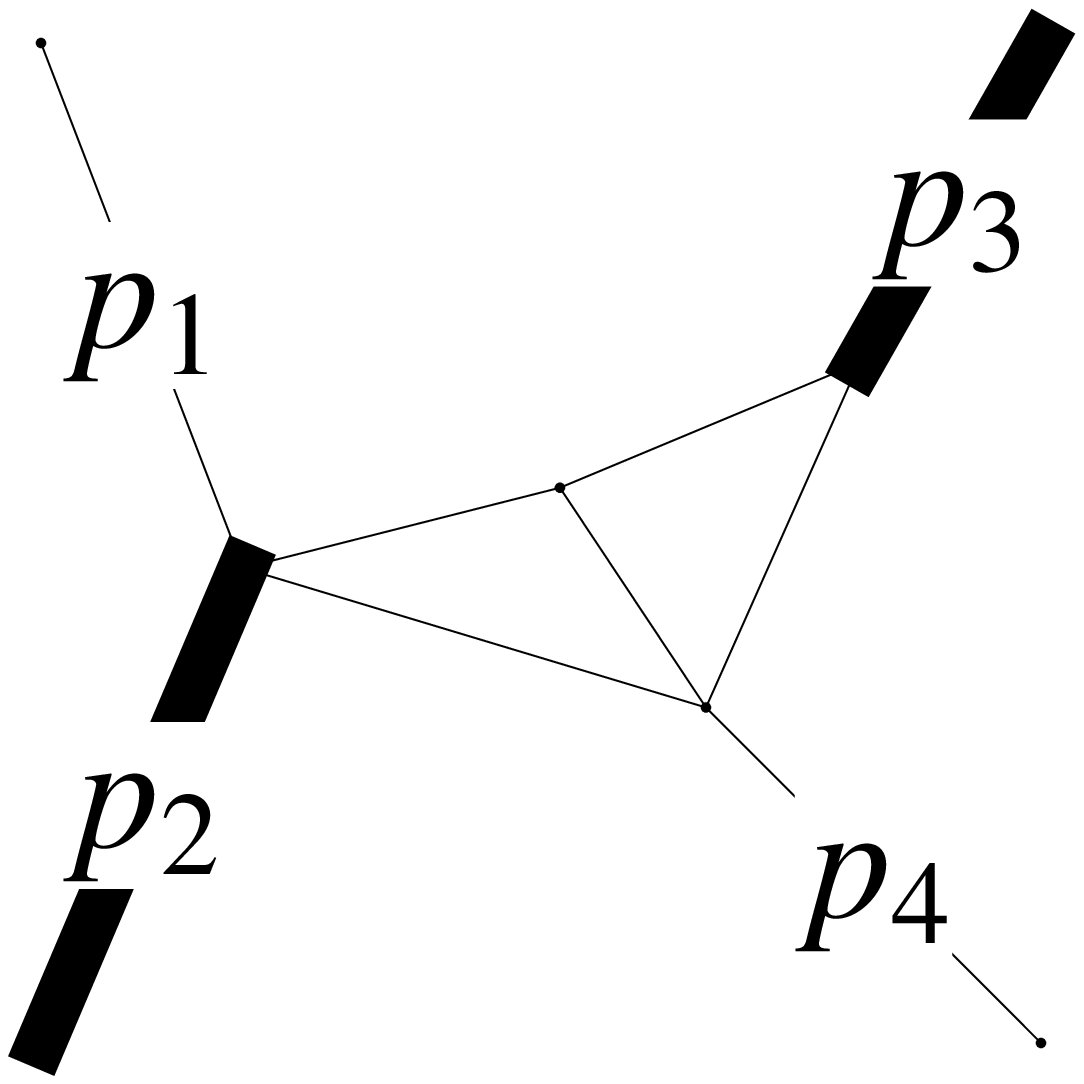}
{  \bea     
  g^{\rm P23}_{24} &=& 4 \eps^4 (p_3^2  - s) G_{1, 1, 1, 0, 0, 1, 1, 0, 0}
\,,~~~~~~~~~~~~~~~~~~~~~~~~~~~~~~~~~~~~~~~~~~~~~~~~~~~~~~~~
  \\
     f^{\rm P23}_{24} &\sim &  -\frac{ \pi^2 \ep^2 }{3} -  \left ( \frac{2i \pi^3}{3} + 2 \zeta_3 \right ) \ep^3 
  +  \left (  \frac{17 \pi^4}{30} - 4 i \pi \zeta_3 \right )  \ep^4 \nn
\\
&  -& x^{-2\ep} \left ( \frac{ \pi^2 \ep^2 }{3}  + 14  \zeta_3 \ep^3    + \frac{2  \pi^4 \ep^4}{3}  \right ) 
\nn
\\
&+& 
 x^{-\ep} \left (  \frac{2  \pi^2 \ep^2 }{3} +  \left (  \frac{2 i \pi^3}{3} + 16 \zeta_3 \right ) \ep^3 
 + \left ( \frac{\pi^4}{6} + 16 i \pi \zeta_3 \right ) \ep^4  \right ),
   \nn 
\eea} \nn \\
\picturepage{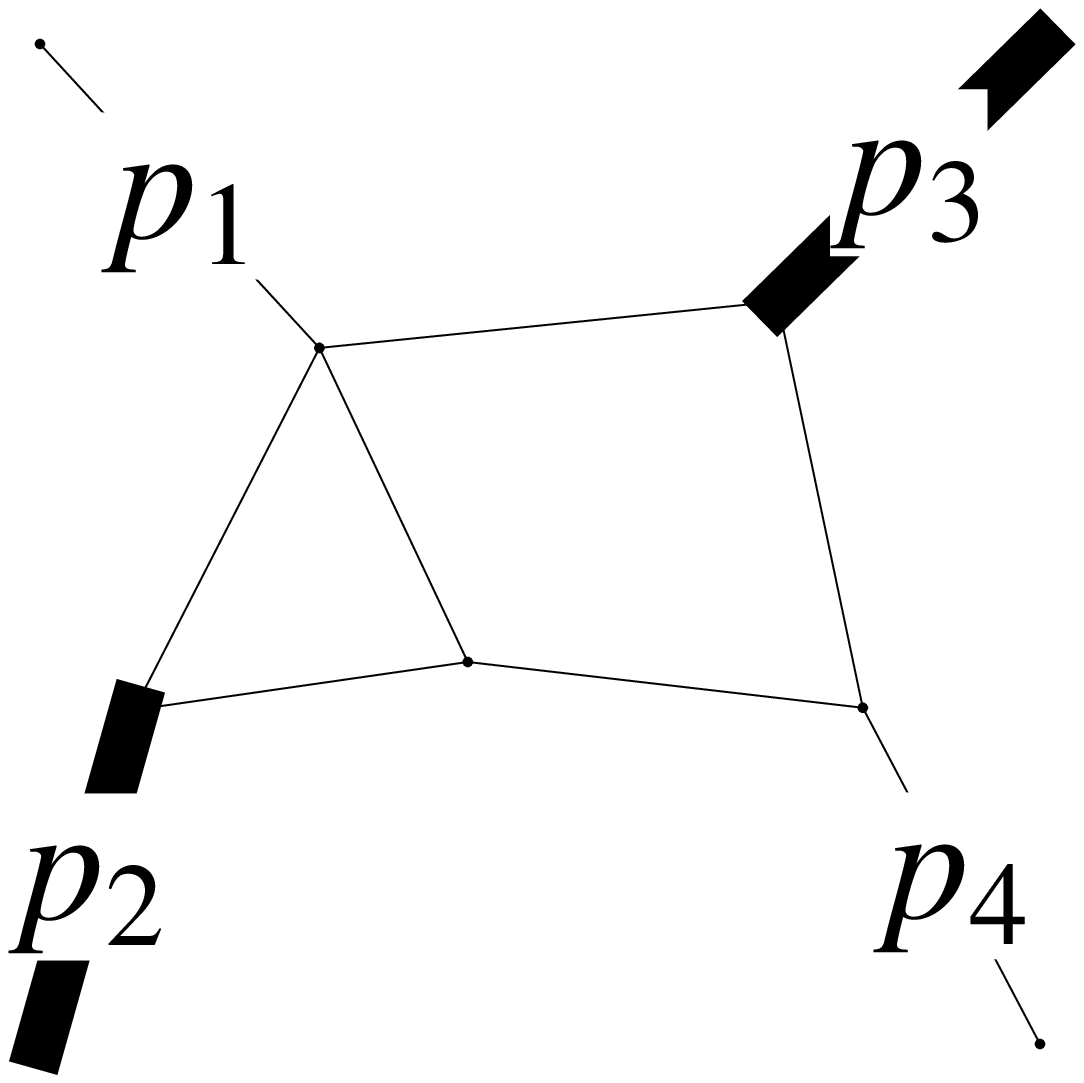}
{  \bea     
  g^{\rm P23}_{25} &=& 4 \eps^4 s (p_2^2  - t) G_{0, 1, 1, 1, 1, 1, 1, 0, 0}
\,,~~~~~~~~~~~~~~~~~~~~~~~~~~~~~~~~~~~~~\\
     f^{\rm P23}_{25} &\sim &  x^{-4\ep} \left ( - \frac{8i \pi \ep}{3} + 5\pi^2\ep^2 
    + \left ( \frac{34i\pi^3}{9} + 10\zeta_3 \right ) \ep^3 + 
     \left ( -\frac{67\pi^4}{45} + \frac{76i \pi \zeta_3}{3} \right )  \ep^4 \right )
\nn 
\\
&+& 4 i \pi \ep x^{-4\ep} [ (z-y) (1-z) ]^{-2\ep}  
   -\frac{4 i\pi \ep}{3} N_3  x^{-4\ep} [ (y-z) (1-z) ]^{-3\ep},   
 \nn 
\eea} \nn \\
\picturepage{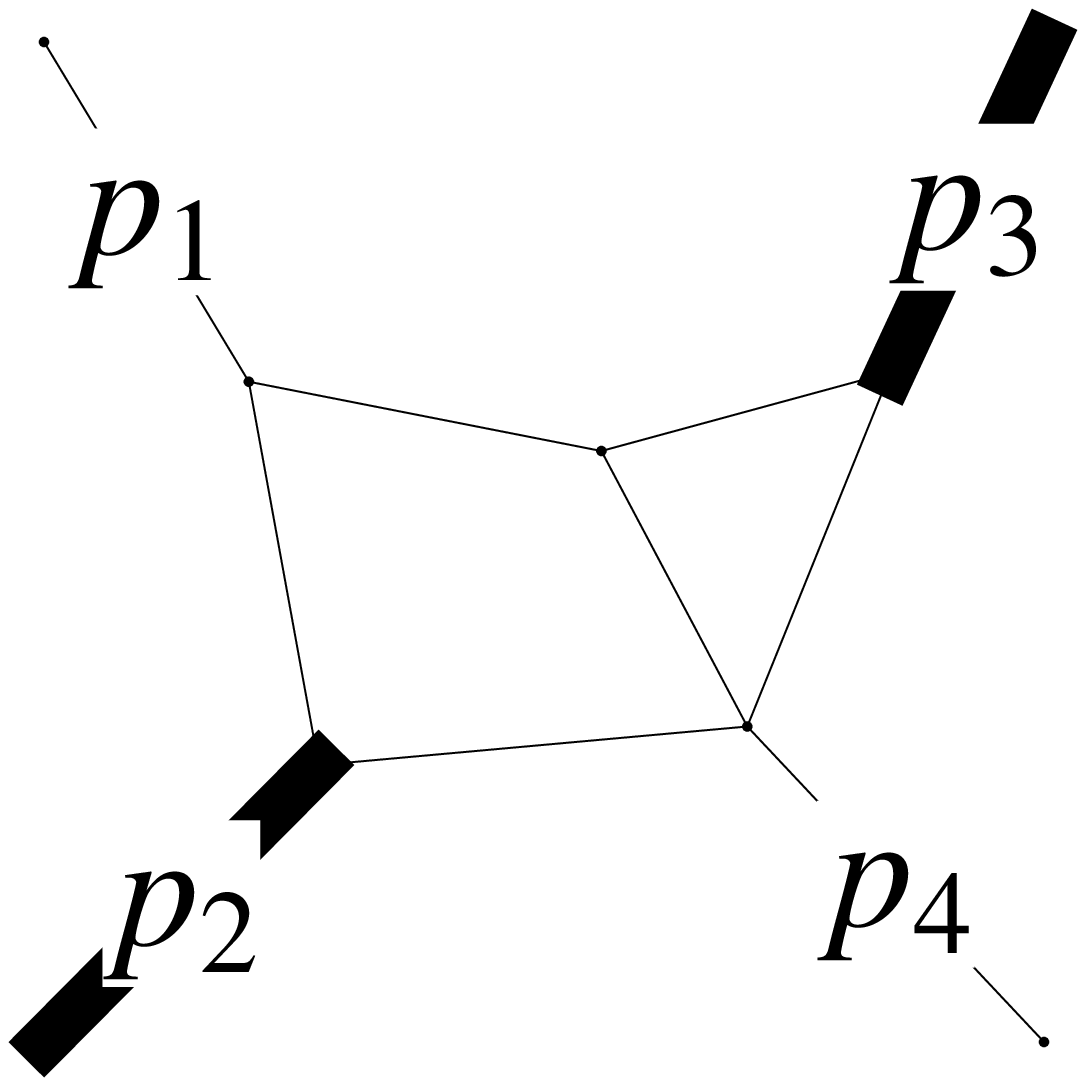}
{  \bea     
  g^{\rm P23}_{26} &=&  4 \eps^4 s (p_3^2  - t) G_{1, 1, 1, 0, 1, 1, 1, 0, 0}
\,,~~~~~~~~~~~~~~~~~~~~~~~~~~~~~~~~~~~~~~~~~~~~~~~~~
  \\
     f^{\rm P23}_{26} &\sim &  
x^{-4\ep}  \left ( -1 - \frac{2 i  \pi \ep }{3} 
 - \frac{ \pi^2 \ep^2}{3} + 
    \left (- \frac{2 i \pi^3}{9} + 2 \zeta_3 \right )  \ep^3
     + 
     \left (  \frac{7 \pi^4}{90} + \frac{4 i \pi  \zeta_3 }{3} \right ) \ep^4  \right ) \nn
\\
&+&  x^{-\ep} \left ( 2 + 2 i  \pi \ep  +  \left ( \frac{2 i \pi^3}{3} + 36 \zeta_3 \right ) \ep^3 + 
     \left (  \frac{8 \pi^4}{15} + 36 i \pi  \zeta_3 \right )  \ep^4 \right ) 
\nn
\\
& -&  x^{-2\ep} \left [ 5 + 8 i  \pi \ep  - \frac{23  \pi^2 \ep^2 }{3}
    - \left ( \frac{16 i \pi^3}{3} - 38 \zeta_3 \right ) \ep^3
    +  \left ( \frac{56 \pi^4}{15} + 48 i \pi \zeta_3 \right ) \ep^4 \right ]
\nn
\\
& +&  4 x^{-3\ep}  N_1  
+4 i \pi \ep x^{-3\ep} \left [ (y-z)(1-z) \right ]^{-2\ep}  
\nn \\
& -&    \frac{4 i \pi \ep }{3} N_3   x^{-4\ep} \left [ (z-y)(1-z) \right ]^{-3\ep},
 \nn 
\eea} \nn  \\
%
%
%\end{align} 
%\end{small}
%
%\begin{small}
%\begin{align}
\picturepage{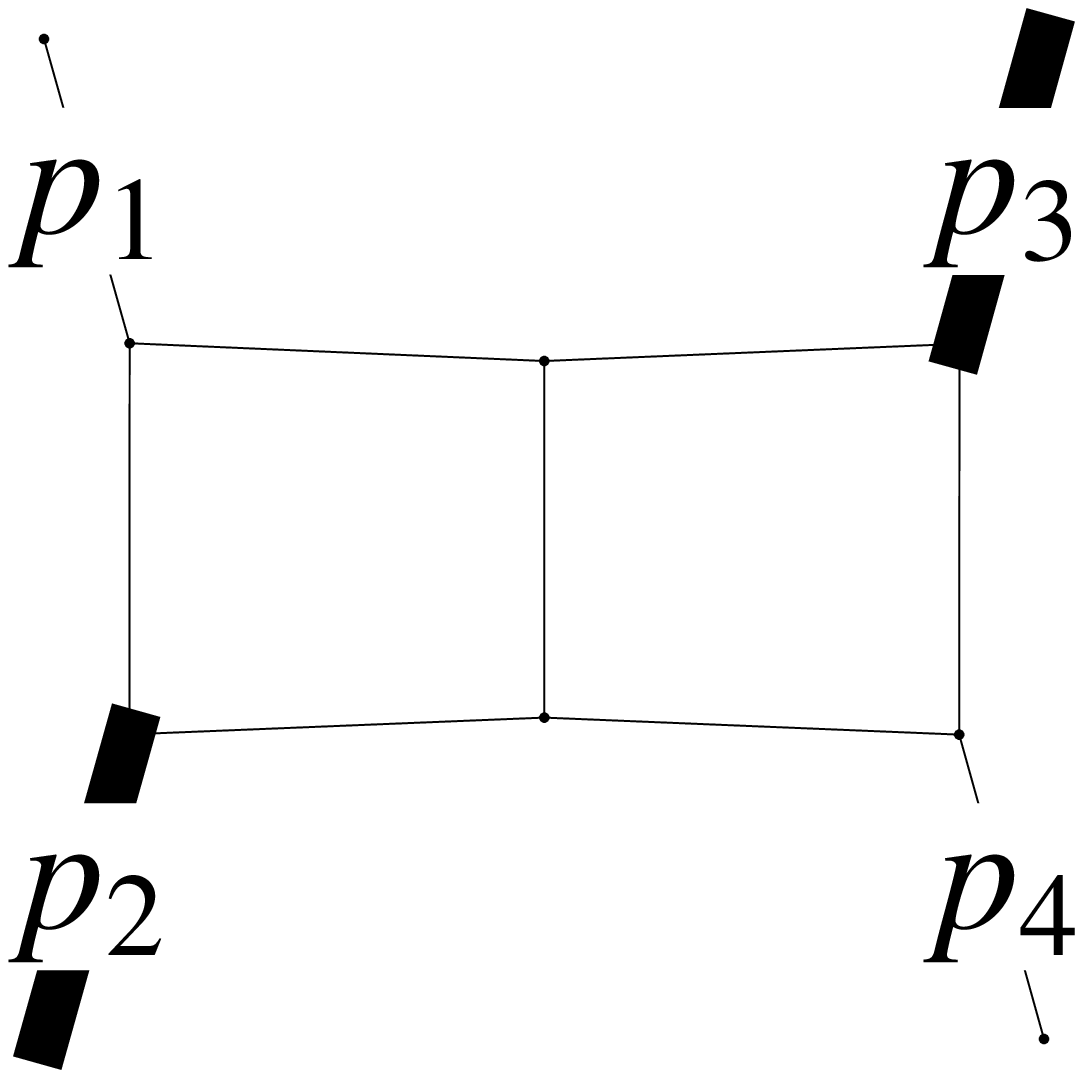}
{  \bea     
  g^{\rm P23}_{27} &=& -4 \eps^4 s (-p_2^2  p_3^2  + s t) G_{1, 1, 1, 1, 1, 1, 1, 0, 0}
\,,~~~~~~~~~~~~~~~~~~~~~~~~~~~~~~~~~~~~~~~~~~~~~~~~~~~~
  \\
     f^{\rm P23}_{27} &\sim & 
-8i \pi \ep \left ( 1-3i\pi \ep  + \frac{\pi^2 \ep^2 }{2} 
-15\zeta(3) \ep^3 \right )((z-y)^{-2\ep}(1-z)^{-2\ep} x^{-4\ep}  \nn
\\
&+&   24 i \pi \ep^2 x^{-4\ep} \left [ (z-y) (1-z) \right ]^{-2\ep} \ln((z-y)(1-z)) \nn
\\
 &+&  8i\pi \ep N_3 x^{-4\ep} \left [ (z-y) (1-z) \right ]^{-3\ep},  \nn 
\eea} \nn \\
\nn \\
\picturepage{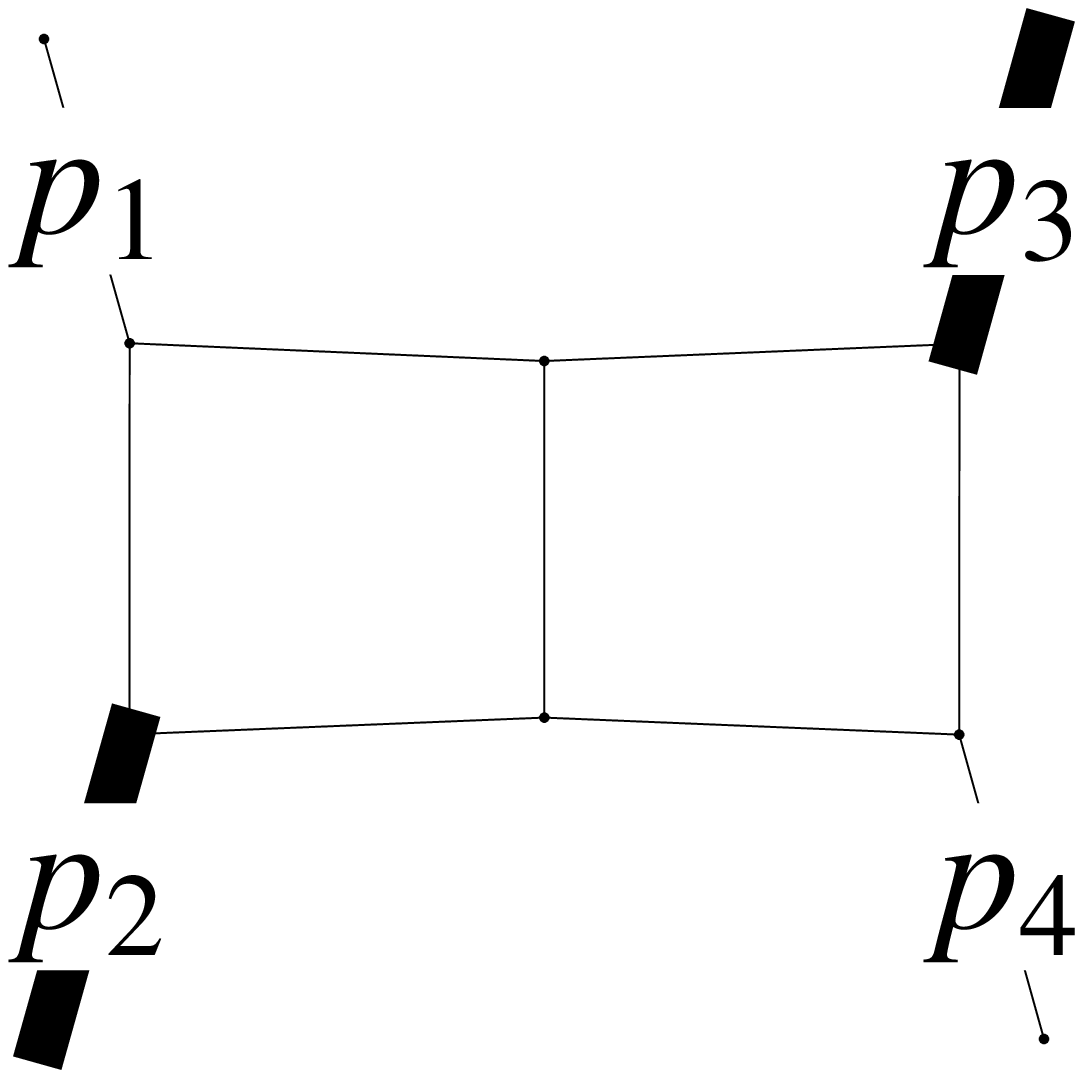}
{  \bea     
  g^{\rm P23}_{28} &=&  \eps^2 \Big(  2 \eps p_2^2  (p_3^2  - s) G_{1, 0, 0, 1, 1, 2, 1, 0, 0} 
 - 4 \eps p_2^2  (p_3^2  - s) G_{1, 1, 0, 0, 1, 1, 2, 0, 0} 
\\
&+&  4 \eps^2 s (-p_3^2  + s) G_{1, 1, 1, 1, 1, 1, 1, -1, 0} \Big ) ,
\nn
  \\
     f^{\rm P23}_{28} &\sim & 
x^{-4\ep}  \left ( 2 + \frac{8 i \pi  \ep }{3} - \frac{4  \pi^2 \ep^2 }{3} 
    +   \left (  - \frac{4 i \pi^3}{9} + 12 \zeta_3 \right ) \ep^3 + 
     \left ( \frac{53 \pi^4}{90} + \frac{80 i \pi \zeta_3}{3} \right )  \ep^4 \right ) 
\nn
\\
&+& 
 x^{-\ep} \left ( -2 - 2 i  \pi \ep   +  \left ( -\frac{ 2 i \pi^3}{3} - 36 \zeta_3 \right ) \ep^3
  +  \left (  - \frac{8 \pi^4}{15} - 36 i \pi \zeta_3 \right )  \ep^4 \right ) 
\nn
\\
& + & x^{-3\ep}  \left ( -6 - 6 i  \pi \ep   + \frac{8  \pi^2 \ep^2 }{3} 
  + 
     \left (  \frac{2 i \pi^3}{3} - 44 \zeta_3 \right ) \ep^3 + 
     \left (  - \frac{14 \pi^4}{15} - 44 i \pi \zeta_3 \right ) \ep^4  \right )  
\nn
\\
&  + & x^{-2\ep} \left ( 6 + 8 i  \pi \ep  - \frac{20  \pi^2 \ep^2 }{3} 
  +  \left (  - \frac{16 i \pi^3}{3} + 68 \zeta_3 \right ) \ep^3 + 
     \left (  \frac{133 \pi^4}{30} + 48 i \pi \zeta_3 \right ) \ep^4 \right )
\nn
\\
& + & \frac{4 i \pi \ep}{3}N_3 \left [ (z-y)(1-z) \right ]^{-3\ep} x^{-4\ep} 
   -12 i \pi \ep x^{-4\ep} [(z-y)(1-z)]^{-2\ep}  
\nn
\\
& -& \frac{8 i \pi \ep}{3} (1-3i\pi \ep  + \frac{\ep^2\pi^2}{2} 
- 15 \zeta_3 \ep^3 ) x^{-4\ep}[(z-y)(1-z)]^{-2\ep} 
\nn
\\
&+&  8 i \pi \ep x^{-3\ep}[(z-y)(1-z)]^{-2\ep}     
 + \frac{8 i \pi \ep }{3} x^{-4\ep}[(z-y)(1-z)]^{-2\ep} 
\nn
\\
&+& 
 8 i \pi \ep^2 x^{-4\ep}[(z-y)(1-z)]^{-2\ep} \ln((y-z)(1-z)).  \nn 
\eea} \nn 
\end{align} 
\end{small}

\endgroup

\section{Checks of the results} 
\label{sec:checks}

In this Section, we describe some checks  of our results. We begin by making  a few nearly self-evident
comments. First, we emphasize  that {\it all} the integrals are computed using one and the same 
method. While this, obviously, does not guarantee that results are correct, it reduces the number 
of issues that can appear if every integral is computed with a new technique.  Second, we stress 
that, once  the choice of master integrals is made and suitable variables are found, the 
integration procedure is  straightforward and can be thoroughly  checked by differentiating
 the obtained  result to ensure that it satisfies the  original differential equations in $x,y,z$ variables. 
Unfortunately, this procedure does not check the boundary conditions which, therefore have to be checked 
in some other way.

As we already mentioned  in the Introduction, when we require external masses to be equal
$M_3^2 = M_4^2 =  M^2$, we  obtain a class of integrals considered recently in Ref.~\cite{tancredi}.
Using the results for the integrals appended to the arXiv submission of Ref.~\cite{tancredi}, 
we have  compared numerical values for a large number  of integrals that we compute in this 
paper with integrals computed 
in Ref.~\cite{tancredi}, finding perfect agreement.\footnote{
For integral $g_{16}^{P23}$, which corresponds to the integral $I_{213,1}^{(B)}$
of Ref.~\cite{tancredi}, $\zeta$ in that reference should be $\zeta + i0$. We thank L.~Tancredi for clarifying 
this point to us.}
As another check, we have computed some of our integrals
% $g_{16}^{P23}$ 
{\it numerically} using the new version 
of the program FIESTA \cite{Smirnov:2013eza}, 
that is capable of calculating  Feynman integrals in the physical region. 
A perfect agreement with our analytic result is  found for a few randomly selected $(x,y,z)$ points. 

Finally, a procedure of analytic continuation discussed in Sec.~\ref{sec:analytic} can also be used 
to independently construct solutions in the physical region for integrals  of the P23 family. 
As we  explained there,  that procedure 
can also be implemented by means of  numerical integration 
over contour in the complex plane starting from a point in unphysical region where the boundary 
conditions are simple. 
We have checked that, for a randomly selected point, this procedure gives results for master integrals  
of family~P23 that are in agreement with our analytic solutions. 

\section{Conclusions} 
\label{sec:concl}

In this paper we reported on the computation of all two-loop planar master  integrals 
that are required to describe production of two {\it  off-shell} vector bosons in hadron collisions. 
We constructed the differential equations for the carefully-chosen 
basis of master integrals following the 
strategy suggested in  Ref.~\cite{jhenn}. We have computed boundary conditions for these 
integrals in the physical region and integrated them to 
obtain analytic results in  terms of Goncharov polylogarithms. The results are fairly  
large. We note, however, that  we did not 
try to simplify these results although such simplifications should be possible. 
Probably the most compact and flexible form can be achieved in terms
of Chen iterated integrals, at the cost of giving up the feature
of a linear parametrization. The matrices $\tilde{A}$ specifying them
are included in the arXiv submission, as well as files with results for the integrals
in terms of Goncharov polylogarithms.

The method for calculating multi-loop master integrals  suggested in Ref.~\cite{jhenn} 
appears to be quite promising. We look forward to  its application to even more complicated 
two-loop integrals and, in particular, to the non-planar ones required for the complete description of the 
off-shell production of two vector bosons at the LHC.

\section*{Acknowledgments} 

K.M. would like to thank Fabrizio Caola for many useful conversations. 
J.M.H. wishes to thank the organizers of RADCOR 2013, where a preliminary version of these results was presented, for their invitation. J.M.H. is supported in part by
the DOE grant DE-SC0009988 and by the Marvin L. Goldberger fund.
The work of K.M.  is partially supported by US NSF under grants PHY-1214000 
and by Karlsruhe Institute of Technology through its distinguished 
researcher fellowship program.  
The work of V.S. was supported by the Alexander von Humboldt Foundation (Humboldt Forschungspreis).
We are grateful to the Institute for Theoretical Particle Physics (TTP) at Karlsruhe Institute of Technology 
where some of the results were obtained.

\end{document}